\documentclass[11pt]{article}
\pdfoutput=1 
\usepackage{amssymb}
\usepackage{amsmath}
\usepackage{amstext}
\usepackage{graphicx,epsfig}
\usepackage{epsfig}
\usepackage{verbatim} 
\usepackage{fancybox}
\usepackage{color}
\usepackage{ulem}
\usepackage{enumitem}
\usepackage{youngtab}
\usepackage{subfigure}
\usepackage{bbm}
\usepackage{parskip}
\usepackage{slashed}
\usepackage[numbers,compress]{natbib}
\usepackage{ytableau}
\usepackage[utf8]{inputenc}
\usepackage{cancel}

\newcommand{\Comment}[1]{{}}
\definecolor{darkblue}{rgb}{0.15,0.35,0.55}
\definecolor{reddish}{rgb}{0.65, 0.2, 0.2}
\usepackage[linktocpage=true]{hyperref}
\hypersetup{
colorlinks=true,
citecolor=darkblue,
linkcolor=reddish,
urlcolor=darkblue,
pdfauthor={},
pdftitle={},
pdfsubject={}
}

\definecolor{green3}{RGB}{44, 160, 44}

\newcommand{\be}{\begin{equation}}
\newcommand{\ee}{\end{equation}}
\newcommand{\bea}{\begin{eqnarray}}
\newcommand{\eea}{\end{eqnarray}}
\newcommand{\beas}{\begin{eqnarray*}}
\newcommand{\eeas}{\end{eqnarray*}}

\newcommand{\vev}[1]{\left\langle #1 \right\rangle}

\definecolor{darkred}{rgb}{0.7,0.3,0.3}
\definecolor{darkgreen}{rgb}{0.2,0.7,0.3}
\definecolor{lightgreen}{rgb}{.816,.94,.753}
\definecolor{greyish}{rgb}{.8,.8,.8}
\definecolor{darkblue2}{rgb}{0.3,0.4,0.9}
\usepackage{pifont}% http://ctan.org/pkg/pifont
\usepackage{xcolor,colortbl}

\def\({\left(}
\def\){\right)}

\newcommand{\cc}{c_{0}}

\def\gsim{ \lower .75ex \hbox{$\sim$} \llap{\raise .27ex \hbox{$>$}} }
\def\lsim{ \lower .75ex \hbox{$\sim$} \llap{\raise .27ex \hbox{$<$}} }

\def\xyma{\xymatrix@M.7em}
\def\xymas{\xymatrix@M.1em}

\numberwithin{equation}{section}

\begin{document}

\vspace*{1cm}
\begin{center}
{\Huge \bf Axions in String Theory} \\
\vspace*{0.3cm}
{\huge \bf --} \\ 
\vspace*{0.3cm}
{\Huge \bf Slaying the Hydra of Dark Radiation}
\end{center} 

\vspace*{1.5cm}
\thispagestyle{empty}
\centerline{\Large 
Michele Cicoli$^{1,2}$, Arthur Hebecker$^{3}$, Joerg Jaeckel$^{3}$ and Manuel Wittner$^{3}$
}

\vspace{0.5cm}

\begin{center}
\small{$^{1}$\it Dipartimento di Fisica e Astronomia, Universit\`a di Bologna,} \\
\small{\it via Irnerio 46, 40126 Bologna, Italy}\\
\small{$^{2}$ \it INFN, Sezione di Bologna, viale Berti Pichat 6/2, 40127 Bologna, Italy}\\
\small{$^{3}$\it Institut f\"ur theoretische Physik, Universit\"at Heidelberg,}
\\
\small{\it Philosophenweg 16+19, 69120 Heidelberg, Germany}\\[0.5ex]  

\vspace{.25cm}
{ \textit{E-Mail:} michele.cicoli@unibo.it, a.hebecker@thphys.uni-heidelberg.de, jjaeckel@thphys.uni-heidelberg.de, }\\  wittner@thphys.uni-heidelberg.de
 \end{center}

\vspace*{1.2cm}
\begin{abstract}
\noindent
It is widely believed that string theory easily allows for a QCD axion in the cosmologically favoured mass range. The required small decay constant, $f_a\ll M_P$, can be implemented by using a large compactification volume. This points to the Large Volume Scenario which in turn makes certain cosmological predictions: First, the closed string axion behaves similarly to a field-theoretic axion in the pre-inflationary scenario, i.e.~the initial value can be tuned but one is constrained by isocurvature fluctuations. In addition, the volume represents a long-lived modulus that may lead to an early matter-dominated phase. Finally, the decay of the volume modulus to its own axion tends to overproduce dark radiation. In this paper we aim to carefully analyze the cosmology by studying models that not only allow for a QCD axion but also include inflation. Quite generally, limits on isocurvature fluctuations restrict us to relatively low-scale inflation, which in the present stringy context points to K\"ahler moduli inflation. As a novel feature we find that the lightest (volume) modulus couples strongly to the Higgs. It hence quickly decays to the SM, thus resolving the original dark radiation problem. This decay is much faster than that of the inflaton, implying that reheating is determined by the inflaton decay. The inflaton could potentially reintroduce a dark radiation problem since it decays to lighter moduli and their axions with equal rates. However, due its mixing with the QCD-saxion, the inflaton has also a direct decay rate to the SM, enhanced by the number of SM gauge bosons. This results in an amount of dark radiation that is consistent with present limits but potentially detectable in future measurements.
\end{abstract}

\newpage

\tableofcontents 

\vspace*{1cm}

\section{Introduction}

The strong CP problem and its possible resolution by a field-theoretic axion~\cite{Peccei:1977hh,Peccei:1977ur,Weinberg:1977ma,Wilczek:1977pj} have become textbook material (e.g.~\cite{Weinberg:1996kr}). Here, `field-theoretic' refers to axions arising as the angular component of a complex scalar field with a spontaneously broken $U(1)$ symmetry of sufficiently high quality and featuring the right couplings to the Standard Model (SM).

Apart from the quality issue, one may question to which extent trading a naturalness problem for an additional layer of model building represents progress. If, instead, one takes the promise of string theory to supply a UV completion to our field theory models seriously, things look different. Here, gauge couplings are necessarily replaced by vacuum expectation values of complex moduli and their imaginary parts (usually originating from 10d $p$-form potentials) automatically supply the desired high-quality axions. Getting the phenomenology right used to be difficult~\cite{Svrcek:2006yi}, but with the advent of the type IIB landscape~\cite{Giddings:2001yu, Kachru:2003aw, Denef:2008wq, Hebecker:2021egx} and the Large Volume Scenario (LVS)~\cite{Balasubramanian:2005zx, Conlon:2005ki}, it has become plausible that a realistic QCD axion is a natural feature of a broad class of string compactifications~\cite{Conlon:2006tq, Conlon:2006ur, Arvanitaki:2009fg, Cicoli:2012sz, Marsh:2015xka, Acharya:2015zfk,Visinelli:2018utg, Broeckel:2021dpz, Demirtas:2021gsq} \footnote{In addition to the proper QCD axion there could be a plethora of stringy ``axions'', often dubbed the string axiverse~\cite{Arvanitaki:2009fg}, that do not necessarily solve the strong CP problem but could form fuzzy dark matter \cite{Hui:2016ltb,Cicoli:2021gss}. For a recent exploration in connection with by means of black hole superradiance~\cite{Arvanitaki:2010sy} see, e.g.~\cite{Mehta:2020kwu,Mehta:2021pwf}.}. 

However, having a more complete theory also has important implications for cosmology.
A by now well-known example is the prediction of a significant amount of dark radiation~\cite{Cicoli:2012aq, 
Higaki:2012ar, Hebecker:2014gka, Angus:2014bia, Allahverdi:2014ppa, Cicoli:2015bpq, Cicoli:2018cgu, Acharya:2019pas, Jaeckel:2021gah, Angus:2021jpr, Jeong:2021yol, Frey:2021jyo, Cicoli:2021tzt, Jaeckel:2021ert}. In the present paper we aim for a consistent cosmological picture by facing the challenge of dark radiation in combination with that of a stringy QCD axion. To do this we try to build a model that includes an inflationary sector as well as a QCD axion. Moreover, we also study interactions with the Higgs sector of the SM. This allows to coherently investigate constraints from cosmology and the interplay of the different sectors. The outcome, somewhat unexpected to us, is that the original dark radiation problem caused by a relatively long-lived volume modulus decaying into axions can be avoided by taking into account natural interactions with the SM Higgs. This rests on the key observation that the Higgs mass is small by fine-tuning. As a result, the Higgs mass term in the lagrangian represents a much stronger portal to the moduli sector than the low Higgs mass scale naively suggests. However, one may worry that dark radiation overproduction will reappear when including the inflaton and its decays. We show that this is not the case. The main inflaton decay channel is into SM gauge bosons thanks to the mixing between the inflaton and the QCD-saxion, and the associated enhancement by the number of SM gauge bosons. Our results, therefore, represent an important step forward in the attempt to realize a `natural' QCD axion in a fully working (string) cosmology.

In practical terms our paper follows a somewhat bottom-up strategy by formulating requirements for our model and then adding ingredients to fulfill these.
Drawing a very rough outline we pursue the following points.
\begin{itemize}
\item We recall that in type II string models and especially in the LVS, an appropriately small axion decay constant requires a very large volume:  $f_a \sim M_P/\mathcal{V}^{-1/2}$ (cf.~Section~\ref{sec:scaling} and Appendix~\ref{appendix_f_a_expression}).
\item We argue that at least in simple string scenarios one expects a so-called pre-inflationary cosmology, where the initial value of the axion field is selected during inflation, axionic cosmic strings are absent, but we have to beware of isocurvature constraints (cf.~Section~\ref{sec:scaling}).
\item Using the LVS expression for $f_a$, as well as dark matter (DM) and isocurvature bounds, we arrive at the result that the scale of inflation must be relatively low, irrespective of whether the axion begins to oscillate during radiation or matter domination (cf.~Section~\ref{sec:cosmo}).
\item Usually it is assumed that an exponentially large volume leads to a long-lived volume modulus that is non-relativistic and comes to dominate the Universe. Its decay into axions causes a significant amount of dark radiation. However, we argue that natural couplings to the Higgs lead to two effects: ($i$) the volume mode decays primarily into the SM, and ($ii$) its decay into Higgses is so fast that no phase of matter-domination from the volume modulus occurs. This drastically reduces the amount of dark radiation produced by the decay of the volume mode, thereby solving the usual dark radiation problem (cf.~Section~\ref{sec:dr1}).
\item As a next step we include inflation. Since the cosmological constraints enforce a relatively low scale of inflation, we pick a suitable specific model, K\"ahler moduli inflation \cite{Conlon:2005jm}, and try to create a realistic scenario (cf.~Section~\ref{sec:inflationmodel}).
\item To realize the SM on D-branes with a light QCD axion, we assume a combination of (at least) two 4-cycles wrapped by an appropriate D7-brane configuration. Following \cite{Cicoli:2012sz,Broeckel:2021dpz,Cicoli:2011qg}, all but one of the corresponding cycle moduli are stabilised via D-terms, the remaining one by loops. Effectively, one may then think of a loop-stabilized cycle supporting the SM brane stack and the light axion (cf.~Section~\ref{sec:inflationmodel}).

\item As the decay of the volume modulus (into Higgses) is now quite fast, the inflaton itself (and its axion) take over the role of the longest-lived particles. We therefore have to carefully check whether their decay to light axions is problematic. Fortunately this is not the case, as the corresponding branching ratio is naturally suppressed by the number of SM gauge degrees of freedom. The resulting amount of dark radiation is consistent with present observations of the CMB and structure formation but could lead to interesting signatures in future observations and may potentially also be detectable with future earthbound axion experiments (cf.~Sections~\ref{sec:dr2}, \ref{sec:cosmofinal} and Appendices~\ref{dtl}, \ref{sec:2_moduli_system}, \ref{appendix_field_dynamics} and~\ref{app:parameters}).
\end{itemize}

\subsection*{Note on conventions}
For easy reference let us specify the conventions used in the following: 

In the main text as well as in the Appendices~\ref{appendix_f_a_expression}, \ref{dtl} and~\ref{app:parameters}, we explicitly spell out the reduced Planck mass $M_P=2.4\times10^{18}\,{\rm GeV}$, whereas in the Appendices~\ref{sec:2_moduli_system} and~\ref{appendix_field_dynamics} we set $M_P=1$ for brevity.

The (large) Calabi-Yau (CY) volume in the 10d string frame, ${\mathcal{V}}_{s}$, and in the 10d Einstein frame, $\mathcal{V}$, are measured in units of the sixth power of the string length $l_s=2\pi\sqrt{\alpha'}$. They are therefore dimensionless. The two volumes are related by
\begin{equation}
    {\mathcal{V}}=\mathcal{V}_{s}/g_{s}^{3/2}\,.
\end{equation}
Moreover,the string length and the Planck scale are related by
\begin{equation}
    M^{2}_{P}=\frac{4\pi {\mathcal{V}_s}}{g_{s}^2l^{2}_{s}}
    \,.
\end{equation}
For the dimensionless K\"ahler moduli and their axions, as they are conventionally used in type IIB model building, we write $\tau$ and $c$ respectively. The corresponding canonically normalized fields with mass dimension 1 are denoted by $\phi$ and $a$. Furthermore, basic equations and conventions regarding the LVS can be found in Section~\ref{sec:basic_definitions}.

\section{Requirements for and first consequences of stringy QCD axions}
\label{sec:scaling}

\subsection{Parametrically small $f_a$}

We require $f_a \ll M_P$ for at least two reasons: First, to avoid DM overproduction (see Section~\ref{sec:cosmo} for details). Second, to protect the quality of our axion, which would otherwise be endangered by non-QCD-related corrections to the potential,  $\Delta V \sim \text{exp}(-M_P/f_a)$ \cite{Demirtas:2021gsq,Rey:1989mg,Kamionkowski:1992mf,Holman:1992us,Kallosh:1995hi,Banks:2003sx,Alonso:2017avz,Hebecker:2018ofv}. The first constraint is typically more severe, demanding $f_a\lesssim 10^{13}\,$GeV. This will be discussed in more detail later on. For now, we only want to note that the most straightforward way for creating the required hierarchy between $f_a$ and $M_P$ is to employ a large compactification volume \cite{Conlon:2006tq} (see also~\cite{Conlon:2006ur, Arvanitaki:2009fg, Cicoli:2012sz, Marsh:2015xka, Acharya:2015zfk, Visinelli:2018utg, Broeckel:2021dpz, Demirtas:2021gsq}). We will argue for this below under the assumption that our axion is the imaginary part of the gauge coupling modulus. This is the standard case in superstring compactifications and such closed string axions are, at the fundamental level, $p$-form axions.

An alternative approach to realizing small $f_a$ would be to have the SM D-brane cycles sit in a strongly warped region \cite{Cascales:2003wn, Cascales:2005rj, Franco:2005fd, Dasgupta:2008hb}. If these cycles are singular, the gauge coupling is governed by the axio-dilaton. The latter is in general stabilized by fluxes, removing the axion. Hence, one needs geometries where the relevant cycles with appropriate brane stacks are stabilized in the geometric regime at the bottom of a warped throat. This is conceivable and an interesting direction for further research. However, to the best of our knowledge, this possibility has not been demonstrated so far and we will not pursue this path in the present paper.

Potentially, another alternative to large volume compactifications with branes is provided by the heterotic string. While achieving small $f_a$ is in general difficult in this context \cite{Svrcek:2006yi}, heterotic M-theory at strong warping may provide a way out. The reader may consult e.g.~\cite{Dasgupta:2008hb, Buchbinder:2014qca, Im:2019cnl} for further progress. Nevertheless, the heterotic path to small $f_a$ remains non-trivial.

Obviously, a field-theoretic axion realized as the phase of a 4d charged field or fermion condensate can also arise in string theory. In this case, there are no stringy obstructions to small $f_a$. For example ref. \cite{Cicoli:2013cha} presented an explicit CY model where the QCD axion is the phase of a charged matter field on D3-branes at singularities which features an intermediate scale $f_a$ by a combination of D-term stabilization and sequestered SUSY breaking. We will not pursue such `open string' axions (see also \cite{Berenstein:2012eg, Honecker:2013mya, Honecker:2015ela}) in the present paper. Apart from the familiar axion-quality issues, there is nothing wrong with such constructions as a matter of principle. However, our goal here is to explore the concept of an `intrinsically stringy' or `fundamental' axion, where the shift symmetry is by definition perturbatively exact.

After these preliminaries let us now explain why we are forced into the large volume regime in our approach \cite{Conlon:2006tq}: Consider a large CY with a small $p$-cycle on which a (spacetime-filling) $D(p+3)$ brane stack is wrapped. Let this brane stack be responsible for QCD and let the Ramond-Ramond (RR) $p$-form $C_p$, integrated over our $p$-cycle, provide the corresponding axion. Then, assuming a weak but not too small string coupling, $g_s\sim 0.1$, and that our $p$-cycle is not much larger than string length, one expects that $f_a\sim 1/l_s$ on dimensional grounds. At the same time, the 4d Planck mass scales as $M_P^2\sim {\cal V}_s/l_s^2$, where ${\cal V}_s$ is the string-frame CY volume measured in string units. Thus
\be
f_a^2/M_P^2\sim 1/{\cal V}_s\,.
\ee

We perform this analysis more carefully in Appendix~\ref{appendix_f_a_expression}. There, we take into account the parametric dependence on $g_s$ and the fact that the QCD cycle, the volume of which is proportional to the high-scale value $\alpha_{s,\,UV}^{-1}$ of the inverse strong coupling $\alpha_s^{-1}$, is larger than the string scale. As it turns out, this latter feature can in principle be helpful in lowering $f_a$, but at the price of a very peculiar profile of the relevant $C_p$-form mode. Yet, the resulting suppression is limited by $\alpha_s$, which cannot be too small near the string scale without excessive model building. The result (see Appendix~\ref{appendix_f_a_expression}) reads
\be
\frac{f_{a,\text{min}}^2}{M_P^2} \sim \frac{g_s \alpha_{s,UV}}{\mathcal V_s} \sim \frac{\alpha_{s,UV}}{\sqrt{g_s}} \frac{1}{\mathcal V} \qquad \text{or} \qquad \frac{f_{a,\text{min}}^2}{M_P^2} \sim \frac{\alpha_{s,UV}}{\mathcal V}.
\label{gbo}
\ee
Here $\mathcal V = \mathcal V_s / g_s^{3/2}$ is the CY volume in the 10d Einstein frame and, in the last expression, we have returned to the choice $g_s \sim 0.1$, which is optimal in our context. We reiterate: To the best of our knowledge, a small axion decay constant of a stringy $p$-form axion comes at the price of a large compact volume.

The only phenomenologically viable and reasonably well-understood class of models in which a sufficiently large volume can be realized is the LVS scenario \cite{Balasubramanian:2005zx}.\footnote{See \cite{Junghans:2022exo, Gao:2022fdi} for the most recent round of criticism and defense as well as the refs.~therein for a wider view on the theoretical status.} The key feature is the K\"ahler potential
\be
K=-2\ln{\cal V} \qquad \mbox{with} \qquad {\cal V}=\tau_b^{3/2}-\sum_{i \neq b} \gamma_i\tau_i^{3/2}\,. \label{volume_swiss_cheese}
\ee
Here the $\tau_i$ are 4-cycle volumes measured in units of $l_s=2\pi\sqrt{\alpha'}$.\footnote{
A
possible constant prefactor of $\tau_b^{3/2}$
is equivalent to an additive constant coming with $\ln{\cal V}$ together with a redefinition of the $\gamma_i$. This constant can be absorbed in the K\"ahler potential for the complex structure moduli.} The corresponding 4-cycles $\Sigma_i$ are Poincare dual to harmonic 2-forms $\omega_i$, which form a basis for the decomposition of the K\"ahler form: $J = t^i \omega_i$. The $t^i$ are the volumes of 2-cycles and are related to $\mathcal V$ and $\tau_i$ via
\be
    \mathcal V = \frac{1}{6} \int_X  J \wedge J \wedge J\,\,\,,\qquad \qquad
    \tau_i = \frac{\partial \mathcal V}{\partial t^i}\,.
    \label{tvdef}
\ee
The K\"ahler potential $K$ has to be interpreted as a function of the complexified 4-cycle moduli $T_i = \tau_i + {\rm i} c_i$, where $c_i$ are the integrals of the RR 4-form $C_4$ over the respective 4-cycles:
\begin{equation}
c_i = \int_{\Sigma_i} C_4.
\end{equation}
Finally, the $\gamma_i$ in \eqref{volume_swiss_cheese} are positive $\mathcal{O}(1)$ parameters related to the triple intersection numbers which are implicitly present in \eqref{tvdef}. For basic LVS relations and our conventions see also Appendix~\ref{sec:basic_definitions}.

We will return to this scenario in detail below, but for now let us assume that $\tau_b\gg \tau_{i\neq b}$ and that one of the so-called `blow-up' cycles $\tau_i$ carries the QCD brane stack. We label this cycle by $i=L$ for ``Loop" because, as we will argue below, the most promising scenario is characterized by a SM cycle stabilized by loop effects. Then a straightforward calculation (cf.~Appendix~\ref{appendix_f_a_expression}) gives the axion decay constant\footnote{
It 
is easy to convince oneself that fibred models, where the total volume is proportional to $\tau_L$, do not allow for sufficiently small $f_a$. Intermediate options, with a more complicated dependence of ${\cal V}$ on $\tau_L$, do in general not provide a more efficient way of lowering $f_a$ either.
}
\be \label{axion_decay_constant}
\frac{f_a^2}{M_P^2} \approx \frac{K_{L \bar{L}}}{2 \pi^2} \simeq \frac{\mathcal O (1)}{2 \pi^2 \sqrt{\tau_L}\,{\mathcal V}}\,. 
\ee
This is consistent with our general bound \eqref{gbo}. Parametrically, the bound is not quite saturated since $\tau_L$ appears under a square root while $\alpha_{s,\,UV}= 1/(2\tau_L)$. Crucially, we have seen that the LVS with the QCD sector on a blow-up cycle realizes an optimal volume suppression of $f_a$.

For the convenience of the reader, we record the volume scaling of some key quantities:
\be
f_a\sim \frac{1}{l_s}\sim \frac{M_P}{{\cal V}^{1/2}}\,\,,\qquad 
m_{3/2}\sim \frac{M_P}{\cal V}\,\,,\qquad
m_{\tau_b} \sim \frac{M_P}{{\cal V}^{3/2}}\,.
\ee
Here $m_{3/2}$ is the gravitino mass and $m_{\tau_b}$
the volume modulus mass.

\subsection{Axion realization and consequence for pre- vs post-inflationary cosmology}

For the cosmological behavior of a field theoretic axion a crucial question is whether the Peccei-Quinn (PQ) symmetry giving rise to the axion is restored (typically by thermal effects) after inflation or not, leading to the so-called post- and pre-inflationary scenarios, respectively. Important consequences of symmetry restoration after inflation (post-inflationary scenario) are:
\begin{itemize}
    \item{} The symmetry restoration effectively removes the axion as light degree of freedom. Any isocurvature fluctuations imprinted by inflation are therefore erased (see e.g.~\cite{Kolb:1990vq, Hertzberg:2008wr, Visinelli:2009zm, Visinelli:2014twa}).
    \item{} The field value is randomly selected in each Hubble volume. As the relevant volumes are relatively small today, the cosmological density is given by an average over the random initial values of different Hubble patches and is therefore fixed. A fine-tuning of the initial value to achieve a smaller density is not possible (see e.g.~\cite{Hertzberg:2008wr, Visinelli:2014twa, Sikivie:2006ni}).
    \item{} The random selection leads to large density fluctuations on very small scales that may lead to axion miniclusters (see e.g.~\cite{Hogan:1988mp, Kolb:1993zz, Kolb:1993hw, Visinelli:2018wza}).
    \item{} There can be additional contributions to the density from topological configurations such as axion strings and domain walls (see e.g.~\cite{Vilenkin:1982ks, Hagmann:2000ja, Kawasaki:2014sqa, Ringwald:2015dsf, March-Russell:2021zfq}). 
\end{itemize}

It is therefore important to ask what situation will be realized in a stringy setting. As we will see, momentarily the answer is essentially that we will be in the pre-inflationary scenario but also that the question is slightly different from the field theoretic situation.

Our specific axion comes from $C_4$ integrated over a 4-cycle of a type IIB CY orientifold. It is a special case of the more general class of string axions which originate in RR-forms $C_p$ or Kalb-Ramond fields $B_2/B_6$ integrated over some internal cycles of the compact space. For such axions the shift symmetry is not the result of spontaneous symmetry breaking but rather it is non-linearly realized at the fundamental level. Hence the standard question of axion cosmology whether the PQ symmetry is or is not restored after inflation can strictly speaking not be asked.

That said, we can nevertheless ask questions about initial conditions but also on the influence of thermal effects. Let us start with the initial conditions.
A first step is to consider whether the axion actually exists at all during inflation.
In the model we consider we expect this to be the case.
However, more generally, during inflation the moduli, most notably the inflaton, are not at their minima and therefore the geometry is not necessarily the same as during later phases of the evolution. It is therefore conceivable that the axion does not exist at all, or its properties are different from those at later times. An example of the latter is the possibility to have a different value of the axion decay constant
that evolves during inflation and/or reheating to today's value. In this case the axion would exist during inflation and be subject to questions about initial conditions and isocurvature fluctuations in a similar way as in the pre-inflationary case. However the quantitative answers to these questions could be strongly modified. 
While we think that this deserves further study, we reiterate that in the simple scenario we are envisioning this does not happen, as we will explain in Section \ref{sec:inflationmodel}.

A second equally crucial issue is what happens after inflation and reheating when the Universe is in a very hot state. Here, the question is whether there are strong modifications to the axion due to the high temperature\footnote{In addition to the thermal production of a (small) number of axions.}.
In lieu of a full treatment let us ask a closely related question. Do we expect axionic strings to be present after inflation? We believe that, in most controlled scenarios that we are aware of, the answer is negative. The reason is very simple: With our fundamental forms in 10d come charged objects (various branes of the fundamental string) with a tension of the order of the string scale or above. The 4d axionic string arises from the magnetic dual of our charged object (again some type of brane). It is wrapped on a cycle which is dual to the cycle defining our axion. In non-degenerate geometries, the tension of our 4d axion string will hence be parametrically higher than the string scale. The latter is in general parametrically higher than the KK scale, which in turn is higher than the moduli mass scale. Reheating to a temperature above that scale would destabilize the geometry and is therefore excluded.\footnote{
Exceptions 
may arise if the SM model cycle can be realized in the geometric regime in a strongly warped region, as noted before. This could allow for cosmologically relevant axionic strings \cite{March-Russell:2021zfq} and also impact the dark radiation problem \cite{Frey:2021jyo}. However, apart from the model building challenge, one presumable faces a considerable tadpole issue related to such an appropriately large throat  \cite{Gao:2022fdi}.}

\section{Cosmological constraints}\label{sec:cosmo}

In this section, we use standard cosmological constraints~(see \cite{Visinelli:2009zm, Visinelli:2009kt, Alvarez:2017kar, Nelson:2018via, Arias:2021rer} for detailed studies), i.e.~measurements of the DM abundance and isocurvature bounds, to set the scene and to obtain restrictions concerning the volume, the decay constant etc. In light of the simple analytical approximations used and the possible presence of ${\mathcal{O}}(1)$ factors, the numbers obtained in the following should be taken as order-of-magnitude estimates and an indication of the different qualitative regimes.

\subsection{Assuming standard cosmology for the expansion history}
\label{subsec:standardexpansion}

\subsubsection*{Dark matter abundance}

Having argued in the previous section that we are in a scenario where the axion is present during inflation, any inhomogeneities of the axion field get smoothened out and the initial misalignment angle $\theta_i$ takes on a single random value in the region that will become today's Hubble volume. We therefore have the usual misalignment production of axion DM~\cite{Preskill:1982cy,Abbott:1982af,Dine:1982ah}: When the temperature of the Universe reaches a value $T_\text{osc}$ such that $H(T_\text{osc})\sim \tilde{m}_{a}(T_\text{osc})$, with $\tilde{m}_{a}(T)$ being the temperature-dependent axion mass, the axion field starts to oscillate. Assuming a standard cosmological history, this onset of oscillations occurs during radiation domination. As is well known, the axion relic density is given approximately by~\cite{Visinelli:2009zm,Visinelli:2009kt,Arias:2021rer}
\begin{equation}
    \Omega_a h^2 = 0.2 \left( \frac{f_a}{10^{12} \, \text{GeV}} \right)^{7/6} \theta_i^2, \label{axion_relic_density_standard_case}
\end{equation}
where the standard relation between $f_a$ and the zero-temperature mass of the axion has been used~\cite{GrillidiCortona:2015jxo},
\begin{equation}
    m_a \simeq 5.7 \, \text{meV} \left( \frac{10^9\, \text{GeV}}{f_a} \right).
\end{equation}
Obviously, the contribution~\eqref{axion_relic_density_standard_case} must respect observational constraints on the DM abundance $\Omega_c h^2 = 0.12$ \cite{Aghanim:2018eyx}, which implies a bound on $\theta_i$ in terms of $f_a$,
\begin{equation}
    \Omega_a \leq \Omega_c \quad \Rightarrow \quad \theta_i \leq 0.8 \left( \frac{10^{12} \, \text{GeV}}{f_a} \right)^{7/12}. \label{dark_matter_abundance_angle_tuning}
\end{equation}

\subsubsection*{Saturating the observed dark matter abundance}

If we want axions to be all of dark matter, $f_{a}$ also cannot be too small. Indeed, \eqref{dark_matter_abundance_angle_tuning} suggests that initial angles $\theta_{i}\sim 1$ are already needed for $f_{a}\sim 10^{12}\,{\rm GeV}$. 
As $\theta_{i}\leq \pi$ this suggests a lower limit on the values of $f_{a}$ that can yield the required dark matter abundance.

Avoiding significant fine-tuning to saturate the DM density, we can require $\theta_{i}\leq 3$. This yields 
\begin{equation}
f_{a}\gtrsim 1\times 10^{11}\,{\rm GeV}\,.
\end{equation}

Allowing for some tuning this can be ameliorated. Using the results of~\cite{Alvarez:2017kar}, that include also corrections due to the anharmonicity of the potential, the dark matter abundance requires
\begin{equation}
\label{eq:darkmatter}
    f_{a}\gtrsim 10^{10.3}\,{\rm GeV}\qquad{\rm for}\qquad H_{I}\gtrsim 10^{4}\,{\rm GeV}.
\end{equation}
Here, the dependence on $H_{I}$ arises because fluctuations during inflation limit the possible amount of fine-tuning as well as causing excessive isocurvature fluctuations.
That being said, in the following we usually only consider the non-tuned region.

In what follows, we will mostly assume that the QCD axion constitutes all of dark matter, i.e. that the bound~\eqref{dark_matter_abundance_angle_tuning} is saturated. 

\subsubsection*{Isocurvature constraints}

Another relevant, observational bound is the one on isocurvature perturbations~\cite{Hertzberg:2008wr,Visinelli:2009zm,Visinelli:2014twa,Alvarez:2017kar}. The latter can arise due to quantum fluctuations of the axion field imprinted during inflation
\begin{equation}
    \langle | \delta a(k) |^2 \rangle = \left( \frac{H_I}{2 \pi} \right)^2 \frac{2 \pi^2}{k^3},
\end{equation}
where $a$ is the canonically normalized axion field. The power spectrum of the axion density fluctuations is given by
\begin{equation}
    \Delta_a^2 = \left. \left\langle \left( \frac{\delta \rho_a}{\rho_a} \right)^2 \right\rangle \right|_{t_\text{CMB}} \approx \left( \frac{\gamma H_I}{\pi f_a \theta_i} \right)^2, \label{axion_power_spectrum}
\end{equation}
where we use $\gamma=2$ as done in~\cite{Hertzberg:2008wr}.
Planck reports upper bounds at 95\% CL for the isocurvature fraction~\cite{Planck:2018jri}
\begin{equation}
    \beta_\text{iso} = \frac{\Delta_a^2 (k_*)}{\Delta_a^2 (k_*) + \Delta_{\mathcal R}^2 (k_*)} < 0.038, \label{isocurvature_fraction}
\end{equation}
where $k_* = 0.050 \, \text{Mpc}^{-1}$ and $\Delta_{\mathcal{R}}^2 (k_*)$ is the curvature power spectrum amplitude at the scale $k_*$. The latter is also given by Planck as $ \Delta_{\mathcal{R}}^2 (k_*) = (2.101^{+ 0.031}_{- 0.034}) \times 10^{-9}$ at 68\% CL \cite{Aghanim:2018eyx}. Inserting~\eqref{axion_power_spectrum} into~\eqref{isocurvature_fraction}, one finds
\begin{equation}
    H_I \lesssim 1.4\times 10^{-5} f_a \theta_i. \label{H_I_isocurvature_bounds}
\end{equation}

\subsubsection*{Implications for the string scenario}

We are now ready to apply these constraints to our string setup. Combining the value of $\theta_i$ which is required to saturate the DM abundance~\eqref{dark_matter_abundance_angle_tuning} with the bound from isocurvature constraints~\eqref{H_I_isocurvature_bounds} and using the expression for the volume dependence of the axion scale~\eqref{axion_decay_constant}, we arrive at a volume dependent constraint for the inflation scale
\begin{equation}
    H_I \lesssim \frac{2 \times 10^9 \, \text{GeV}}{\mathcal{V}^{5/24}}\,. \label{H_I_standard_case}
\end{equation}
This implies a very low inflation scale, in particular since the LVS requires ${\mathcal{V}}\gg 1$. In consequence this also implies a very low tensor-to-scalar ratio.\footnote{One might argue that the bound~\eqref{H_I_isocurvature_bounds} from isocurvature constraints can be loosened or even evaded by considering a scenario where the QCD axion constitutes only a minor fraction of the total DM density. However, this would require that either $\theta_i$ or $f_a$ are very small. The former cannot be tuned to arbitrarily low values because the emergence of quantum fluctuations would spoil such a tuning, whereas a smaller $f_a$ in the LVS context is only achieved by an even larger volume, which would again imply a small inflation scale. We therefore believe that a small $H_I$ is a general and hardly circumvented feature of a stringy axion. Note also that the relevance of very low-scale string inflation has recently been emphasized in \cite{Bedroya:2019tba}, though from a rather different perspective.}

\bigskip

In string-theoretic constructions, especially if the inflaton is a modulus, one expects the inflationary potential to be comparable to the potential stabilizing the moduli: If it is higher, one faces the danger of moduli destabilization, whereas if it is much lower, more tuning is in general required. Concretely in the LVS, we then expect
\begin{equation}
    H_I^2 \simeq \beta \frac{|W_0|^2 M_P^2}{\mathcal{V}^3}, \label{H_I_LVS_scaling}
\end{equation}
where $\beta$ is a model-dependent $\mathcal{O}(1)$ parameter. In fact (for more details see Section~\ref{sec:inflationmodel}), we will later focus on LVS K\"ahler (or `blow-up') moduli inflation \cite{Conlon:2005jm}, where the above estimate holds.

With that, we can solve~\eqref{H_I_standard_case} explicitly for $\mathcal{V}$ and obtain an estimate for a lower bound on $\mathcal V$, which translates into upper bounds on $H_I$ and $f_a$. Moreover, we can estimate a lower bound on $f_a$ by demanding that the axion relic density saturates the DM density without fine-tuning $\theta_i \approx \pi$, that is by saturating~\eqref{dark_matter_abundance_angle_tuning} for $\theta_i \leq 3$. The resulting bounds are given by
\begin{alignat}{2}
\label{eq:volumeconstrad}
   (\kappa^{24/31})\,1 \times 10^7 \;\; &\lesssim  \mathcal{V} &&\lesssim \;\; 9\times 10^{12}, \\
  (\kappa^{-5/31}) \, 7 \times 10^7 \, \text{GeV}  \;\;&\gtrsim   H_I &&\gtrsim \;\; 0.1\,{\rm GeV} \, \kappa,\\
(\kappa^{-12/31}) \, 9 \times 10^{13} \, \text{GeV} \;\;&\gtrsim   f_a && \gtrsim\;\; 1\times 10^{11}\,{\rm GeV}, \\
 (\kappa^{7/31})\, 0.1  \;\;&\lesssim   \theta_i && \lesssim \;\;3,
\end{alignat}
where we have defined $\kappa^2 \equiv \beta |W_0|^2$ and in the penultimate line used~\eqref{axion_decay_constant} with the ${\mathcal{O}}(1)$ factor taken to be equal to unity and $\tau_L = 1/ (2\alpha_{s,UV}) = 25/2$.
Here, the left-hand side corresponds to the bounds from isocurvature constraints and the right-hand side to those from DM saturation.
Allowing for some tuning, the upper limit on the volume relaxes slightly, but as already mentioned, eventually this becomes a strict limit due to isocurvature fluctuations.

\subsection{Assuming early matter domination}
\label{subsec:matterdomination}

It is far from clear that cosmologies following from string models result in a standard expansion history.
Indeed string models often feature long-lived moduli that lead to a phase of matter domination\footnote{These particles may already be produced with low temperature, but their long life-time also allows them to further cool by expansion.} before decaying to reheat the Universe~\cite{Cicoli:2012sz,Banks:2002sd}.

\subsubsection*{Dark matter abundance}

Such an Early Matter Dominated (EMD) phase may have significant impact on the predictions for axion dark matter~\cite{Visinelli:2009kt,Nelson:2018via,Arias:2021rer,Banks:2002sd}. 
In particular, if a modulus $\phi$ decays very late, the axion may begin to oscillate during a phase of matter domination. The modified expansion history leads to a different value of the axion dark matter density and the above analysis changes in this scenario.
The modified axion relic density is approximately given by~\cite{Visinelli:2009kt,Arias:2021rer}\footnote{Note that the exact numerical prefactor depends on the number of relativistic degrees of freedom. Here we have used that in the EMD scenario $T_\text{osc}$ is typically still high enough such that $g_*(T_\text{osc}) \gtrsim 60$. For the numerical value of the prefactor we simply use $g_*(T_\text{osc}) = 70$.}
\begin{equation}
\label{eq:densitymatter}
    \Omega_a h^2 = 6 \times 10^{-5} \left(\frac{f_a}{10^{12} \, \text{GeV}} \right)^{3/2} \left( \frac{T_\text{end}}{10\, \text{MeV}} \right)^2 \theta_i^2,
\end{equation}
where $T_\text{end}$ is the temperature at which the $\phi$-modulus ceases to dominate the energy content of the Universe. Explicitly, it is defined using the modulus decay rate: $\Gamma_\phi=H(T_{\rm end})$. As is commonly done, we will also refer to $T_{\rm end}$ as the reheating temperature, $T_r\equiv T_{\rm end}$. However, it is important to remember that the SM sector itself thermalizes already much earlier and at a higher temperature~\cite{Visinelli:2009kt,Nelson:2018via,Arias:2021rer}\footnote{Going back to earlier times the temperature actually increases because the energy density of the modulus increases as $\rho_\phi \sim a^{-3}$, of which a fraction $\sim \Gamma_\phi/H\sim a^{3/2}$ decays during one Hubble time. Then the energy in SM radiation scales as $\rho_{\textrm{SM}}\sim T^4\sim \rho_\phi\Gamma_\phi/H\sim a^{-3/2}$ such that the temperature is $T\sim a^{-3/8}$.}, based on the energy input from early $\phi$-decays. Because of this, one can use the formula for the axion potential valid at temperatures above the QCD phase transition underlying \eqref{eq:densitymatter}.

\subsubsection*{Isocurvature constraints}

We expect that isocurvature constraints from CMB measurements for the EMD scenario are the same  as for a standard cosmology. This is because the observed CMB modes have entered the horizon shortly before (dark) matter-radiation equality or later but long after the EMD phase. They have therefore experienced only a standard cosmological evolution so that~\eqref{H_I_isocurvature_bounds} is still valid if used together with the modified axion relic density~\eqref{eq:densitymatter}, which we again assume to saturate the DM density, $\Omega_a h^2 = \Omega_c h^2 = 0.12$. We obtain
\begin{equation}
\label{eq:isomat}
    H_I \lesssim \frac{1 \times 10^{10}\, \text{GeV}}{\mathcal{V}^{1/8}} \left( \frac{10\,\text{MeV}}{T_\text{end}} \right).
\end{equation}
As before this implies a relatively low tensor component in the CMB. 

\subsubsection*{Implications in the string scenario}

Relating \eqref{eq:isomat} to the typical order of the LVS potential~\eqref{H_I_LVS_scaling}, we again obtain a lower bound on $\mathcal V$ or, equivalently, upper bounds on $H_I$ and $f_a$. In addition we can follow the same logic as above to obtain bounds by demanding that the initial misalignment angle is not fine-tuned, $\theta_i \leq 3$. This gives
\begin{alignat}{2}
    \left( \frac{\kappa T_\text{end}}{10 \, \text{MeV}} \right)^{8/11} \, 9 \times 10^5 \;\;&\lesssim \mathcal{V} &&\lesssim \;\; 6\times 10^7 \, \left( \frac{T_\text{end}}{10\,\text{MeV}} \right)^{8/3}, \label{eq:vlimit}\\
    \kappa^{-1/11} \left( \frac{10\,\text{MeV}}{T_\text{end}} \right)^{12/11} \, 3 \times 10^{9} \, \text{GeV}  \;\;&\gtrsim H_I &&\gtrsim \;\; 5 \times 10^{6} \, \text{GeV} \, \left( \frac{10\, \text{MeV}}{T_\text{end}} \right)^4 \kappa, \\\label{eq:falimit}
    \left( \frac{10\,\text{MeV}}{\kappa T_\text{end}} \right)^{4/11} \, 3 \times 10^{14} \, \text{GeV} \;\;&\gtrsim f_a &&\gtrsim \;\; 4 \times 10^{13}\,{\rm GeV} \, \left( \frac{10\, \text{MeV}}{T_\text{end}} \right)^{4/3},
\end{alignat}
where again the left-hand side corresponds to bounds from isocurvature constraints and the right-hand side to DM saturation. We see that for small $T_\text{end}$ the window that is given by the above bounds becomes very narrow and, depending on $\kappa$, might even close.

To make this more explicit, we make use of the fact that the reheating temperature is given by the decay of the longest-lived modulus. For the usual case of this being the volume modulus, we have
\be \label{reheating_temperature}
T_r = \left(\frac{90}{g_* \pi^2}\right)^{1/4} \sqrt{\Gamma_{\tau_b} M_P},
\ee
where $\Gamma_{\tau_b}$ is its decay constant and $g_*(T)$ the effective number of relativistic degrees of freedom. The decay constant and volume modulus mass are typically given by
\be
\Gamma_{\tau_b} \sim \frac{m_{\tau_b}^3}{M_P^2}, \quad m_{\tau_b} \sim \frac{W_0 M_P}{\mathcal V^{3/2} }, \label{mass_volume_modulus}
\ee
where we use simplistic formulae ignoring prefactors that cannot be parametrically small or large and even potential logarithms of the volume. With this, we are ready to re-derive the bounds~\eqref{eq:vlimit} -- \eqref{eq:falimit} with $T_\text{end}$ eliminated. 
However, it turns out that for the small volume region affected by the isocurvature bound, the reheating temperature~\eqref{reheating_temperature} is usually so large that we are back to the radiation dominated case discussed in Section~\ref{subsec:standardexpansion}. We therefore only give the upper bounds on the volume arising from the requirement of saturating the DM density
\begin{alignat}{2}
   \mathcal{V}& \lesssim  7\times 10^8 \, \left(W_0^{4/7} \right), \label{eq:vlimit_explicit}\\
     H_I& \gtrsim 1 \times 10^{5} \, \text{GeV} \left( \beta^{1/2} W_0^{1/7} \right), \\\label{eq:falimit_explicit}
     f_a &\gtrsim 1 \times 10^{13}\,{\rm GeV} \left( W_0^{-2/7} \right), \\
    T_r& \gtrsim 30 \,{\rm MeV} \left( W_0^{3/14} \right).
\end{alignat}

\subsubsection*{Bounds from BBN}

At this point, we make use of the fact that another constraint comes in. In order to not spoil successful BBN, $T_r$ cannot be smaller than $\mathcal{O}(1\,  \text{MeV})$~\cite{Kawasaki:1999na, Kawasaki:2000en,Hannestad:2004px, Ichikawa:2006vm, DeBernardis:2008zz, deSalas:2015glj, Kawasaki:2017bqm, Hufnagel:2018bjp, Forestell:2018txr, Hasegawa:2019jsa, Kawasaki:2020qxm,Depta:2020zbh}. We can use this together with~\eqref{reheating_temperature} and~\eqref{mass_volume_modulus} to derive another upper bound on $\mathcal{V}$, which again translates into bounds on $H_I$ and $f_a$,
\begin{align}
    \mathcal V &\lesssim 3 \times 10^{9} \left( W_0^{2/3} \right), \\
    H_I &\gtrsim 1 \times 10^4 \, \text{GeV} \, (\beta^{1/2}), \\
    f_a &\gtrsim 5 \times 10^{12} \, \text{GeV} \, (W_0^{-1/3}).
\end{align}
Note that the above bounds from BBN follow directly from the assumption that the SM is reheated by the volume modulus decay. These bounds are hence independent of the onset of axion oscillations, i.e.~they apply to both the standard scenario and the EMD scenario. As a result, they imply a stronger bound than DM saturation in the former scenario but a weaker bound in the latter one.

Let us stress, however, that the assumption of a late-decaying volume modulus that reheats the SM will be challenged in this paper. As we will see in the next section, it is conceivable that there exists an additional decay channel, dramatically enhancing the decay rate of the volume modulus and therefore increasing the reheating temperature. In this case the BBN bound will be invalidated or at least strongly modified and the expansion history during the time relevant for axions may be closer to the standard scenario discussed in Section~\ref{subsec:standardexpansion}.

\section{Dark radiation I: A new solution to an old problem by Higgs-mass-mediated decays}
\label{sec:dr1}

Significant dark radiation abundance is a familiar problem or, more optimistically, a prediction of LVS cosmology~\cite{Cicoli:2012aq, 
Higaki:2012ar, Hebecker:2014gka, Angus:2014bia, Allahverdi:2014ppa, Cicoli:2015bpq, Cicoli:2018cgu, Acharya:2019pas, Jaeckel:2021gah, Angus:2021jpr, Jeong:2021yol, Frey:2021jyo, Cicoli:2021tzt, Jaeckel:2021ert}. It arises in the simplest, sequestered setting because the light volume modulus is the last one to decay. Its ${\cal O}(1)$ branching fraction into its own, essentially massless, axion is the source of the issue \cite{Cicoli:2012aq, Higaki:2012ar}. The dark radiation problem persists in more general LVS implementations \cite{Hebecker:2014gka}, including SM-realizations on D7-branes, loop-stabilized cycles or additional flavor branes. Interesting proposals to ameliorate the problem use a sequestered setting with an enhanced decay rate to light scalar superpartners \cite{Cicoli:2015bpq} or reheating after Fibre Inflation~\cite{Cicoli:2008gp} together with a large flux on the SM cycle~\cite{Cicoli:2018cgu}. Unfortunately, the former is not suitable for our purposes since we require a D7-brane SM for our stringy QCD axion realization. The latter is excluded due to the Fibre Inflation scale being too high.

Using the standard definition of the effective number of neutrino species, the axionic decay of a modulus, $\phi\to aa$, contributes approximately\footnote{This formula actually assumes that the energy density in $a$ is subdominant.}~\cite{Cicoli:2012aq,Higaki:2012ar,Jaeckel:2021ert,Choi:1996vz,Conlon:2013isa}
\begin{equation}
\label{eq:deltaneff}
    \Delta N_{{\rm eff}}\sim 6.1\left(\frac{11}{g^{4}_* g^{-3}_{*,S}}\right)^{1/3}BR(\phi\to aa)\sim 6.1\left(\frac{11}{g^{4}_* g^{-3}_{*,S}}\right)^{1/3}\frac{\rho_\text{DR}}{\rho_\text{SM}+\rho_\text{DR}} \bigg|_{T=T_r},
\end{equation}
where $g_*$ and $g_{*,S}$ are respectively the
relativistic degrees of freedoms of the energy density and the entropy density at the reheating temperature $T_r$.

This has to be compared with the limits placed by observations of the CMB and large scale structure that usually lie in the region (depending on the data sets used)~\cite{Aghanim:2018eyx}
\begin{equation}
\label{eq:deltaneffob}
    \Delta N_{\rm eff}\lesssim 0.2-0.4\,.
\end{equation}
In the following two subsections, we will first recall how the LVS dark radiation problem usually arises and then show why it is generically avoided in situations with high-scale SUSY breaking.

\subsection{Decay of $\tau_b$ to its axion $a_b$}

We follow~\cite{Cicoli:2012aq, Higaki:2012ar} in using the standard no-scale K\"ahler metric derived from \eqref{volume_swiss_cheese} and neglecting small-cycle effects. This immediately gives the operators relevant for the decay of $\tau_b$ into its own axion $c_b$
\be
{\mathcal L}/M_P^2 \,\,\supset\,\, \frac{3}{4\tau_b^2} \partial_\mu \tau_b \partial^\mu \tau_b + \frac{3}{4\tau_b^2} \partial_\mu c_b \partial^\mu c_b\,.\label{ttl}
\ee
It is then straightforward to obtain the canonical volume and axion fields, $\phi_b/M_P \equiv \sqrt{3/2} \ln \tau_b$ and $a_b/M_P = \sqrt{3/2} \,c_b/ \langle \tau_b \rangle$. One then determines the trilinear coupling of the volume fluctuation $\delta\phi_b$ to two axions and hence the decay rate \cite{Cicoli:2012aq, Higaki:2012ar}
\be
\Gamma_{\phi_b \rightarrow a_b a_b} = \frac{1}{48 \pi} \frac{m_{\tau_b}^3}{M_P^2}\,.
\label{adr}
\ee

This has to be compared with the decay rate to the SM, which in the simplest (sequestered) setting is dominated by decays to Higgs fields. The latter follow from the appropriately extended K\"ahler potential
\be
K = -3\ln\left[T_b+\overline{T}_b + \frac{1}{3}(H_u\overline{H}_u+H_d\overline{H}_d+zH_u H_d+\mbox{h.c.})\right]\,,
\ee
where the small cycle has been disregarded. Expanding to leading order in the Giudice-Masiero-type term $zH_uH_d$ one finds an operator which involves the two SUSY Higgs fields $H_{u,d}$, the volume modulus and two derivatives. This induces a decay rate~\cite{Cicoli:2012aq, Higaki:2012ar}
\begin{equation}
    \Gamma^{\rm SUSY}_{\tau_{b}\to SM}\sim     \Gamma^{\rm SUSY}_{\tau_{b}\to H_{u}H_{d}}=\frac{2z^2}{48\pi}\frac{m_{\tau_b}^3}{M_P^2}\,.
    \label{hdr}
\end{equation}
Note that $z=1$ is a special point distinguished by a shift symmetry in the Higgs sector~\cite{Hebecker:2012qp, Hebecker:2013lha}. It is not understood whether $z\gg 1$ and a corresponding enhancement can be realized. We have given the decay rate above the index `SUSY' since this setting and this $\Gamma$ apply most directly to the case of low-scale supersymmetry. However, this result and in particular the parametric similarity of \eqref{adr} and \eqref{hdr} apparently persist in many variations of the simplest setting~\cite{Hebecker:2014gka,Cicoli:2015bpq, Cicoli:2018cgu}, including SUSY breaking above the scale $m_{\tau_b}$ or additional decays to SM gauge bosons. It is non-trivial to make the prefactor of the SM decay rate larger and, as result, the production of too much dark radiation is widely accepted to be a generic issue.

However, as we will argue next, in the case of a high SUSY breaking scale a different coupling of the volume modulus to the SM Higgs takes over, inducing a parametrically larger decay rate.

\subsection{Mass-induced rapid decay of $\tau_b$ to Higgses}
\label{subsec:rapid}

The basic idea is that the Higgs mass depends on $\tau_b$, giving rise to a trilinear coupling between the volume mode and the Higgs fields. Naively, one would not hope for a large effect since the Higgs mass squared is small,\footnote{To be precise, when we specify the phenomenologically interesting volume range in Section~\ref{sec:cosmofinal}, it will turn out that for the largest volumes the mass $m_{\tau_b}$ comes dangerously close to $m_H$. One may avoid this situation by assuming $|W_0|\gg 1$ or one may study this possibly interesting regime in the future. For our present analysis, we mostly assume that $|m_H^2|\ll m^{2}_{\tau_b}$. We briefly look into lower masses in Section~\ref{sec:lowmassb}.} 
$|m_H^2|\ll m^{2}_{\tau_b}$. One then expects the amplitude to be suppressed by $|m_H^2|/m_{\tau_b}^2$ compared to what we found in the last subsection. However, it turns out that the fine-tuning of the Higgs mass, which is usually seen as a problem of high-scale SUSY, is in this case advantageous. Namely, the strength of the mass-induced coupling between the Higgses and the volume mode is governed by the untuned, high value that the Higgs mass would naturally have. The actual tuning of the Higgs mass to a small value in the SM vacuum does not diminish this coupling.

To be specific, let us start from the Higgs mass matrix at the KK scale $m_{\rm KK}$ of the SM brane(s). Below this scale, we may use a 4d supersymmetric EFT and run this matrix down to the SUSY breaking scale $m_{3/2}$. In our scenario, the $F$-terms of the K\"ahler moduli set the scale of SUSY breaking: $m_{3/2}/M_P\sim F_T/T$. In particular, the gaugino masses are comparable to the mass of the gravitino, $m_{1/2}\sim m_{3/2}$.\footnote{
Note that the masses of gauginos related to non-perturbatively stabilized cycles are suppressed compared to the gravitino mass by a factor $\sim \ln(M_P / m_{3/2}) \sim \ln \mathcal{V}$ due to a leading-order cancellation of $F$-terms \cite{Conlon:2006us}. If however, as in our case, the relevant cycle is stabilized by loop corrections, the cancellation is avoided and the corresponding gaugino has a mass comparable to the gravitino \cite{Conlon:2006ur, Krippendorf:2009zza}.} 
Independently of the details of how the soft Higgs masses arise, it is then guaranteed that at least some of the entries of the Higgs mass matrix are of order $m_{3/2}^2$, possibly suppressed by a loop factor but logarithmically enhanced due to the running. For example, gaugino masses contribute a term 
$\sim c_\text{loop} m^2_{1/2} \ln(m_\text{KK}/m_{3/2})$ to the squared soft Higgs masses (see e.g.~\cite{Antoniadis:1994hg, Brignole:1997wnc, Martin:1997ns} for some of the classic results and \cite{Hebecker:2012qp, Hebecker:2013lha} for a more recent discussion in the present context).

Now, after running down to the scale $m_{3/2}$, SUSY breaks and the Higgs mass matrix removes one linear combination of the scalars in $H_u$ and $\overline{H}_d$. The mass squared of the remaining (SM) Higgs doublet is then set by the determinant of the Higgs mass matrix, which is fine-tuned to a very small value. According to what was said above, this fine-tuning involves the running of some of the entries of the Higgs mass matrix from $m_{\rm KK}$ to $m_{3/2}$. Thus, symbolically we have
\be \label{eq:Higgs_mass_loop_corrected}
m_H^2 \sim m_{3/2}^2 \left[ \cc + c_\text{loop} \ln\left( \frac{m_\text{KK}}{m_{3/2}} \right) \right]
\ee
and $|m_H^2|\ll m_{3/2}^2$.
Since the SM lives on one (or several intersecting) brane-stack(s) wrapping a small cycle of size $\mathcal{O}(1-10)$ in string units, we have $m_\text{KK} \sim M_s \sim M_P \mathcal{V}^{-1/2}$. Together with $m_{3/2}/M_P\sim W_0/{\cal V}$, this gives
\be
m_H^2\sim \left(\frac{W_0}{\mathcal V}\right)^2 \left[  \cc + c_\text{loop} \ln \left( \frac{\mathcal V^{1/2}}{W_0} \right) \right]
\label{mhv} M_P^2\,.
\ee
To determine the coupling between the canonical volume modulus $\phi_b$ and two Higgs fields, we now recall that ${\cal V}\sim \tau_b^{3/2}$ and $\sqrt{3/2}\ln\tau_b =\phi_b/M_P$. We write $\phi_b=\langle\phi_b\rangle+ \delta \phi_b$ and expand $m_H^2$ to linear order in $\delta \phi_b$. Due to the fine-tuning between $c_0$ and the logarithmic term, the effect of expanding the log dominates and we find
\be
\label{eq:trilinear}
{\cal L}\,\,\supset\,\,\sim\,\,\left(
m_{3/2}^2\frac{c_\text{loop}}{2} \sqrt{\frac{3}{2}}\right)
\,h^2\,\frac{\delta \phi_b}{M_P}\,\,\sim\,\,
m_{3/2}^2 \,c_\text{loop}\,h^2\,\frac{\delta \phi_b}{M_P}\,,
\ee
where $h$ is the Higgs scalar. Parametrically, the decay width for the $\tau_b \rightarrow hh$ decay is then given by
\be
\label{eq:volumedecayfinal}
\Gamma_{\phi_b \rightarrow hh} \sim \frac{m_{3/2}^4 c^{2}_{\rm loop}}{m_{\tau_b} M_P^2} \sim \mathcal (c_{{\rm loop}}{\cal V})^2 \frac{m_{\tau_b}^3}{M_P^2} \gg \Gamma_{\phi_b \rightarrow a_b a_b}\,.
\ee

The last relation in \eqref{eq:volumedecayfinal} assumes $\mathcal V\gg 1/c_{{\rm loop}}\sim (16\pi^2)$, implying that the decay of the volume modulus into SM fields is much stronger than that into volume axions. The corresponding contribution to dark radiation, $\Delta N_\text{eff} \sim \Gamma_{\phi_b \rightarrow a_b a_b} / \Gamma_{\phi_b \rightarrow hh}$, is then negligible and the standard dark radiation problem of the LVS is solved.

We stress that this drastic enhancement of the decay rate is a positive result of the large fine-tuning required to achieve an acceptable Higgs mass/vacuum expectation value. In the scenario considered, this is due to the large SUSY breaking contribution of the order of $m_{3/2}$. If the fine-tuning were to be smaller, e.g. due to cancellations occurring in sequestered scenarios where soft scalar masses are hierarchically smaller than $m_{3/2}$ \cite{Blumenhagen:2009gk,Aparicio:2014wxa}, the enhancement could be significantly smaller and in this case the dark radiation problem may not be fully addressed by the Higgs decays. In our present work, we do not consider such settings since they lack our desired $p$-form QCD axion of naturally high quality.

In the above analysis we have assumed $\phi_b$ to be at its post-inflationary minimum. However, during inflation the volume modulus is generically shifted from its minimum due to the inflationary energy density. Let us denote this displacement as $\hat\phi_b$ which in K\"ahler moduli inflation (the inflationary model we will focus on) has been computed to be of order $\hat\phi_b/M_P\sim \mathcal{O}(0.1)$ \cite{Cicoli:2016olq}. Thus during inflation there is no fine-tuning of the Higgs mass which scales instead as $|m_H^2| \sim c_{\rm loop} m_{3/2}^2 \hat\phi/M_P$. This is indeed higher than the mass squared of the volume modulus $m_{\tau_b}^2\sim m_{3/2}^3/M_P$ if $\mathcal{V} > M_p/(c_{\rm loop}\hat\phi_b) \sim \mathcal{O}(10^3)$. Hence, one might be worried that the volume modulus decay into Higgses is kinematically forbidden in the early Universe. Fortunately, this does not happen for the following reason: The volume modulus initially starts oscillating when $H_{\rm ini}\sim m_{\tau_b}$ and it decays when $H_{\rm dec}\sim \Gamma_{\phi_b\to hh}$. In a background which is matter-dominated due to oscillations of the inflaton $\tau_I$, the amplitude of $\phi_b$ oscillations redshifts as $H$. This allows one to determine the Hubble parameter $H= H_{\rm eq}$ which corresponds to the moment when $|m_H^2|\simeq m_{\tau_b}^2$. The volume modulus amplitude at that moment is $\hat\phi_{b,{\rm eq}}/M_P\simeq 1/(\mathcal{V} c_{\rm loop})$. This gives
\begin{equation}
H_{\rm eq} = \left(\frac{\hat\phi_{b,{\rm eq}}}{\hat\phi_{b,{\rm ini}}}\right) H_{\rm ini} \simeq \left(\frac{M_P}{c_{\rm loop}\hat\phi_{b,{\rm ini}}}\right) \frac{M_P}{\mathcal{V}^{5/2}}\,,
\end{equation}
which is well above $H_{\rm dec}$ since
\begin{equation}
\frac{H_{\rm eq}}{H_{\rm dec}} \simeq \left(\frac{M_P}{c_{\rm loop}^3 \hat\phi_{b,{\rm ini}}}\right) \sim \mathcal{O}(10^7)\qquad\text{for}\qquad c_{\rm loop}\sim \frac{1}{16\pi^2}\quad \text{and}\quad \hat\phi_{b,{\rm ini}}\sim 0.1\, M_P\,.    
\end{equation}

In this analysis we have ignored the potential non-perturbative production of Higgs particles due to pre-heating effects. If relevant, they would however not change our results qualitatively since they convert the energy stored in $\phi_b$ into Higgs particle production even faster.

In summary, we have shown that the volume modulus has a large decay rate to the Standard Model. But, more than that, this decay rate is so fast that the volume modulus may, contrary to what is usually assumed, never get to dominate the total energy density. We therefore need to investigate whether other moduli have longer lifetimes and hence play a central role for the energy budget, including through their possible decays to axionic dark radiation. In particular, such moduli may be part of concrete realizations of inflation. We will therefore consider possible inflaton sectors and specify a suitable inflation scenario in Section~\ref{sec:inflationmodel} before returning to dark radiation in Section~\ref{sec:dr2}.

\section{Combining the QCD axion with a suitable inflation model}
\label{sec:inflationmodel}

Our logic has led us to focus on stringy QCD axions in the LVS, and we are hence interested in a consistent cosmological history in this particular framework. The two most popular inflation models in the LVS context are K\"ahler moduli \cite{Conlon:2005jm} and Fibre \cite{Cicoli:2008gp} inflation (see \cite{Cicoli:2011zz} for a review). However, Fibre Inflation comes with a relatively high inflation scale, hence falling victim to the isocurvature constraints discussed earlier.\footnote{In Fibre Inflation better DM candidates seem to be ultralight axions behaving as fuzzy DM \cite{Cicoli:2021gss} and/or primordial black holes \cite{Cicoli:2018asa, Cicoli:2022sih}.}

Thus we focus on K\"ahler moduli inflation~\cite{Conlon:2005jm}, which in the simplest case requires a volume of the form
\be 
\label{volume_blowup_inflation}
\mathcal V = \tau_b^{3/2} - \gamma_s \tau_s^{3/2} - \gamma_I \tau_I^{3/2}\,.
\ee
We see that there are two blow-up cycles, the LVS typical small cycle $\tau_s$ and the additional inflaton cycle $\tau_I$. During inflation, $\tau_I$ is so large that its non-perturbative effects $\sim \exp(-\mathfrak{a}_I\tau_I)$ are tiny and the slow-roll conditions are obeyed. Eventually $\tau_I$ rolls to its minimum where it is stabilized by the competition of potential terms $\sim -\exp(-\mathfrak{a}_I\tau_I)/{\cal V}^2$ and $\sim \exp(-2\mathfrak{a}_I\tau_I)/{\cal V}$. The non-perturbative effects just discussed come either from E3 instantons ($\mathfrak{a}_I=2\pi)$
or gaugino condensation on a stack of $N$ D7 branes ($\mathfrak{a}_I=2\pi/N_I$). The latter case, considered in \cite{Allahverdi:2020uax}, seems however disfavored since loop effects, if not tuned, might spoil slow-roll. The former possibility also has issues since the vacuum value of the volume
\be
\label{eq:blowup}
\mathcal V \sim \exp({\cal O}(1)\mathfrak{a}_I/g_s)\,,
\ee
tends to become too large at small $g_s$. However, we take the attitude that $g_s$ does not have to be extremely small but can instead take `smallish' ${\cal O}(0.1)$ values, as is common in F-theory models.

The small cycle $\tau_s$ is stabilized by non-perturbative corrections to the superpotential $\sim \exp(-\mathfrak{a}_s\tau_I)$, with $\mathfrak{a_s}=2\pi/N_s$, and guarantees that the volume is kept almost constant during inflation. The value of the volume during inflation $\mathcal{V}_I$ has been computed in \cite{Cicoli:2016olq} and reads
\begin{equation}
\mathcal{V}_I \simeq \mathcal{V}^{1+2\delta}
\quad\text{where}\quad 
\delta =\frac{\tau_I^{3/2}}{\tau_I^{3/2} + \tau_s^{3/2}}\simeq \frac{\mathfrak{a}_I^{-3/2}}{\mathfrak{a}_I^{-3/2} + \mathfrak{a}_s^{-3/2}} = \frac{1}{N_s^{3/2}+1}\ll 1 \quad\text{for}\quad N_s\gg 1\,.
\end{equation}
It is easy to see that already values $N_s\sim \mathcal{O}(5-10)$ are large enough to keep the volume approximately constant, so that inflation is almost single-field and the properties of the QCD axion do not change significantly between inflation and today. Having a hidden sector D7-stack on the $\tau_s$-cycle is not a problem since the corresponding string loops would be inflaton-independent. Moreover, as shown in \cite{Cicoli:2010ha}, the scale of strong dynamics is higher than both the Hubble scale during inflation and the inflaton mass after the end of inflation. This ensures that non-perturbative effects have already been generated at the inflationary scale, and that the inflaton decay to hidden sector degrees of freedom, like glueballs, at the end of inflation is kinematically forbidden. Note also that, for $N_s=1$, one could still obtain $\delta\ll 1$ due to $\tau_I\ll \tau_s$ via a large hierarchy among the prefactors of the corresponding non-perturbative effects \cite{Cicoli:2021dhg}.

So far, everything is fine, but of course we need to also implement the QCD axion. Most naively, one might try to stabilize the visible sector cycle via non-perturbative effects. But this is ruled out since the axion, being the SUSY partner of the corresponding K\"ahler modulus, would obtain a mass of the same order of magnitude and thus be too heavy. Instead, we adapt a proposal from Section~4.3 of \cite{Cicoli:2012sz} to our case. Namely, we add a further blow-up sector consisting of two small cycles which, when both shrinking to zero volume, produce a codimension-3 singularity. One combination of the two corresponding K\"ahler moduli is stabilized by D-terms, obtains a high mass and is integrated out. The remaining modulus is stabilized by loop effects \cite{vonGersdorff:2005bf, Berg:2005ja, Berg:2007wt, Cicoli:2007xp}.\footnote{Another option, which is qualitatively similar to this one, would be follow \cite{Cicoli:2011qg} where the authors considered just a single blow-up mode. D-terms fix the radial part of a charged open string, while the blow-up mode is fixed by a combination of loops and $F$-terms of the matter field.} We hence denote it by $\tau_L$ and use from now on the effective volume formula
\be
\label{eq:inflatonfinal}
\mathcal{V} = \tau_b^{3/2} - \gamma_s \tau_s^{3/2} - \gamma_I \tau_I^{3/2} - \gamma_L \tau_L^{3/2} \,.
\ee
The visible sector lives on intersecting branes wrapping the blow-up sector governed by $\tau_L$. 

Crucially, perturbative effects, and hence the loop potential for $\tau_L$, respect axionic shift symmetries, so the $C_4$-based QCD axion associated with $\tau_L$ remains light in this setting. As argued in \cite{Cicoli:2012sz} using the results of the explicit CY model built in \cite{Cicoli:2011qg}, the part of the loop potential relevant to $\tau_L$ may plausibly take the form
\be
V_\text{loop} = \left( \frac{\mu_1}{\sqrt{\tau_L}} - \frac{\mu_2}{\sqrt{\tau_L} - \mu_3} \right) \frac{|W_0|^2 M_P^4}{\mathcal{V}^3},
\label{lco}
\ee
where the $\mu_i$ are positive constants. Specifically, $\mu_1$ and $\mu_2$ depend only on the complex structure moduli and are hence fixed below the flux scale. By contrast, $\mu_3$ must scale like the square root of a 4-cycle to conform with the known scaling of such loop corrections. We assume that the geometries of the blow-ups governed by $\tau_L$ and $\tau_s$ are related just in the right way for a term $\sim 1/(\sqrt{\tau_L}-c\sqrt{\tau_s})$ to arise. Below the mass scale of $\tau_s$ we then have a constant $\mu_3=c\sqrt{\tau_s}$ and a loop correction as in \eqref{lco}. Moreover, if $c$ is a `largish' ${\cal O}(1)$ number and $\mu_{1,2}$ are similar in magnitude, then $\tau_L$ is fixed at a size somewhat larger than $\tau_s$. This is precisely what one would like for the cycles supporting the SM brane stacks. The pseudoscalar $a_L$ coming with $\tau_L$ now becomes a suitable QCD axion.

With this, we are ready to study reheating.

\section{Dark radiation II: Absence of overproduction from the decay of other moduli}\label{sec:dr2}
In Section~\ref{sec:dr1}, we have seen that the strong coupling of the volume modulus $\phi_b$ to Higgses leads to a dramatically increased decay rate\footnote{Remember that we denote the canonically normalized moduli and axion fields by $\phi_i$ and $a_i$, respectively.} 
\begin{equation}
    \Gamma_{\phi_b\to hh}\sim c^2_{\rm loop}\, \frac{M_P}{{\mathcal{V}}^{5/2}}\,.
    \label{tpf}
\end{equation}
This is fast enough to call into doubt the standard paradigm by which $\phi_b$, being the lightest modulus, gets to dominate the total energy density and, through its branching ratios, determines the late-time composition of the Universe \cite{Cicoli:2016olq}. Instead, the decay rates and branching ratios of heavier moduli become essential. If they decay more slowly than $\phi_b$, then their decay to $\phi_b$ is followed by an essentially instantaneous decay of $\phi_b$ to Higgses. This is basically equivalent to a direct decay to the SM. As a result, relative late-time abundances are determined by the branching ratios of such heavier moduli to $\phi_b$, to the SM and to other particles.

To analyze such a situation, one needs to specify an inflationary scenario, which we have tried to do in the last section. We will now focus on the subsequent period of reheating, starting with a discussion of the relevant decay rates. 

Jumping ahead, let us state right away that the main players will be the inflaton $\phi_I$, the lightest modulus $\phi_b$ and its axionic partner $a_b$, the loop-stabilized SM-cycle modulus $\phi_L$ and its partner, the QCD axion $a_L$. The inflaton decays to the SM, not just via $\phi_b$ and $\phi_L$, which has a fast decay rate to the SM as well, but also via direct decay to SM gauge bosons due to the mixing between $\phi_I$ and $\phi_L$. In addition, dark radiation arises because $\phi_I$ also decays to $a_b$ and $a_L$, and $\phi_L$ has a non-negligible decay width into $a_L$. However, the total amount of dark radiation turns out to be in agreement with present observational bounds.

We will not discuss the decay rates of the small cycle modulus $\phi_s$ since in K\"ahler moduli inflation its energy density is subdominant with respect to the one of the inflaton $\phi_I$, and in addition its decay rates are identical to those of $\phi_I$. It hence suffices to understand the decay chain starting from $\phi_I$. The question to which extent the inflaton axion $a_I$ is excited after inflation is an interesting one \cite{Bond:2006nc, Barnaby:2009wr}, but we will not discuss it below. However, we will see that the branching ratio of $a_I$ to dark radiation coincides with that of $\phi_I$, so that this issue does not affect our predictions.

In what follows, we only sketch the derivation of the decay rates and present the final results. More detailed calculations can be found in Appendix~\ref{appendix_field_dynamics}.

\subsection{Decay rates}

\subsubsection{The underlying mass hierarchy}

First, we discuss the relevant mass hierarchy. It follows from the K\"ahler potential 
\begin{equation}
K=-2 \ln \left( \mathcal{V} + \xi/2\right)\,,
\end{equation}
with ${\cal V}$ given in \eqref{eq:inflatonfinal}, together with the scalar potential
\begin{equation}
V= V_\text{LVS}^I (\tau_I, c_I,\mathcal V) + 
V_\text{loop} (\tau_L, \mathcal{V})\,.
\end{equation}
Here $\xi\sim 1/g_s^{3/2}$ parameterizes the leading $\alpha'$ correction \cite{Balasubramanian:2005zx, Becker:2002nn}, $V_\text{loop}$ is given in~\eqref{lco} and $V_\text{LVS}^I$ is the usual LVS $F$-term potential, simplified to serve our present purposes. More precisely, we need this scalar potential near the minimum of $\tau_I$ such that we may think of $\tau_I$ as just another small cycle modulus. In addition, we do not need the $\tau_s$ dependence, as argued above. Thus, assuming that the relevant part of the superpotential takes the form 
\be
W=W_0+A_I e^{-\mathfrak{a}_I T_I}\,,
\ee
we have
\begin{equation}
\frac{V_\text{LVS}^I}{M_P^4} = \frac{1}{\mathcal{V}^2} \left[ \frac{8 \tau_b^{3/2} \sqrt{\tau_I}}{3 \gamma_I} \mathfrak{a}_I^2 |A_I|^2 \text{e}^{-2\mathfrak{a}_I \tau_I} +4 \mathfrak{a}_I \tau_I \text{e}^{-\mathfrak{a}_I \tau_I} |A_I W_0| \cos\left( \mathfrak{a}_I c_I \right) \right] + \frac{3 |W_0|^2 \xi}{4 \mathcal{V}^3}\,.
\end{equation}
With this, all relevant masses can be calculated in a straightforward manner (they are of course also well-known in the literature \cite{Balasubramanian:2005zx, Conlon:2005ki}). We summarize them in Table~\ref{tab:masses}. The heaviest fields are the inflaton $\phi_I$ and its axion $a_I$, with equal masses due to their supersymmetric, non-perturbative stabilization. The next-lightest field is the loop-stabilized modulus $\phi_L$, governing the size of the SM cycle. We emphasize that it is heavier than $\tau_b$ for ${\cal V}\gg \tau_L^2$, so that $\tau_b$ is the lightest K\"ahler modulus, even if the potential $V_{\rm loop}$, which fixes $\tau_L$, is subdominant with respect to $V_{\rm LVS}^I$ which fixes $\tau_b$. Finally, during the reheating era both the volume axion $a_b$ and the QCD axion $a_L$ are effectively massless.

{\renewcommand{\arraystretch}{1.8}
\begin{table}[ht]
\centering
\begin{tabular}{ccc}
\hline\hline
Field & scaling of $m_i^2$ & explicit expression of $m_i^2$ \\
\hline
$\phi_I$ & $(\mathfrak{a}_I \tau_I)^2 M_P^2/ \mathcal V^3$ & $\frac{4 |W_0|^2 \mathfrak{a}_I^2 \tau_I^2}{\mathcal{V}^2} M_P^2$  \\
$a_I$ & $(\mathfrak{a}_I \tau_I)^2 M_P^2/\mathcal V^2$ & $\frac{4 |W_0|^2 \mathfrak{a}_I^2 \tau_I^2}{\mathcal{V}^2} M_P^2$ \\
$\phi_b$ & $M_P^2/\mathcal{V}^3$ &  $\frac{m_{11} m_{22} m_{33}- m_{13} m_{22} m_{31} - m_{12} m_{21} m_{33}}{m_{22} m_{33}} \frac{M_P^2}{\mathcal{V}^3}$ \\
$\phi_L$ & $M_P^2/ (\tau_L \mathcal V)^2$ & $\frac{|W_0|^2 \left( 3 \tilde{\mu}^3 \mu_1 + \mu_2 \tau_L \left( - \mu_3 + 3 \sqrt{\tau_L} \right) \right)}{3 \gamma_L \tilde{\mu}^3 \tau_L^2 \mathcal{V}^2}M_P^2$ \\
$a_b$ & $0$ & 0 \\
$a_L$ & $0$ & 0 \\
\hline
\end{tabular}
\caption{Mass squareds of canonical moduli and axion fields. The parameters $\tilde{\mu}$ and $m_{ij}$ are defined below \eqref{second_derivative_matrix} and in~\eqref{expression_m11}~$-$~\eqref{expression_m33}.}
\label{tab:masses}
\end{table}
}

\subsubsection{Inflaton decay to the volume modulus}

Having reviewed the masses, we now turn to the decays. Due to mixing effects, obtaining even the leading order rates is computationally involved. Therefore, we dedicate the following subsections to a qualitative derivation of the main results. In particular these include that kinetic-term-induced inflaton decays dominate over potential-induced decays, that decay rates to axions and corresponding saxions are equal, and that the decays to the SM-cycle (s)axions and the volume (s)axions are comparable. The reader prepared to take this on trust may jump to our summary of decay rate results in Section~\ref{sumdr} and Table~\ref{tab:decay_rates}.

Let us start with the, perhaps simplest, decay rate of the inflaton -- the decay rate to the volume modulus. The relevant lagrangian terms can be obtained by taking derivatives of the (simplified) K\"ahler potential
\be
K=-2\ln(\tau_b^{3/2}-\gamma_I\tau_I^{3/2}).
\ee
With this we find the diagonal part of the kinetic lagrangian
\be
\frac{\cal L}{M_P^2}\,\,\supset \,\,\frac{\tau_b}{{\cal V}^2}\,(\partial \tau_b)^2 + \frac{1}{{\cal V}\sqrt{\tau_I}}(\partial \tau_I)^2\,.
\label{kte}
\ee
Here we disregard ${\cal O}(1)$ factors for brevity, referring to the appendix for more precise formulae.

Expanding the first term of \eqref{kte} in leading order in $\delta\tau_I$, we find the 3-vertex
\be
\frac{\cal L}{M_P^2} \,\,\supset\,\,\frac{\tau_b \sqrt{\tau_I}}{{\cal V}^3}\,\delta\tau_I(\partial \tau_b)^2\,.\label{ktex}
\ee
Taking care of canonical normalization factors and replacing $\partial^2$ by $m_{\tau_I}^2$, the decay amplitude is estimated as
\be
A M_P \,\,\sim\,\, \frac{\tau_b \sqrt{\tau_I}}{{\cal V}^3}\,\frac{{\cal V}^2}{\tau_b}\,\sqrt{\cal V}\tau_I^{1/4}\,m_{\tau_I}^2
\,\,\sim\,\, \frac{\tau_I^{3/4}}{\sqrt{\cal V}}\,m_{\tau_I}^2 \,.
\ee
This gives the rate 
\be
\Gamma_{\phi_I\to \phi_b\phi_b} \sim \frac{A^2}{m_{\tau_I}}\sim \frac{\tau_I^{3/2}}{\cal V}\,\frac{m_{\tau_I}^3}{M_P^2}\sim \frac{\tau_I^{9/2}}{{\cal V}^4}\,M_P\,,
\ee
where we have used $m_{\tau_I}\sim \tau_I/{\cal V}$. A more precise derivation requires not only keeping all the ${\cal O}(1)$ factors but also a diagonalization of the kinetic terms and of the mass matrix. These mixing effects modify the result at an ${\cal O}(1)$ level in many cases. We present the corresponding analysis, following the procedure explained in \cite{Cicoli:2010ha}, in our Appendices~\ref{sec:2_moduli_system} and~\ref{appendix_field_dynamics}. Crucially, we extend that analysis to include 3-vertices arising from the higher-order expansion of the kinetic lagrangian, as was just seen in \eqref{ktex}. In fact, the kinetic-term-induced decays will turn out to be dominant.

Before closing, let us note that in our discussion above, as well as in the following subsections, we work only at leading order in the logarithmically large quantity $\tau_I\sim \tau_s\sim \ln{\cal V}\gg 1$. Thus, further significant corrections may be expected from an even more precise analysis. However, we believe that what we have done is sufficient to support the qualitative conclusions of this paper.

\subsubsection*{Dominance of kinetic over potential terms in the inflaton decay to the volume modulus}

The decay rate of the previous subsection was based on a trilinear term coming from the kinetic lagrangian. Clearly, similar trilinear terms can be derived from the scalar potential. Let us check that their contribution to the decay rate is indeed negligible. In doing so, we will disregard mixing and suppress ${\cal O}(1)$ factors. We start by giving the already familiar contribution from the kinetic term in a slightly different form
\begin{equation}
    \mathcal{L}/M_P^2 \,\,\supset \,\,\sim\,\, (\partial_{\tau_I} K_{jk})\, \delta \tau_I \partial_\mu \delta \tau_j \partial^\mu \delta \tau_k \,\,\sim\,\, m_{\tau_I}^2\, (\partial_{\tau_I} K_{jk})\, \delta \tau_I \delta \tau_j \delta \tau_k\,.
    \label{akin}
\end{equation}
Here $\tau_I$ is the inflaton and $\tau_i,\tau_k$ are its decay products. We have also used the substitution $\partial^2\to m_{\tau_I}^2$, where $m_{\tau_I}$ is the inflaton mass. It can be evaluated as
\begin{equation}
    m_{\tau_I}^2 \sim \frac{\partial_{\tau_I} \partial_{\tau_I} V}{K_{II} M_P^2}\,,
\end{equation}
where $K_{II}$ corrects for the non-canonical normalization of $\tau_I$.

The competing coupling from the scalar potential reads
\begin{equation}
    \mathcal{L} \,\,\supset \,\,\sim\,\, (\partial_{\tau_I} \partial_{\tau_j} \partial_{\tau_k} V) \delta \tau_I \delta \tau_j \delta \tau_k\,.
    \label{apot}
\end{equation}
To compare the amplitudes resulting from \eqref{akin} and \eqref{apot} we now focus on the big cycle, i.e. we set  $\tau_j=\tau_k=\tau_b$, and make use of the approximate relations
\begin{equation}
    \partial_{\tau_I} \partial_{\tau_I} V \sim \mathfrak{a}_I^2 V\,,
    \qquad 
    \partial_{\tau_I} \partial_{\tau_b} \partial_{\tau_b} V \sim \frac{\mathfrak{a}_I}{\tau_b^2} V\,,
    \qquad
    \partial_{\tau_I} K_{bb} \sim \frac{\sqrt{\tau_I}}{\tau_b^{7/2}}\,, \qquad K_{II} \sim \frac{1}{\tau_b^{3/2}\sqrt{\tau_I}}\,.
\end{equation}
For the first two relations we used the fact that $V$ depends on $\tau_I$ primarily through $\exp(-\mathfrak{a}_I\tau_I)$, while its $\tau_b$ dependence is power-like. With this, the ratio of the amplitudes may be estimated as
\begin{equation}
    \frac{A^\text{kin}_{\phi_I \rightarrow \phi_b \phi_b}}{A^\text{pot}_{\phi_I \rightarrow \phi_b \phi_b}} 
    \,\,\sim\,\, 
    \frac{m_{\tau_I}^2 M_P^2 (\partial_{\tau_I} K_{bb})}{(\partial_{\tau_I} \partial_{\tau_b} \partial_{\tau_b} V)}
    \,\,\sim\,\, 
    \frac{(\partial_{\tau_I} \partial_{\tau_I} V) (\partial_{\tau_I} K_{bb})}{K_{II} (\partial_{\tau_I} \partial_{\tau_b} \partial_{\tau_b} V)} \,\,\sim\,\, 
    \mathfrak{a}_I \tau_I \,\,\gg\,\, 1\,. \label{amplitude_ratio_kinetic_potential_decay_into_volume}
\end{equation}

\subsubsection{Inflaton decay to the SM modulus}

An analysis similar to that of the previous subsection can be performed for the decay $\phi_I \rightarrow \phi_L \phi_L$ to the loop-stabilized modulus $\tau_L$ which supports the SM sector. The ratio of kinetic-term- and potential-induced amplitudes follows from the general formula given before after the substitution $\tau_j=\tau_k=\tau_L$. For the purpose of our estimate, we may disregard $\tau_s$ and work with
\be
K=-2\ln{\cal V}\,\,,\qquad {\cal V} = \tau_b^{3/2} - \gamma_L \tau_L^{3/2} - \gamma_I \tau_I^{3/2}\,.
\ee
For the kinetic-term induced decay amplitude, we need
\begin{equation}
    \partial_{\tau_I} K_{LL} \sim \frac{\sqrt{\tau_I}}{\tau_b^3 \sqrt{\tau_L}}\,.
\end{equation}
For the potential-induced couplings, we report the contributions of the LVS and loop potentials, $V_\text{LVS}$ and $V_\text{loop}$, separately. 

Concerning the contribution from $V_{\rm LVS}$, there is a small complication: $V_{\rm LVS}$ is merely a function of ${\cal V}$ and $\tau_I$. In other words, $V_{\rm LVS}$ possesses an exactly flat direction and this direction corresponds to fluctuations of the `loop-stabilized modulus' $\tau_L$. However, it would be too naive to conclude from this that $V_{\rm LVS}$ does not contribute to $\phi_I\to \phi_L \phi_L$. Indeed, let us carefully specify the fluctuation associated with the loop modulus:
As stated above, the LVS potential takes the form $V_\text{LVS} (\mathcal{V} (\tau_b, \tau_I, \tau_L), \tau_I)$. The flat direction is then specified by the two conditions $\delta \tau_I = 0$ and $\delta {\cal V} = 0$. Using the first condition, the second becomes
\begin{equation}
    \delta \mathcal{V} = (\partial_{\tau_b} \mathcal{V}) \delta \tau_b + (\partial_{\tau_L} \mathcal{V}) \delta \tau_L = 0\,.
\end{equation}

Hence, to be aligned with the flat direction at linear order, the variation $\delta \tau_L$ must always be accompanied, at the linear level, by a variation of $\delta \tau_b$,
\be
\delta \tau_b = \delta \tau_b(\delta \tau_L) = \frac{\gamma_L \tau_L^{1/2}}{\tau_b^{1/2}}\,\delta \tau_L\,.
\ee
While this ensures that the corresponding fluctuation of ${\cal V}$ vanishes at linear order, at second order a non-zero fluctuation persists
\be
\delta {\cal V}(\delta \tau_L)= \left( \frac{3 \gamma_L^2 \tau_L}{8 \tau_b^{3/2}} - \frac{3 \gamma_L}{8 \tau_L^{1/2}}\right)\delta \tau_L^2\,.
\ee
For $\tau_b\gg \tau_L$, this is the same as the naive second-order fluctuation 
\be
\frac{1}{2}\,\frac{\partial^2{\cal V}}{\partial \tau_L^2} \delta \tau_L^2 =- \frac{3 \gamma_L}{8 \tau_L^{1/2}}\,\delta \tau_L^2\,.
\ee
Thus, we may proceed by analogy to \eqref{apot}, using the third partial derivative of the potential. Naively, we would expect
\begin{equation}
    \partial_{\tau_I} \partial_{\tau_L} \partial_{\tau_L} V_\text{LVS} \sim \frac{\mathfrak{a}_I}{\tau_b^{3/2} \sqrt{\tau_L}} V_\text{LVS}\,.
\end{equation}
However, the terms where the $\tau_I$-derivative acts directly on the two exponential functions $\sim \partial_{\tau_I} \exp(-2\mathfrak{a}_I \tau_I)$ and $\sim \partial_{\tau_I} \exp(-\mathfrak{a}_I \tau_I)$ cancel. As a result, the leading-order contribution due to the LVS potential suffers a slight further suppression
\begin{equation}
    \partial_{\tau_I} \partial_{\tau_L} \partial_{\tau_L} V_\text{LVS} \sim \frac{1}{\tau_b^{3/2} \tau_I \sqrt{\tau_L}} V_\text{LVS}\,.
\end{equation}

The loop potential manifestly depends of $\tau_L$, so we do not have to be concerned about small corrections of the loop-modulus direction. The corresponding contribution is simply given by
\begin{equation}
    \partial_{\tau_I} \partial_{\tau_L} \partial_{\tau_L} V_\text{loop} \sim \frac{\sqrt{\tau_I}}{\tau_b^{3/2} \tau_L^2} V_\text{loop}\,.
\end{equation}
Since $V_\text{LVS} \sim \tau_I^{3/2} \sqrt{\tau_L} V_\text{loop}$, the third-order derivatives of the individual potentials scale like
\begin{equation}
    \frac{\partial_{\tau_I} \partial_{\tau_L} \partial_{\tau_L} V_\text{LVS}}{\partial_{\tau_I} \partial_{\tau_L} \partial_{\tau_L} V_\text{loop}} \sim \tau_L^2 \gg 1. \label{inflaton_to_loop_decay_scaling_LVS_vs_loop_potential}
\end{equation}
Hence, even though the loop cycle $\tau_L$ is stabilized by $V_\text{loop}$, the potential decay rate into its modulus is dominated by $V_\text{LVS}$. Nevertheless, as in the previous subsection, this potential-induced decay amplitude is too small to compete with the kinetic-term effect,
\begin{equation}
    \frac{A^\text{kin}_{\phi_I \rightarrow \phi_L \phi_L}}{A^\text{pot}_{\phi_I \rightarrow \phi_L \phi_L}} \sim \frac{(\partial_{\tau_I} \partial_{\tau_I} V) (\partial_{\tau_I} K_{LL})}{K_{II} (\partial_{\tau_I} \partial_{\tau_L} \partial_{\tau_L} V)} \sim \mathfrak{a}_I^2 \tau_I^2 \gg 1.
\end{equation}

Before closing, let us also compare the two dominant, kinetic-term-induced decay amplitudes to volume and SM-cycle modulus. This is easily done using \eqref{akin}, with the result that they are parametrically the same
\be
    \frac{A^\text{kin}_{\phi_I \rightarrow \phi_b \phi_b}}{A^\text{kin}_{\phi_I \rightarrow \phi_L \phi_L}} 
\sim \frac{K_{LL} (\partial_{\tau_I} K_{bb})}{K_{bb} (\partial_{\tau_I} K_{LL})} \sim \mathcal{O}(1)\,.
\ee
Here the factor $K_{LL} / K_{bb}$ arises from the transformation of $\delta \tau_b$ and $\delta \tau_L$ to canonical fields.

A more careful and accurate analysis can be found in the Appendices~\ref{sec:2_moduli_system} and~\ref{appendix_field_dynamics}, where we also show that decays of the inflaton into two different decay products are suppressed by powers of $\tau_b^{-1}$ and hence irrelevant.

\subsubsection{Inflaton decay to axions: equality with the decay rate to saxions}

We have seen that, in transitions between K\"ahler moduli, kinetic-term-induced decays dominate over potential-induced decays. For the dominant, kinetic-term-induced rates we can make the following important observation: The decay rate of a certain K\"ahler modulus into two lighter K\"ahler moduli is equal to the decay rate of the same K\"ahler modulus into the two axions associated with these lighter moduli. While this appears natural due to supersymmetry, it is technically not immediately obvious if one considers the explicit amplitudes. 

To make our point, we focus on the most important case of the decaying inflaton $\tau_I$. As decay products, consider either the big-cycle modulus $\tau_b$ or the loop-stabilized modulus $\tau_L$. Let us refer to them collectively as $\tau_i$ (with $i=b$ or $i=L$) and to their axions as $c_i$. Thus, we claim that the kinetic-term-induced rates $\phi_I \rightarrow \phi_i \phi_i$ and $\phi_I \rightarrow a_i a_i$ coincide. The relevant trilinear couplings are
\begin{align}
    \mathcal{L}_{\phi_I \rightarrow \phi_i \phi_i}^\text{kin} / M_P^2 &= \partial_{\tau_I} K_{ii} \delta \tau_I \partial_\mu \tau_i \partial^\mu \tau_i + \partial_{\tau_i} K_{Ii} \delta \tau_i \partial_\mu \tau_I \partial^\mu \tau_i + \partial_{\tau_i} K_{iI} \delta \tau_i \partial_\mu \tau_i \partial^\mu \tau_I, 
    \label{kinf}
    \\
    \mathcal{L}_{\phi_I \rightarrow a_i a_i}^\text{kin} / M_P^2 &= \partial_{\tau_I} K_{ii} \delta \tau_I \partial_\mu c_i \partial^\mu c_i\,\,,
    \label{kins}
\end{align}
where no sum over $i$ is implied. As just noted, the equality of the resulting rates is not obvious since the last two terms of \eqref{kinf} arise due to the dependence of the K\"ahler metric on $\tau_i$. These terms are missing in \eqref{kins} because, by shift symmetry, no $c_i$-dependence is possible.

It is, however, straightforward to convince oneself that the decay rates nevertheless agree. To do so, replace the derivatives $\partial^2$ by the inflaton mass squared $m_{\tau_I}^2$, making use of the fact that the $\tau_i$ mass is negligible in this context (cf.~\eqref{eliminate_partial_derivatives} and the subsequent discussion). This gives
\begin{align}
\frac{\mathcal{L}_{\phi_I \rightarrow \phi_i \phi_i}^\text{kin}}{ M_P^2} &= \frac{m_{\tau_I}^2}{2} \left( \partial_{\tau_I} K_{ii}  \delta \tau_I \delta \tau_i \delta \tau_i - \partial_{\tau_i} K_{Ii} \delta \tau_i \delta \tau_I \delta \tau_i - \partial_{\tau_i} K_{iI} \delta \tau_i \delta \tau_i \delta \tau_I \right) \nonumber\\
        &= -\frac{m_{\tau_I}^2}{2} \partial_{\tau_I} K_{ii}  \delta \tau_I \delta \tau_i \delta \tau_i, 
        \label{maf}
        \\
\frac{\mathcal{L}_{\phi_I \rightarrow a_i a_i}^\text{kin}}{M_P^2} &= \frac{m_{\tau_I}^2}{2} \partial_{\tau_I} K_{ii} \delta \tau_I \delta c_i \delta c_i,
    \label{mas}
\end{align}
where we also used the invariance of $\partial_{\tau_i} K_{jk}$ under permutations of $i,j$ and $k$. Now note that the kinetic terms of $\delta \tau_i$ and $\delta c_i$ have the same prefactor, in other words, the corresponding canonical fields arise from an identical rescaling: $\delta \phi_i \sim \sqrt{K_{ii}} \delta \tau_i$ and $\delta a_i \sim \sqrt{K_{ii}} \delta c_i$. Thus, the amplitudes following from \eqref{maf} and \eqref{mas} differ only by a minus sign, leading to identical decay rates.

In our argument, we have, so far, disregarded the fact that $\tau_I$, $\tau_b$ and $\tau_L$ together with their axions do not represent the field basis relevant for scattering since neither their kinetic nor their mass terms are precisely diagonal. Concerning the saxions, this is of course easily remedied by a linear field redefinition $\tilde{\tau}_\alpha\equiv C_{\alpha\beta}\tau_\beta$, where $\alpha,\beta$ take values in $\{I,b,L\}$ or in any other set of the K\"ahler moduli. Now, in the present subsection we may treat all fields except $\tau_I$ and $c_I$ as massless.\footnote{Note that the mass for $\tau_I$ comes from a superpotential term and hence singles out $\tau_I$ together with its axion $c_I$.} Thus, we may promote our proposed field redefinition above to the set of complex fields: $\tilde{\tau}_\alpha + {\rm i} \tilde{c}_\alpha \equiv C_{\alpha\beta}(\tau_\beta+{\rm i} c_\beta)$, still obtaining a diagonal kinetic and mass lagrangian in the new basis. But such a linear coordinate change on the K\"ahler manifold does not interfere with our earlier analysis leading to \eqref{maf} and \eqref{mas}. Thus, our conclusion about equal rates for decays to saxions and axions stands up also in the presence of mixing.

Finally, note that potential-induced decays to axions are irrelevant: The potential of the QCD axion is $\sim \Lambda_{\rm QCD}^4$, that of the big-cycle axion even smaller. This is negligible in our context. Moreover, decays of the inflaton to its own axion are kinematically forbidden since their masses are approximately equal. By contrast, the small-cycle axion could be lighter. Indeed, while $\tau_I$ and $\tau_s$ are qualitatively equivalent in our minimalist setting, their masses depend on $\gamma_I,A_I$ and $\gamma_s,A_s$ respectively. However, without loss of generality we can assume that the inflaton is the lighter of the two. It is then also lighter than the small-cycle axion, preventing any decays $\phi_I\to a_s a_s$.

\subsubsection{Inflaton decay to SM gauge bosons via mixing with the SM-cycle modulus}

In addition to the decays of the inflaton into light moduli and axions which have been discussed above, a direct coupling of the inflaton to SM gauge bosons $A_\mu$ is also present. Such a coupling arises from the mixing between the inflaton modulus $\tau_I$ and the loop-stabilized modulus $\tau_L$. The latter couples to SM gauge bosons through
\begin{equation}
    \mathcal{L} \supset \,\,\sim\, \tau_L\, \text{tr} F_{\mu \nu} F^{\mu \nu}.
\end{equation}
Writing $\tau_L=\langle\tau_L\rangle + \delta \tau_L$ and normalizing the gauge fields canonically, one finds the relevant trilinear coupling
\begin{equation}
\mathcal{L} \supset \frac{1}{2} \text{tr} F^\text{can}_{\mu \nu} F_\text{can}^{\mu \nu}
\,+\, \frac{1}{2} \frac{\delta \tau_L}{\vev{\tau_L}} \text{tr} F^\text{can}_{\mu \nu} F_\text{can}^{\mu \nu}\,.
\end{equation}
We keep working with these canonical gauge fields below but drop the label `can' for brevity. From the mixing effects specified in~\eqref{transformation_to_canonical} and~\eqref{transformation_vector_v_I}, we infer the advertised direct coupling between the canonical inflaton $\delta \phi_I$ and the canonical gauge bosons
\begin{align}
    \mathcal{L} &\supset \frac12 \frac{(\vec{v}_I)_L \delta \phi_I}{\sqrt{2} \vev{\tau_L}} \text{tr} F_{\mu \nu} F^{\mu \nu} \nonumber \\
    &= \frac{1}{2 \sqrt{2} \vev{\tau_L}} \left(\frac{4 (m_{12} m_{31} + m_{22} m_{32}) \vev{\tau_I}^{1/4}}{\sqrt{6 \gamma_I} m_{22} (m_{22} - m_{33} ) \vev{\tau_b}^{3/4}}\right) \delta \phi_I \text{tr} F_{\mu \nu} F^{\mu \nu} \\
    &\approx \frac{ m_{32} \vev{\tau_I}^{1/4}}{\sqrt{3 \gamma_I} m_{22} \vev{\tau_L} \vev{\tau_b}^{3/4}} \delta \phi_I \text{tr} F_{\mu \nu} F^{\mu \nu}
    \approx \frac{  \sqrt{3 \gamma_I} \vev{\tau_I}^{3/4}}{2 \vev{\tau_b}^{3/4}} \delta \phi_I \text{tr} F_{\mu \nu} F^{\mu \nu}\,.\nonumber
\end{align}
Here the $m_{ij}$ are given in~\eqref{expression_m11}~$-$~\eqref{expression_m33} and, in the step from the second to the last line, we made use of the approximate relations $m_{22} m_{32} / (m_{12} m_{31}) \sim \mathfrak{a}_I^2 \tau_I^2 \tau_L^2\gg 1$ and $m_{22} / m_{33} \sim \mathfrak{a}_I^2 \tau_I^2 \tau_L^2\gg 1$. The resulting decay rate reads
\begin{align}
    \Gamma_{\phi_I \rightarrow A A} \,\approx\, \frac{3 \gamma_I N_g}{64 \pi} \left(\frac{\vev{\tau_I}}{\vev{\tau_b}} \right)^{3/2} \frac{m_{\phi_I}^3}{M_P^2}
    \,\approx\, \frac{3 \gamma_I N_g |W_0|^3 \mathfrak{a}_I^3 \tau_I^{9/2}}{8 \pi \mathcal{V}^4} M_P\,.
\end{align}

\subsubsection{Subdominant inflaton decays to Higgses}

One may wonder whether the direct coupling of $\tau_I$ to the Higgs drastically changes the decay, in analogy to what happened with the enhanced $\tau_b$ decay of Section~\ref{subsec:rapid}. To understand this, recall how the relevant decay amplitude of $\tau_b$ arose from \eqref{mhv}: The fluctuation of the canonical field $\phi_b$ induces a fluctuation of the $\ln{\cal V}$ term in that equation, which is described by the simple relation
\be
\delta \ln{\cal V}\sim \delta\phi_b / M_P\,. \label{dvb}
\ee
The fluctuation of $\phi_I$ analogously induces a fluctuation of $\ln{\cal V}$, with the relevant relation this time being
\be
\delta \ln{\cal V} \sim 
\delta \ln(\tau_b^{3/2}-\gamma_I\tau_I^{3/2}) 
\sim (\sqrt{\tau_I}/{\cal V}) \,\delta \tau_I 
\sim (\tau_I^{3/4}/\sqrt{\cal V}) \,\delta \phi_I / M_P\,.
\label{dvi}
\ee
Comparing \eqref{dvb} and \eqref{dvi}, we see that the amplitude squared for the decay of $\phi_I$ to two Higgs fields is suppressed by a factor $\tau_I^{3/2}/{\cal V}$ relative to the corresponding amplitude squared for $\phi_b$. In the ratio of decay rates, a further suppression factor $m_{\tau_b}/m_{\tau_I}\sim 1/(\tau_I \sqrt{\cal V})$ comes in. Thus, on the basis of \eqref{tpf} we conclude that
\be
\Gamma_{\phi_I\to hh} \sim c^{2}_{\rm loop}\frac{\sqrt{\tau_{I}}}{{\mathcal V}^4}M_{P}\,.
\ee
Compared with the kinetic-term- and mixing-induced decay rates of the inflaton, this rate is down by $c^2_{\rm loop}/\tau_I^4$. Thus, it only has a small effect on the branching ratios.

\subsubsection{Summary of decay rates}
\label{sumdr}

We summarize the main findings of this section by noting that kinetic decays of the inflaton dominate the potential ones and that their branching ratios into light moduli fields and their respective axions are equal. The results for several decay channels of the inflaton and its axion are summarized in Table~\ref{tab:decay_rates}. To keep the table of manageable size we normalize the decay rates to two suitable channels for the inflaton decay,
\begin{align}
    \Gamma_1 &\equiv \Gamma^\text{kin}_{\phi_I \rightarrow \phi_b \phi_b} \approx \frac{3 \gamma_I |W_0|^3 \mathfrak{a}_I^3 \tau_I^{9/2}}{64 \pi \mathcal V^4} M_P, \\
    \Gamma_2 &\equiv \Gamma^\text{pot}_{\phi_I \rightarrow \phi_L \phi_L} \approx \frac{ 3 \gamma_I \sqrt{\tau_I} \left[ - 3 |W_0|^2 \gamma_L \tilde{\mu}^4 \tau_L^2 + W_0^2 \left(  -4 \mu_1 \tilde{\mu}^4 + \mu_2 ( \mu_3^2 - 4 \mu_3 \sqrt{\tau_L} + 4 \tau_L) \tau_L \right)  \right]^2}{64 \pi \gamma_L^2 |W_0| \tilde{\mu}^8  \mathfrak{a}_I \tau_L^4 \mathcal{V}^4} M_P.
\end{align}
Note that $\Gamma_1 / \Gamma_2 \sim \mathfrak{a}_I^4 \tau_I^4 \sim (\ln \mathcal V)^4 \gg 1$. We emphasise the appearance of the number $N_g=12$ of SM gauge bosons in the fourth-to-last and the last line, making these the dominant decay channels.

{\renewcommand{\arraystretch}{1.8}
\begin{table}[t]
\centering
\begin{tabular}{clc}
\hline\hline
Decay rate & \; scaling & explicit \\
\hline
$\Gamma^\text{kin}_{\phi_I \rightarrow \phi_b \phi_b}$ & $\sim (\ln \mathcal{V})^{9/2} \mathcal{V}^{-4} M_P $ & $\Gamma_1$ \\
$\Gamma^\text{pot}_{\phi_I \rightarrow \phi_b \phi_b}$ & $\sim (\ln \mathcal{V})^{5/2} \mathcal V^{-4} M_P$ & $4 \Gamma_1 / (\mathfrak{a}_I \tau_I)^2$  \\
$\Gamma^\text{kin}_{\phi_I \rightarrow \phi_L \phi_L}$ & $\sim (\ln \mathcal{V})^{9/2} \mathcal{V}^{-4} M_P $ & $4 \Gamma_1$ \\
$\Gamma^\text{pot}_{\phi_I \rightarrow \phi_L \phi_L}$ & $\sim (\ln \mathcal{V})^{1/2} \mathcal{V}^{-4} M_P $ & $\Gamma_2$ \\
$\Gamma^\text{kin}_{\phi_I \rightarrow a_b a_b}$ & $\sim (\ln \mathcal{V})^{9/2} \mathcal{V}^{-4} M_P $ & $\Gamma_1$ \\
$\Gamma^\text{kin}_{\phi_I \rightarrow a_L a_L}$ & $\sim (\ln \mathcal{V})^{9/2} \mathcal{V}^{-4} M_P $ & $4\Gamma_1$ \\
$\Gamma_{\phi_I \rightarrow A A}$ & $\sim (\ln \mathcal{V})^{9/2} \mathcal{V}^{-4} M_P $ &  $8 N_g \Gamma_1$ \\
$\Gamma^\text{kin}_{a_I \rightarrow \phi_b a_b}$ & $\sim (\ln \mathcal{V})^{9/2} \mathcal{V}^{-4} M_P $ & $2\Gamma_1$ \\
$\Gamma^\text{kin}_{a_I \rightarrow \phi_L a_L}$ & $\sim (\ln \mathcal{V})^{9/2} \mathcal{V}^{-4} M_P $ & $8\Gamma_1$ \\
$\Gamma_{a_I \rightarrow AA}$ & $\sim (\ln \mathcal{V})^{9/2} \mathcal{V}^{-4} M_P $ & $8N_g\Gamma_1$ \\
\hline
\end{tabular}
\caption{Decay rates of inflaton into moduli and axion fields. The explicit decay rates are defined as $\Gamma_1 \equiv \Gamma^\text{kin}_{\phi_I \rightarrow \phi_b \phi_b}$ and $\Gamma_2 \equiv \Gamma^\text{pot}_{\phi_I \rightarrow \phi_L \phi_L}$ where $\Gamma_1 \gg \Gamma_2$.}
\label{tab:decay_rates}
\end{table}

\subsection{A non-vanishing but acceptable amount of dark radiation}

As we can see from Table~\ref{tab:decay_rates}, all these decay rates are much smaller than those for the rapid decays of the volume modulus $\phi_b$ into Higgses discussed in Section~\ref{subsec:rapid}. We are therefore in a situation where after inflation the inflaton itself is the longest-lived modulus\footnote{Also the decay of the SM-cycle modulus is much faster: $\Gamma_{\phi_L} / \Gamma_{\phi_I} \sim \mathcal{V}^2$.} and therefore its decays determine the composition of the energy density into Standard Model parts and those in dark radiation.

Note that this would not be the case if the inflaton 4-cycle were wrapped by a hidden sector D7-stack, as in the model considered in \cite{Allahverdi:2020uax}, where the inflaton decays before the volume mode. In \cite{Allahverdi:2020uax} dark matter is a WIMP living on the SM-cycle with mass of order $m_{3/2}\sim \mathcal{O}(10^{10})$ GeV. Such a superheavy WIMP would normally be overproduced. However this is not the case since its relic abundance is suppressed by the tiny initial branching ratio from the inflaton decay (which decays primarily to the hidden sector wrapping $\tau_I$) and the subsequent dilution due to the volume mode decay. However, in light of the new results of Section \ref{subsec:rapid} which imply a faster decay of the volume modulus, and so less dilution, the WIMP DM abundance computed in \cite{Allahverdi:2020uax} could very likely be underestimated. 

The comment above implies that, if stable, neutralinos with a mass of order $m_{3/2}$ produced from the inflaton decay, would also overproduce DM in our model since their relic abundance would not be diluted by the decay of any light modulus. We therefore consider a realization of the SM where $R$-parity is badly broken, so that neutralinos are unstable.

\bigskip 
Let us now turn to the main point. Based on Table~\ref{tab:decay_rates}, it is straightforward to obtain the branching ratio of $\phi_I$ to dark radiation: From the decays of $\phi_I$ to $a_b, \phi_b, a_L, \phi_L,$ and to gauge bosons $A$, we have
\be
\frac{\Gamma_{\phi_I\to {\rm DR}}}{\Gamma_1}=1+4\left(1+\frac{1}{8N_g+1}\right)\,\,,\qquad 
\frac{\Gamma_{\phi_I\to {\rm SM}}}{\Gamma_1}=1+4\left(1-\frac{1}{8N_g+1}\right)+8N_g\,.
\ee
Here the corrections $\pm 1/(8N_g+1)$ account for the fact that $\phi_L$ branches to the SM and the SM axion in the ratio $8N_g\, :\,1$, with $N_g$ the number of SM gauge bosons. It now follows that
\begin{equation}
    BR(\phi_{I}\to {\rm DR})
    = \frac{5+4/(8N_g+1)}{8N_g+10}
    \simeq \frac{5}{8N_{g}} \simeq 0.05\,.
\end{equation}

The branching ratio $BR(a_{I}\to {\rm DR})$ of the inflaton-axion is identical. This result can be immediately understood from Table~\ref{tab:decay_rates} as follows: On the one hand, the decay rates of $a_I$ to $\phi_b a_b$ and $\phi_L a_L$ are twice the corresponding decay rates of $\phi_I$ to $\phi_b \phi_b$ and $\phi_L \phi_L$. On the other hand, the energy fraction from these decays going directly to dark radiation is only one half. Since the decay rate to SM gauge bosons is equal between $\phi_I$ and $a_I$ (both rates come from mixing with $\phi_L/a_L$ and the equivalent couplings to tr$FF$ and tr$F\tilde{F}$ respectively), our initial claim follows.

Using formula~\eqref{eq:deltaneff} for the effective numbers of neutrino species, and the minimal $N_g=12$, we then have
\begin{equation}
    \Delta N_{\rm eff}\simeq 0.3 \left(\frac{11}{g_* ^4g^{-3}_{*,S}}\right)^{1/3}\simeq 0.14\,.
\end{equation}
Here the final numerical value is obtained for $g_* = g_{*,S} = 106.75$, as appropriate~\cite{Husdal:2016haj} for the typically moderately high reheating temperatures in our scenario (see below). But the result remains consistent with the observational bound~\eqref{eq:deltaneffob} even for the lowest reheating temperatures and corresponding small values of the effective number of degrees of freedom. Yet, the produced amount of dark radiation is close to the boundary of the region allowed by observation. Therefore, it is well within the reach of future CMB experiments~\cite{CMB-S4:2016ple}.

\section{Resulting axion dark matter cosmology}
\label{sec:cosmofinal}

The discussion of the previous sections suggests a significantly changed cosmological scenario.
It still features a long-lived modulus, but this is now the inflaton $\phi_I$ and not the volume modulus. In the following we want to briefly collect the main results for the relevant axion phenomenology and cosmology.

For the sake of simplicity and concreteness, we will ignore some of the $\mathcal{O}(1)$ parameters by fixing them to unity. The more general analysis including the dependence on these parameters is sketched in Appendix~\ref{app:parameters}. We will also again make use of the approximate analytical formulae of~\cite{Visinelli:2009kt,Arias:2021rer} to obtain order-of-magnitude estimates for the most important quantities.

As already discussed in Section~\ref{sec:cosmo}, the axion dark matter abundance depends significantly on whether the axion starts its oscillations before or after reheating. The crucial ingredient that determines the reheating temperature is the decay rate of the longest-lived modulus, which we now have identified to be the inflaton $\phi_I$. Its decay rates are given in Table~\ref{tab:decay_rates}. Choosing $\gamma_{I}=1$, $\mathfrak{a}_{I}=2\pi$, $W_{0}=1$ and $\mathfrak{a}_{I}\tau_{I}\approx\ln({\mathcal{V}}/W_{0})$ (cf.~\eqref{replacement_relations}) and $N_g=12$, we have
\begin{equation}
\label{eq:totdecay}
\Gamma_{\phi_I}^{\rm  tot}\approx \left(10+8N_g\right)\,\Gamma_1\sim 0.1  \times {\mathcal{V}}^{-4}(\ln{\mathcal{V}})^{9/2} M_P\,.
\end{equation}

For the expressions of $f_a$ and $H_I$ we use~\eqref{axion_decay_constant} and~\eqref{H_I_LVS_scaling} respectively,
\begin{equation}
    f_a \simeq \frac{M_P}{\sqrt{2}\pi \tau_L^{1/4} \sqrt{\mathcal{V}}}\,\,,  \qquad H_{I}\simeq \frac{M_P}{{\mathcal{V}}^{3/2}}\,.
\end{equation}
Here $\tau_L = 1/ (2 \alpha_{s,UV}) = 25/2$ and we have set the $\mathcal{O}(1)$ prefactor and $\beta |W_0|^2$ to unity.

The decay rate \eqref{eq:totdecay} is still slow enough that we expect $\phi_I$ to become non-relativistic and to dominate the Universe before it decays. Then $\phi_I$ reheats the Universe to a temperature
\begin{equation}
\label{eq:reheatingT}
    T_r \sim\left(\frac{g_*\pi^2}{90}\right)^{-1/4}\sqrt{\Gamma_{\phi_I}^{\rm tot} M_P}\sim 1\,{\rm GeV}\left(\frac{g_*}{80}\right)^{-1/4} \left(\frac{2.4\times 10^{10}}{{\mathcal{V}}}\right)^{2}\left(\frac{\ln\mathcal{V}}{\ln(2.4\times 10^{10})}\right)^{9/4}.
\end{equation}

In the following, we want to assess the phenomenological consequences of two different scenarios: One is characterized by a high reheating temperature implying that the axion starts oscillating only after reheating during a radiation-dominated Universe, whereas the other scenario describes a low reheating temperature where the axion starts oscillating already before reheating during a period of early matter domination. 

\bigskip

The calculation in the next two subsections follows essentially the same steps as in equivalent subsections of Section~\ref{sec:cosmo}.

\subsection{High reheating temperatures, standard radiation-dominated cosmology}

Using the results of~\cite{Visinelli:2009kt,Arias:2021rer} we know that for a reheating temperature
\begin{equation}
    T_r\gg 1\,{\rm GeV}
\end{equation}
the axion indeed starts oscillating in a radiation dominated phase, and hence follows a more or less standard cosmology.
From~\eqref{eq:reheatingT} we can see that being in this regime requires the volume not to be too large and thus provides us an upper bound for $\mathcal V$.

If the axions constitute all of dark matter, we also need to avoid excessive isocurvature fluctuations.
This requires the Hubble scale of inflation not to be too large and accordingly the volume should not be too small, which yields a lower bound on $\mathcal{V}$.

Combining the two requirements and using the relevant equations of Section~\ref{subsec:standardexpansion} to translate them into bounds on the other parameters, we have (setting $\kappa=1$)
\begin{alignat}{2}
    1\times 10^7 \;\;&\lesssim {\mathcal{V}}&&\lesssim \;\;2\times 10^{10},\label{eq_constraints_final_high_V}\\
    9\times 10^{13}\,{\rm GeV} \;\;& \gtrsim f_{a} &&\gtrsim\;\; 2\times 10^{12}\,{\rm GeV},\\
    6\times 10^{-8}\,{\rm eV}\;\;&\lesssim m_{a}&&\lesssim \;\; 3\times 10^{-6}\,{\rm eV},\\
    2\times 10^6\,{\rm GeV}\;\; &\gtrsim T_r &&\gtrsim \;\; 1\,{\rm GeV},\\
    7\times 10^7\,{\rm GeV}\;\; &\gtrsim H_I &&\gtrsim \;\; 1 \times 10^3\,{\rm GeV},\\
     0.1\;\; &\lesssim \theta_i&&\lesssim \;\; 0.5 .\label{eq_constraints_final_high_theta}
\end{alignat}
Here the constraints on the left-hand side arise due to isocurvature fluctuations.
The right-hand side is the requirement for being in the regime of high reheating temperature. It does therefore not represent an actual limit.

\subsection{Low reheating temperatures, axion oscillates during matter-domination}

Let us now turn to lower temperatures, for which the
axion starts oscillating already during the $\phi_I$-dominated phase where the equation of state is matter-like.

The dark matter density is then given by~\eqref{eq:densitymatter} (adapted from~\cite{Visinelli:2009kt,Arias:2021rer}).
For \eqref{eq:densitymatter} to be applicable the reheating temperature needs to be sufficiently low, which is why we consider
\begin{equation}
    T_r\ll 300\,{\rm MeV}.
\end{equation}

Analogously to before, being in this regime of low temperature implies a lower boundary on the volume. Moreover, requiring that the axion constitutes all of dark matter for an initial misalignment angle that is not tuned large, $\theta_i\leq 3$, provides an upper bound on $\mathcal{V}$. Using the relevant equations in Section~\ref{subsec:matterdomination}, we then find
\begin{alignat}{2}
     5\times 10^{10} \;\;&\lesssim {\mathcal{V}}&&\lesssim\;\; 8\times 10^{10},\label{eq_constraints_final_low_V}\\
    1.4\times 10^{12} \,{\rm GeV}\;\;& \gtrsim f_a && \gtrsim\;\; 1.0 \times 10^{12}\,{\rm GeV},\\
    4\times 10^{-6}\,{\rm eV}\;\;&\lesssim m_a&&\lesssim \;\; 6 \times 10^{-6}\,{\rm eV},\\
    300\,{\rm MeV} \;\;&\gtrsim T_r &&\gtrsim \;\; 150 \,{\rm MeV}, \\
      250\,{\rm GeV} \;\;&\gtrsim H_I &&\gtrsim \;\;100\,{\rm GeV}, \\
    1\;\;&\lesssim \theta_i&&\lesssim \;\; 3 .\label{eq_constraints_final_low_theta}
\end{alignat}
Similarly to the previous case, the constraints on the left-hand side arise from requiring that the approximation of low $T_r$ is valid, whereas 
the right-hand side arises from the requirement of a sufficiently large dark matter density without too much tuning of $\theta_i$.

\subsection{Low mass of the volume modulus}
\label{sec:lowmassb}

In the above we have assumed that the volume modulus decays rapidly into Higgses.  Let us briefly check whether this is always the case, i.e.~whether $m_{b}>2m_{h}$, and what happens if this conditions is violated.

Naively applying the simple estimate~\eqref{mass_volume_modulus} at the lower temperature boundary of the high reheating temperature case the volume modulus has a mass
\begin{equation}
    m_{b}\sim 660\,{\rm GeV}\,. 
\end{equation}
While this indicates that the volume modulus is still sufficiently heavy that it can decay into Higgses, its mass is nevertheless relatively close to the threshold.
In the low reheating temperature region the situation is even more uncertain:
For the largest volumes we have
\begin{equation}
    m_{b}\sim 110\,{\rm GeV}\,.
\end{equation}
This is below twice the Higgs mass. Therefore, in this region the rapid decay into Higgses is excluded (unless the neglected O(1) factors take the mass close to or above threshold). In this case the volume modulus may again start to play a significant role. Moreover, it is crucial to remember that \eqref{mass_volume_modulus} is at best a rough estimate and it is far from given that $W_0=1$. Let us therefore at least briefly comment on what happens if the mass is close to or even below the two Higgs threshold.

For this let us first note that the trilinear term \eqref{eq:trilinear} responsible for the decay into two Higgses, is essentially a linear Higgs portal term (cf., e.g.,~\cite{Agrawal:2021dbo} and references therein for investigations of this). Due to electroweak symmetry breaking, this term therefore leads to a mixing between the Higgs and the volume modulus
\begin{equation}
    \tan(\vartheta)\sim 
    \begin{cases}
    \frac{c_{\rm loop}v m_{3/2}^2}{m^{2}_{\tau_b}M_P}\sim c_{\rm loop}\frac{v}{M_P}{\mathcal{V}}
    & {\rm for}\;\;m_{\tau_b}\gg m_H
    \\
    1
    &{\rm for}\;\; m_{\tau_b}\simeq m_H\\
    \frac{c_{\rm loop}v m_{3/2}^2}{m^{2}_{h}M_P} \sim c_{\rm loop}\frac{M_P}{v}{\mathcal{V}}^{-2}
    &{\rm for}\;\; m_{\tau_b}\ll m_H
    \end{cases}\, .
\end{equation}
Note that in the intermediate region we have a resonance which allows for ${\mathcal{O}}(1)$ mixing.

Well below the threshold $m_{\tau_b}\ll 2m_{H}$ we therefore expect
\begin{equation}
    \Gamma_{\phi_b} \sim \Gamma_{H}(m_{\tau_b}) \sin^{2}(\vartheta)\sim 10^{-16} \left(\frac{c_{\rm loop}}{1/100}\right)^2 \left(\frac{10^{11}}{\mathcal{V}}\right)^4 \Gamma_{H}(m_{\tau_b}({\cal V})),
\end{equation}
where $\Gamma_H(m_{\tau_b})$ is the decay rate which the SM Higgs would have if its mass were equal to $m_{\tau_b}$. One can convince oneself that, at our level of precision, this is the right quantity to consider for estimating the decay rate of a light state whose decay is dominated by its mixing with the Higgs.

We also have to account for the fact that the volume moduli are produced relativistically from the decays of the much heavier inflaton. The typical $\gamma$ factor is given by
\begin{equation}
    \gamma_{\rm typ}\sim \frac{m_{\tau_I}}{m_{\tau_b}}\sim {\mathcal{V}}^{1/2}.
\end{equation}
This implies a typical decay time
\begin{equation}
    t_{\rm typ}\sim \frac{\gamma_{\rm typ}}{\Gamma_{\phi_b}}\sim 1\,{\rm s}\left(\frac{1/100}{c_{\rm loop}}\right)^{2}\left(\frac{{\mathcal{V}}}{10^{11}}\right)^{4.5}\left(\frac{\Gamma_{H}(50\,{\rm GeV})}{\Gamma_{H}(m_{\tau_b}({\cal V}))}\right)\,,
\end{equation}
where we have used $\Gamma_{H}(50\,{\rm GeV})\sim 1.5\,{\rm MeV}$ from~\cite{Gomez-Bock:2007azi}. Hence, if the mass of the volume modulus is suppressed by a modest factor of $2m_H/(50\,{\rm GeV})\sim 1/5$ we are already in danger of spoiling BBN due to late decays of the volume modulus.

\bigskip

In addition to the cosmology discussed here, it may also be interesting to study experimental probes of Higgs mixing, cf., e.g.~\cite{Beacham:2019nyx,Agrawal:2021dbo}.

\section{Discussion and Conclusions}
\label{sec:conclusions}

To fully describe the cosmology of axions in a stringy setup, we not only need to realize a QCD axion, but also have a model that at the same time allows for a description of important events in the cosmological history, in particular inflation and the subsequent reheating. Additional information may arise from taking into account electroweak symmetry breaking and the resulting couplings to the SM Higgs.   

As is well known, the QCD axion can be realized in the Large Volume Scenario. Insisting on an acceptable cosmology with QCD axion dark matter imposes a strong constraint not only on the axion, but also on inflation. As simple stringy scenarios favor a situation where the axion is already present during inflation, isocurvature constraints require a rather low scale of inflation, leading us to use K\"ahler moduli inflation.

The Large Volume Scenario generically features moduli with couplings that are suppressed by powers of the large volume.
These moduli are then long-lived and thus become non-relativistic, typically leading to a matter dominated phase in the cosmological evolution. Standard Cosmology, with its early radiation dominated phase only starts with the decay of the last of these moduli.

Up to now the leading candidate for the longest-lived modulus was the volume modulus, with its decay to axions leading to the usual problem of excessive amounts of dark radiation. However, taking into account the volume dependence of the gaugino loop corrections to the Higgs potential, as well as the significant tuning that has to take place in scenarios where the natural scale for the Higgs mass is the supersymmetry breaking scale $m_{3/2}$,
we find that the resulting couplings of the Higgs to the volume modulus lead to fast decays of the volume modulus to SM Higgses. This solves the original dark radiation problem.

At this point, however, we may want to take the next step and also include a model of inflation. In the context of string moduli inflation, the obvious candidates are K\"ahler moduli and Fibre inflation. As already mentioned, taking account of isocurvature constraints from axion dark matter guides us to a lower inflation scale and hence, K\"ahler moduli inflation. In this model it is then the inflaton that features the longest lifetime. 
It, too, can decay to axions and thereby contribute to dark radiation. In fact, it decays with equal rates to the QCD axion and its saxion, with the latter immediately decaying to the SM. Similarly, the inflaton features equal decay rates to the volume modulus and its axion. It would then appear that half of the inflaton energy goes to dark radiation, such that the dark radiation hydra raises another one of its ugly heads. Fortunately, this is not the end of the story: The inflaton mixes with the QCD-saxion, i.e.~the modulus governing the SM gauge couplings. Since the latter couples to the 12 SM gauge bosons, the inflaton acquires a significant branching fraction into those vector fields. This eventually leads to the demise of the dark radiation hydra. Nevertheless, the amount of dark radiation is typically non-negligible, opening the possibility to see it in astrophysical observations~\cite{CMB-S4:2016ple}, and to directly detect it in experiments such as the International Axion Observatory (IAXO)~\cite{Armengaud:2019uso}, potentially opening the possibility to test its origin from reheating~\cite{Conlon:2013isa,Jaeckel:2021gah}.

Let us note that the viable volume range we found is only marginally consistent with the constraints on the volume derived from the CMB normalization of K\"ahler inflation in~\cite{Conlon:2005jm}. This is not surprising since, for the upper end of our volume range, $H_I$ is exceptionally low and the potential must then be very flat indeed to account for the observed scalar perturbations. While in K\"ahler moduli inflation exponential flatness arises by construction, one would presumably need to invoke fine-tuning (maybe including more than one instanton or the interplay with a loop effect) to make the extremely large ${\cal V}$ regime phenomenologically viable. 

The requirements inherent in our desire to realize a stringy QCD axion, including also the inflationary constraints of the last paragraph, have led us to a `sweet-spot' string cosmology scenario. It involves the LVS framework, K\"ahler moduli inflation with a volume parameter ${\cal V}\sim 10^7$ (for $W_0\sim 1$), a reheating temperature $T_r\sim 10^6\,$GeV and, most importantly, a potentially observable dark radiation abundance of order $\Delta N_{\rm eff}\simeq 0.14$.

\subsection*{Acknowledgements}

We are grateful to Paola Arias for quick and helpful answers on the axion density formulas. J.J. would also like to thank Wen Yin for discussion and collaboration on related subjects. We also express our gratitude to Robert Brandenberger for very useful comments. 
We would also like to thank the anonymous referee for their insightful comments that prompted us to clarify several aspects in our discussion. A.H. was supported by Deutsche Forschungsgemeinschaft (DFG, German Research Foundation) under Germany’s Excellence Strategy EXC 2181/1 - 390900948 (the Heidelberg STRUCTURES Excellence Cluster).
J.J. acknowledges support within the EU network HIDDEN (No 860881). Furthermore, M.W. thanks the DFG for support through the Research Training Group ``Particle Physics beyond the Standard Model'' (GRK 1940).

\appendix

\section*{Appendix}

\section{Realizing small $f_a$} \label{appendix_f_a_expression}
\subsection{Small $f_a$ in type II in general}
\label{fage}

A $p$-form-derived axion $c$ may be defined as
\be
C_p(x,y)=c(x) \omega_p(y),
\ee
where $C_p$ is one of the RR forms of type II string theory and $\omega_p$ a harmonic $p$-form on the compact space $X$. The coordinates $(x^\mu,y^m)$ parameterise $\mathbb{R}^{1,3} \times X$. We follow the conventions of \cite{Giddings:2001yu}, setting in addition $l_s \equiv 2 \pi \sqrt{\alpha'} = 1$. The coupling to euclidean brane instantons on a cycle $\Sigma$ then reads
\be
S \supset 2 \pi \int_\Sigma C_p = 2 \pi c \int_\Sigma \omega_p,
\ee
with $c \equiv c + 1$ if $\omega_p$ is chosen integral. The part of the 10d lagrangian relevant for the size of $f_a$ is
\be
S \supset 2 \pi \int \text{d}^4 x \text{d}^6 y \sqrt{-g} \left\{ \frac{1}{g_s^2} \mathcal{R}_{10} - \frac{1}{2(p+1)!} |\text{d}C_p|^2 \right\}.
\ee
Ignoring $\mathcal O(1)$ constants, one reads off
\be \label{f_a_scaling}
\frac{f_a^2}{M_P^2} \sim \frac{g_s^2}{\mathcal V_s} \int_X \omega_p \wedge *\omega_p \sim \frac{g_s^2}{\mathcal V_s} \int_X \text{d}^6 y \sqrt{-g} (\omega_p)_{m_1 \cdots m_p} (\omega_p)_{n_1 \cdots n_p} g^{m_1 n_1} \cdots g^{m_p n_p}\,,
\ee
where ${\cal V}_s$ is the CY volume in units of $l_s$.

To estimate the smallest achievable $f_a$, let us assume that $\omega_p$ has support only in a tubular neighbourhood of $\Sigma$ with diameter $R$. Let $\Sigma$ have typical size $L$. The integral in \eqref{f_a_scaling} evaluated with only $\sqrt{-g}$ in the integrand would then be $\sim L^p R^{6-p}$. But in addition there are the inverse metric factors which, assuming that $\omega_p$ is directed primarily parallel to $\Sigma$, give a factor $\sim 1/ L^{2p}$. This  results in
\be
\frac{f_a^2}{M_P^2} \sim \frac{g_s^2}{\mathcal V_s} \frac{R^{6-p}}{L^p}.
\ee
Next, we should include the constraint that the QCD gauge coupling at the string scale is in general small. Together with $g_s$, this sets the volume of the corresponding brane stack. Most naturally, the relevant minimal-volume cycle is identical to $\Sigma$, such that the brane instantons discussed above are actually the UV cousins of our SM QCD instantons.\footnote{
We cannot rule out models where this identification is broken. For example, this could be because there are different homologically equivalent cycles which are locally minimal-volume or because QCD arises as the diagonal subgroup of several $SU(3)$s. But we do not see an immediate way to make use of this to lower $f_a$ parametrically below our estimate.
} 
The familiar form of the DBI action then implies $\alpha_{s,UV} \sim g_s/L^p$ and hence (setting also $R \sim 1$),
\be
\frac{f_{a,\text{min}}^2}{M_P^2} \sim \frac{g_s \alpha_{s,UV}}{\mathcal V_s} \sim \frac{\alpha_{s,UV}}{\sqrt{g_s}} \frac{1}{\mathcal V} \qquad \text{or} \qquad \frac{f_{a,\text{min}}^2}{M_P^2} \sim \frac{\alpha_{s,UV}}{\mathcal V}. \label{f_a_scaling_type_IIB}
\ee
Here $\mathcal V = \mathcal V_s / g_s^{3/2}$ is the CY volume in the 10d Einstein frame and, in the last expression, we have chosen $g_s \sim 1$, which is optimal in our context.

\subsection{Small $f_a$ in the LVS}

In the previous subsection, we have argued how the decay constant of the QCD axion scales in type-II models in general. Let us specify this behavior, displayed in~\eqref{f_a_scaling_type_IIB}, for the LVS with the SM realized on a small blowup cycle $\tau_L$. In the simplest case, where the $SU(3)$ brane stack directly wraps this cycle,\footnote{
The 
detailed model building will in general involve several K\"ahler moduli with their ratios fixed by gauge fluxes, leading to ${\cal O}(1)$ correction factors.
} 
we have $\alpha_{s, UV}^{-1}=2\tau_L$. Here the index ``$L$" is chosen because this cycle has to be stabilized by loop effects. The kinetic terms for the axions are
\begin{equation}
    \mathcal{L}/M_P^2 \,\,\supset\,\, K_{i \bar{j}} \partial_\mu c_i \partial^\mu c_{\bar{j}}\,,
\end{equation}
and their periodicity is set by $c_i = c_i + 1$. After a rotation to a diagonal basis $c'_i$, we obtain for the QCD axion $c_L'$
\begin{equation}
    \mathcal{L}/M_P^2 \,\,\supset\,\, \lambda_L \partial_\mu c'_L \partial^\mu c'_L,
\end{equation}
where $\lambda_L$ is the appropriate eigenvalue of $K_{i \bar{j}}$. The canonically normalized axion $a_L/M_P \equiv \sqrt{2 \lambda_L} c'_L$ then obeys $a_L = a_L + \sqrt{2 \lambda_L}M_P$. From this, we can read off the axion decay constant (see also~\cite{Cicoli:2021gss} and refs.~therein)
\begin{equation}
    f_{a_L} = \frac{\sqrt{2 \lambda_L} M_P}{2 \pi}\,.
\end{equation}
Since in the LVS we have $\mathcal{V} \gg 1$, the K\"ahler metric is almost diagonal such that $\lambda_L \simeq K_{LL} = 3\gamma_L/8\mathcal{V} \sqrt{\tau_L}$ (cf.~\eqref{33m}). With that we obtain
\begin{equation}
    \frac{f_{a_L}^2}{M_P^2} \,\,\simeq\,\, \frac{3\gamma_L}{16\pi^2\sqrt{\tau_L}{\cal V}}
    \,,
    \label{fasl}
\end{equation}
consistently with our lower-bound estimate in \eqref{f_a_scaling_type_IIB}. (In fact,
\eqref{fasl} scales as $\sqrt{\alpha_{s, UV}}/{\cal V}$ and is hence slightly larger than \eqref{f_a_scaling_type_IIB}. This is not surprising since we made the most optimistic assumptions about the relevant harmonic form in Appendix~\ref{fage} to make $f_a$ as small as possible.)

\section{Decays of \texorpdfstring{$\tau_L$}{}}
\label{dtl}

Decay rates of the modulus $\tau_L$, which governs the SM gauge couplings, could be obtained by a detailed calculation, analogous to that presented below for $\tau_I$. However, for our purposes a shortcut suffices: Since $\tau_L$ is a (relatively small) blowup cycle, we may analyze the physics in terms of the limit where the large volume decouples, ${\cal V}^{1/6} \gg \,$\{all local length scales\}. Then the coefficient of the operator $\phi_L F^2$, with $F_{\mu\nu}$ the SM field strengths, is of the order of the inverse local mass scale, i.e. the string scale $M_s$, on dimensional grounds. This gives a decay rate to SM gauge bosons
\be
\Gamma_{\phi_L\to AA} \sim m_{\phi_L}^3/M_s^2 \sim M_P/{\cal V}^2\,.
\ee
Here in the first step $m_{\phi_L}^3$ appears for dimensional reasons and we then used that $m_{\phi_L}\sim M_P/{\cal V}$ and $M_s\sim M_P/\sqrt{\cal V}$. This is so much faster than all the inflaton decay rates, which are $\sim M_P/{\cal V}^4$, that we may leave it at this very rough estimate.

We need to check whether a significant axion energy density is created while $\tau_L$ reheats the SM as above. Both the decay to gauge bosons and to axions arise through the $\tau_L$ dependence of the respective kinetic terms:
\be
{\cal L}\,\,\,\supset\,\,\,\sim\, \tau_L\, \mbox{tr}F_{\mu\nu}F^{\mu\nu} 
\qquad\mbox{and}
\qquad {\cal L}\,\,\,\supset\,\,\,\sim\, \frac{1}{{\cal V}\sqrt{\tau_L}}\,(\partial_\mu c_L)(\partial^\mu c_L)\,.
\label{decl}
\ee
This follows from the standard DBI action and the K\"ahler potential $-2\ln{\cal V}$ together with \eqref{eq:inflatonfinal}.
Now, since the decay originates from the fluctuation $\tau_L\sim \langle\tau_L\rangle + \delta\tau_L$, it is clear that the amplitude for the rate to gauge bosons is enhanced by a factor $2$ due to the higher-power of $\tau_L$ in \eqref{decl},
\be
A_{\phi_L\to AA}\, /\,A_{\phi_L\to a_L a_L} = 2\,.
\ee
Taking into account that each gauge boson has two polarizations and the $N_g$ active gauge bosons (with $N_g$ at least $1+3+8=12$), we have
\be
\Gamma_{\phi_L \rightarrow A A} \,/\, \Gamma_{\phi_L \rightarrow a_L a_L} = 8 N_g \gg 1\,.
\ee
In summary, $\phi_L$ decays for our purposes instantaneously and without aggravating the dark radiation problem.

\section{Dynamics of the two-moduli system $\tau_b$ and $\tau_I$} \label{sec:2_moduli_system}

In this section, we estimate the decay rates of the inflaton and its axion into the volume modulus and its respective axion. This analysis represents a significant simplification relative to the realistic case, which must include the loop-stabilized modulus $\tau_L$ as well as one (or several) small cycles $\tau_{s,i}$. The latter are needed to keep the volume stabilized during inflation. A more general treatment, including $\tau_L$, is presented in Appendix~\ref{appendix_field_dynamics}. The presentation of the 2-moduli case in the present appendix serves merely to build intuition and to allow the interested reader to start with a very similar but less complex analysis. The fact that we disregard additional small cycles $\tau_{s,i}$ introduces only a negligible error, as we explain below. Our analysis follows the procedure of~\cite{Cicoli:2010ha}, adapted to our purposes: We expand the $F$-term potential $V$ and (going beyond \cite{Cicoli:2010ha}) also the K\"ahler potential up to third order in $\tau_b$ and $\tau_I$ about their respective vacuum expectation values, denoted by $\langle \, \cdot \, \rangle$. This provides us with the (mixed) kinetic and mass terms as well as trilinear couplings. We then diagonalize and canonically normalize all dynamical fields so that we can read off the coupling strengths and hence obtain the decay rates. Note that in what follows we set $M_P = 1$ for brevity.

\subsection{Basic definitions}
\label{sec:basic_definitions}

We consider the following volume and K\"ahler potential:
\begin{equation}
\mathcal{V} = \tau_b^{3/2} - \gamma_I \tau_I^{3/2}\,\,, \quad\qquad K = - 2 \ln \left( \mathcal V + \frac{\xi}{2} \right) - \ln (S + \bar{S}) + K_\text{cs},
\label{kkor}
\end{equation}
where $S$ is the axio-dilaton, $K_\text{cs}$ depends on the complex-structure moduli and we have absorbed a factor $g_s^{-3/2}$ into $\xi$. This gives the leading-order K\"ahler-moduli K\"ahler metric and its inverse
\begin{equation}
K_{ij} = \frac{\partial^2 K}{\partial T_i \partial \bar{T}_j} \approx \begin{pmatrix}
 \frac{3 }{4 \tau _b^2} & -\frac{9 \gamma _I
   \sqrt{\tau _I} }{8 \tau _b^{5/2}} \\
 -\frac{9 \gamma _I \sqrt{\tau _I} }{8 \tau _b^{5/2}} & \frac{3 \gamma
   _I }{8 \tau _b^{3/2} \sqrt{\tau _I}}
\end{pmatrix}, \quad
(K^{-1})^{i j} \approx
\begin{pmatrix}
\frac{4 \tau_b^2}{3} & 4 \tau_b \tau_I \\
4 \tau_b \tau_I & \frac{8 \tau_b^{3/2} \sqrt{\tau_I}}{3 \gamma_I}
\end{pmatrix},
\label{ma2}
\end{equation}
where $T_i = \tau_i + {\rm i} c_i$ with $i \in \{b,I\}$.
The superpotential, corrected by the non-perturbative term due to D3-brane instantons, reads
\begin{equation}
W = W_0 + A_I \text{e}^{- \mathfrak{a}_I T_I}, \label{superpotential_2_moduli}
\end{equation}
with $\mathfrak{a}_I = 2 \pi$.\footnote{
Obviously, 
in the full system, there are also corrections due to instantons on the other small cycles $\tau_{s,i}$. However, the resulting $F$-term potential consists of merely a sum of terms analogous to~\eqref{LVS_potential_non-perturbative_terms} over all $\tau_{s,i}$. These additional terms are irrelevant for the decays of the inflaton and its axion. This is because decays into $\tau_b$ and $\tau_L$ via these terms are highly suppressed whereas decays into the $\tau_{s,i}$ themselves are kinematically forbidden.
}
Since $S$ and the complex structure moduli are fixed by fluxes, the contribution $- \ln (S + \bar{S}) + K_\text{cs}$ in \eqref{kkor} represents merely a constant. We absorb this constant into a redefinition of $A_I$ and $W_0$.

The non-perturbative correction to the LVS potential with one small, non-perturbatively stabilised cycle is given by \cite{Balasubramanian:2005zx,Conlon:2005ki} (see also \cite{Hebecker:2021egx}),
\begin{equation}
V_\text{LVS} = \text{e}^K \left[ (K^{-1})^{22} \mathfrak{a}_I^2 |A_I|^2 \text{e}^{-2 \mathfrak{a}_I \tau_I} + 2 \mathfrak{a}_I \tau_I \left( A_I \bar{W}_0 \text{e}^{- \mathfrak{a}_I T_I} + \bar{A}_I W_0 \text{e}^{- \mathfrak{a}_I \bar{T}_I} \right) \right]. \label{LVS_potential_non-perturbative_terms}
\end{equation}
Restoring the explicit axion dependence in $T_I$ and absorbing the constant phases $\arg W_0$ and $\arg A_I$ into a redefinition of $ c_I$, we arrive at
\begin{equation}
V_\text{LVS} = \mathcal{V}^{-2} \left[ \frac{8 \tau_b^{3/2} \sqrt{\tau_I}}{3 \gamma_I} \mathfrak{a}_I^2 |A_I|^2 \text{e}^{-2\mathfrak{a}_I \tau_I} +4 \mathfrak{a}_I \tau_I \text{e}^{-\mathfrak{a}_I \tau_I} |A_I W_0| \cos\left( \mathfrak{a}_I c_I \right) \right] + \frac{3 |W_0|^2 \xi}{4 \mathcal{V}^3}\,, \label{LVS_potential_full}
\end{equation}
where we have used the explicit formula for $(K^{-1})^{22}$ from \eqref{ma2} and approximated the K\"ahler potential as $K \approx - 2 \ln \mathcal{V}$. We have also added the contribution induced by the $\alpha'$-corrections to $K$. This represents the total potential that we use for our analysis in this section. As is well known, the minimum of this potential is defined by the following equations~\cite{Balasubramanian:2005zx},
\begin{equation}
\xi = 2 \gamma_I \langle\tau_I\rangle^{3/2}, \quad \text{e}^{\mathfrak{a}_I \langle \tau_I \rangle} = \frac{4 \langle \mathcal{V} \rangle |A_I| \mathfrak{a}_I}{3 \gamma_I |W_0| \sqrt{\langle \tau_I \rangle}}, \quad \cos(\mathfrak{a}_I \langle c_I \rangle) = -1. \label{replacement_relations}
\end{equation}
Both the kinetic term as well as the scalar potential can now be expanded around this minimum. The relevant lagrangian for us is the truncation of this expansion at cubic order,
\begin{align}
\mathcal L &= \langle K_{ij} \rangle \partial_\mu \delta \tau_i \partial^\mu \delta \tau_j + \langle \partial_{\tau_i} K_{jk} \rangle \delta \tau_i \partial_\mu \delta \tau_j \partial^\mu \delta \tau_k + \langle K_{ij} \rangle \partial_\mu \delta c_i \partial^\mu \delta c_j + \langle \partial_{\tau_i} K_{jk} \rangle \delta \tau_i \partial_\mu \delta c_j \partial^\mu \delta c_k   \nonumber \\
&\quad - \langle V_\text{LVS} \rangle - \frac{1}{2} \left\langle \frac{\partial^2 V_\text{LVS}}{\partial \tau_i \partial \tau_j} \right\rangle \delta \tau_i \delta \tau_j - \frac{1}{6} \left\langle \frac{\partial^3 V_\text{LVS}}{\partial \tau_i \partial \tau_j \partial \tau_k} \right\rangle \delta \tau_i \delta \tau_j \delta \tau_k  - \frac{1}{2} \left\langle \frac{\partial^2 V_\text{LVS}}{\partial c_i \partial c_j} \right\rangle \delta c_i \delta c_j \nonumber \\
&\quad - \frac{1}{2} \left\langle \frac{\partial^3 V_\text{LVS}}{\partial \tau_i \partial c_j \partial c_k} \right\rangle \delta \tau_i \delta c_j \delta c_k. \label{perturbed_lagrangian_2_moduli}
\end{align}

\subsection{Decay into volume modulus}

\subsubsection*{Diagonalization of fields}

Following~\cite{Cicoli:2010ha}, we must first transform the $\delta \tau_i$ into canonical fields. On this account, we need the second-derivative matrix w.r.t. the moduli, which at leading order is given by
\begin{equation}
\left\langle V_{i j} \right\rangle = \left\langle \frac{\partial^2 V_{LVS}}{\partial \tau_i \partial \tau_j} \right\rangle \approx \frac{3 \gamma_I |W_0|^2 }{\tau_b^{9/2} }
\begin{pmatrix}
\frac{9 \tau_I^{3/2}}{4 \tau_b^{2}} & -\frac{3 \mathfrak{a}_I \tau_I^{3/2}}{2 \tau_b} \\
-\frac{3  \mathfrak{a}_I \tau_I^{3/2}}{2 \tau_b} & \mathfrak{a}_I^2 \tau_I^{3/2}
\end{pmatrix}.
\end{equation}
Here we have used the relations~\eqref{replacement_relations}
after applying the second derivatives.\footnote{Note that after including the other small cycles $\tau_{s,i}$ the expression for $\xi$ becomes a sum over all such non-perturbatively stabilized cycles. However, it turns out that our results for the decay rates of $\tau_I$ and $c_I$ remain unaltered under this modification of $\xi$.}

The transformation to the canonical fields reads
\begin{equation}
\begin{pmatrix}
\delta \tau_b \\
\delta \tau_I
\end{pmatrix}
=
\begin{pmatrix}
\vec{v}_b \vphantom{\begin{pmatrix}
\delta \tau_b \\
\delta \tau_I
\end{pmatrix}}
\end{pmatrix} \frac{\delta \phi_b}{\sqrt{2}}
+
\begin{pmatrix}
\vec{v}_I \vphantom{\begin{pmatrix}
\delta \tau_b \\
\delta \tau_I
\end{pmatrix}}
\end{pmatrix} \frac{\delta \phi_I}{\sqrt{2}}\, , \label{transformation_to_canonical_2_moduli}
\end{equation}
or $\delta \tau_i = P_{i j} \delta \phi_j / \sqrt{2}$ where $P$ is the matrix that contains the vectors $\vec{v}_j$ as columns. These vectors are the eigenvectors of the matrix $(M^2)_{ij} \equiv \left\langle \left(K^{-1}\right)_{i k} V_{k j} \right\rangle / 2$, whose eigenvalues $m_i^2$ are the masses of the canonical fields $\delta \phi_i$, and they fulfill the normalization condition
\begin{equation}
\label{eigenvector_normalisation_2_moduli}
\vec{v}^\mathsf{T} _i \cdot \langle K \rangle \cdot \vec{v}_j \equiv P_{k i} \langle K_{k l} \rangle P_{l j} = \delta_{ij}.
\end{equation}

Next we have to calculate the eigenvectors $\vec{v}_j$. The $M^2$ matrix at leading order reads
\begin{equation}
(M^2)_{i j} \approx \frac{|W_0|^2}{ \tau_b^3}
\begin{pmatrix}
- \frac{9 \gamma_I \mathfrak{a}_I \tau_I^{5/2} }{ \tau_b^{3/2}} & \frac{6 \gamma_I  \mathfrak{a}_I^2 \tau_I^{5/2}}{\sqrt{\tau_b}} \\
- \frac{6  \mathfrak{a}_I \tau_I^2 }{\tau_b} & 4 \mathfrak{a}_I^2 \tau_I^2 
\end{pmatrix}. \\
\end{equation}
The eigenvalues and eigenvectors of this matrix are given by
\begin{align}
m_{\tau_b}^2 &= 0, &&\vec{v}_1 =
\begin{pmatrix}
\frac{2 \mathfrak{a}_I \tau_I}{3} \\
1
\end{pmatrix}, \\
m_{\tau_I}^2 &= \frac{4 |W_0|^2 \mathfrak{a}_I^2 \tau_I^2}{\tau_b^3}, &&\vec{v}_2 =
\begin{pmatrix}
\frac{3 \gamma_I \sqrt{\tau_I}}{2 \sqrt{\tau_b}} \\
1
\end{pmatrix}. \label{inflaton_mass_2_moduli}
\end{align}
To fulfill the normalization conditions \eqref{eigenvector_normalisation_2_moduli}, we rescale the above eigenvectors
\begin{align}
\vec{v}_b &\equiv \frac{\vec{v}_1}{\sqrt{\vec{v}^\mathsf{T}_1 \cdot \langle K \rangle \cdot \vec{v}_1}} \approx \frac{\sqrt{3}}{\mathfrak{a}_I} \vec{v}_1 = \begin{pmatrix}
\frac{2 \tau_b}{\sqrt{3}} \\
\frac{\sqrt{3}}{\mathfrak{a}_I}
\end{pmatrix},  \label{eigenvector_v_b_2_moduli_system}\\
\vec{v}_I &\equiv \frac{\vec{v}_2}{\sqrt{\vec{v}^\mathsf{T}_2 \cdot \langle K \rangle \cdot \vec{v}_2}} \approx  \frac{4\tau_b^{3/4} \tau_I^{1/4}}{\sqrt{6 \gamma_I}} \vec{v}_2 = \begin{pmatrix}
\sqrt{6 \gamma_I} \tau_I^{3/4} \tau_b^{1/4} \\
\frac{2 \sqrt{2} \tau_b^{3/4} \tau_I^{1/4}}{\sqrt{3 \gamma_I}}
\end{pmatrix} \label{eigenvector_v_I_2_moduli_system}.
\end{align}
With these eigenvectors, the transformation to the canonical fields, in analogy to \eqref{transformation_to_canonical_2_moduli}, is given by
\begin{align}
\delta \tau_b &= \sqrt{\frac{2}{3}} \tau_b \delta \phi_b + \sqrt{3 \gamma_I} \tau_I^{3/4} \tau_b^{1/4} \delta \phi_I, \label{b_canonical}\\
\delta \tau_I &= \frac{\sqrt{3}}{\sqrt{2} \mathfrak{a}_I} \delta \phi_b + \frac{2 \tau_b^{3/4} \tau_I^{1/4}}{\sqrt{3 \gamma_I}} \delta \phi_I. \label{I_canonical}
\end{align}

\subsubsection*{Coupling terms}

The kinetic and potential trilinear coupling terms can be read off from~\eqref{perturbed_lagrangian_2_moduli} and are given by
\begin{align}
\mathcal{L}_\text{int,kin} &= K_{mnp} \delta \tau_m (\partial_\mu \delta \tau_n) (\partial^\mu \delta \tau_p)\,, \label{kinetic_coupling_general_2_moduli}\\
\mathcal{L}_\text{int,pot} &= -\frac{1}{6} V_{mnp} \delta \tau_m \delta \tau_n \delta \tau_p\,, \label{potential_coupling_general_2_moduli}
\end{align}
respectively. Here we defined $ K_{mnp} \equiv \langle \partial_{\tau_m} K_{n p} \rangle$ and $ V_{mnp} \equiv \langle \partial_{\tau_m} \partial_{\tau_n} \partial_{\tau_p} V \rangle$.
Let us first calculate these third order derivatives at leading order,
\begin{align}
&K_{bbb} = -\frac{3}{2 \tau_b^3} , \quad K_{bbI} = \frac{45 \gamma_I \sqrt{\tau_I}}{16 \tau_b^{7/2}} , \quad K_{bII} = - \frac{9 \gamma_I}{16 \sqrt{\tau_I} \tau_b^{5/2}} , \quad K_{III} = - \frac{3 \gamma_I}{16 \tau_I^{3/2} \tau_b^{3/2}} , \\
&V_{bbb} = -\frac{81 \gamma_I |W_0|^2 \tau_I^{3/2}}{ \tau_b^{15/2}}, \quad V_{bbI} = \frac{99 \gamma_I |W_0|^2 \mathfrak{a}_I \tau_I^{3/2}}{4 \tau_b^{13/2}}, \\
&V_{bII} = - \frac{27 \gamma_I |W_0|^2 \mathfrak{a}_I \sqrt{\tau_I}}{2 \tau_b^{11/2}}, \quad V_{III} = -\frac{9 \gamma_I |W_0|^2 \mathfrak{a}_I^3 \tau_I^{3/2}}{ \tau_b^{9/2}}.
\end{align}
Here we have again used the relations \eqref{replacement_relations}, however, this time only after forming the third derivatives.

Let us first focus on the kinetic couplings. Inserting the canonical fields \eqref{b_canonical} and \eqref{I_canonical} into \eqref{kinetic_coupling_general_2_moduli}, we have
\begin{equation}
\mathcal{L}_\text{int,kin} = \frac{1}{2^{3/2}} K_{m n p} P_{mi} P_{nj} P_{pk} \delta \phi_i (\partial_\mu \delta \phi_j) (\partial^\mu \delta \phi_k).
\end{equation}
To eliminate the derivatives, we use the relation
\begin{equation}
\delta \phi_i (\partial_\mu \delta \phi_j) (\partial^\mu \delta \phi_k) = \frac{1}{2} \left( m_i^2 - m_j^2 - m_k^2 \right) \delta \phi_i \delta \phi_j \delta \phi_k, \label{eliminate_partial_derivatives}
\end{equation}
which is obtained by partial integration and making use of the free Klein-Gordon equation.\footnote{An analogous relation also holds if we replace one or more of the moduli fields $\delta \phi_i$ by axion fields $\delta a_i$.} Thus we obtain
\begin{equation}
\mathcal{L}_\text{int,kin} = \frac{1}{2^{5/2}} K_{m n p} P_{mi} P_{nj} P_{pk} \left( m_i^2 - m_j^2 - m_k^2 \right) \delta \phi_i \delta \phi_j \delta \phi_k. \label{coupling_term_2_moduli}
\end{equation}
We are only interested in the terms in $\mathcal{L}_\text{int,kin}$ for which one of the three indices $i$, $j$, $k$ is an ``$I$" while the other two are a ``$b$". From \eqref{coupling_term_2_moduli} we see that in $\mathcal{L}_\text{int,kin}$ all terms are invariant under permutation of these indices except for the factor $(m_i^2 - m_j^2 - m_k^2)$. However, since $m_{\tau_I}^2 \gg m_{\tau_b}^2$, this factor is dominated by $m_{\tau_I}^2$ and therefore only changes by a minus sign under permutation of $i$, $j$ and $k$, depending on which of the three indices takes on the value ``$I$". There are in total three coupling terms corresponding to either $i = I$ or $j = I$ or $k = I$. The two terms with $j = I$ and $k = I$ carry a minus sign compared to the term with $i = I$ so that w.l.o.g. we can just write
\begin{equation}
\mathcal{L}^{(\phi_I \rightarrow \phi_b \phi_b)}_\text{int,kin} = -\frac{1}{2^{5/2}} K_{m n p} P_{mI} P_{nb} P_{pb} m_{\tau_I}^2 \delta \phi_I \delta \phi_b \delta \phi_b, \label{coupling_term_kinetic_bbI_2_moduli}
\end{equation}
where the minus sign results from the sum of the three terms. From this we can easily get the coupling by computing the contractions
\begin{align}
K_{m n p} P_{mI} P_{nb} P_{pb} &= K_{b b b} P_{bI} P_{bb} P_{bb} + 2 K_{b b I} P_{bI} P_{bb} P_{Ib} + K_{b I I} P_{bI} P_{Ib} P_{Ib} \nonumber\\
&\quad + K_{I b b} P_{II} P_{bb} P_{bb} + 2 K_{I b I} P_{II} P_{bb} P_{Ib} + K_{I I I} P_{II} P_{Ib} P_{Ib} \\
&= -\frac{2 \sqrt{6 \gamma_I} \tau_I^{3/4}}{\tau_b^{3/4}} + \frac{45 \sqrt{6} \gamma_I^{3/2} \tau_I^{5/4} }{4 \mathfrak{a}_I \tau_b^{9/4}}  - \frac{27 \sqrt{6 } \gamma_I^{3/2} \tau_I^{1/4}}{16 \mathfrak{a}_I^2  \tau_b^{9/4}}  \nonumber\\
&\quad + \frac{5 \sqrt{6 \gamma_I} \tau_I^{3/4}}{2 \tau_b^{3/4}} - \frac{3 \sqrt{6 \gamma_I}}{2 \mathfrak{a}_I \tau_I^{1/4} \tau_b^{3/4}} - \frac{3  \sqrt{6\gamma_I} }{8 \mathfrak{a}_I^2 \tau_I^{5/4} \tau_b^{3/4}}.
\end{align}
We see that the dominating contributions are
\begin{equation}
K_{m n p} P_{mI} P_{nb} P_{pb} \approx K_{b b b} P_{bI} P_{bb} P_{bb} + K_{I b b} P_{II} P_{bb} P_{bb} = \frac{\sqrt{6 \gamma_I} \tau_I^{3/4}}{2 \tau_b^{3/4}}.
\end{equation}
Inserting this and \eqref{inflaton_mass_2_moduli} into \eqref{coupling_term_kinetic_bbI_2_moduli}, we obtain
\begin{equation}
\mathcal{L}^{(\phi_I \rightarrow \phi_b \phi_b)}_\text{int,kin} \approx - \frac{\sqrt{3 \gamma_I} |W_0|^2 \mathfrak{a}_I^2 \tau_I^{11/4}}{2 \tau_b^{15/4}} \delta \phi_I \delta \phi_b \delta \phi_b. \label{coupling_kinetic_bbI_final_2_moduli}
\end{equation}

Now we focus on the potential couplings. Inserting the canonical fields \eqref{b_canonical} and \eqref{I_canonical} into \eqref{potential_coupling_general_2_moduli}, we obtain
\begin{equation}
\mathcal{L}_\text{int,pot} = -\frac{1}{12 \sqrt{2}} V_{m n p} P_{mi} P_{nj} P_{pk} \delta \phi_i \delta \phi_j \delta \phi_k.
\end{equation}
Again we are only interested in terms where one index of $i$, $j$, $k$ takes on the value $I$ while the other two take on the value $b$. There are in total three such terms. Due to the invariance of $\mathcal{L}_\text{int,pot}$ under permutation of $i$, $j$, $k$ are all the same. Hence, we can account for them by a factor 3,
\begin{equation}
\mathcal{L}^{(\phi_I \rightarrow \phi_b \phi_b)}_\text{int,pot} = -\frac{1}{4 \sqrt{2}} V_{m n p} P_{mI} P_{nb} P_{pb} \delta \phi_I \delta \phi_b \delta \phi_b. \label{coupling_term_potential_bbI_2_moduli}
\end{equation}
Calculating the contractions, we obtain
\begin{align}
V_{m n p} P_{mI} P_{nb} P_{pb} &= V_{b b b} P_{bI} P_{bb} P_{bb} + 2 V_{b b I} P_{bI} P_{bb} P_{Ib} + V_{b I I} P_{bI} P_{Ib} P_{Ib} \nonumber\\
&\quad + V_{I b b} P_{II} P_{bb} P_{bb} + 2 V_{I b I} P_{II} P_{bb} P_{Ib} + V_{I I I} P_{II} P_{Ib} P_{Ib} \\
&= -\frac{108 \sqrt{6} \gamma_I^{3/2} |W_0|^2 \tau_I^{9/4}}{ \tau_b^{21/4}} + \frac{99 \sqrt{6 } \gamma_I^{3/2} |W_0|^2 \tau_I^{9/4}}{ \tau_b^{21/4}} - \frac{81 \sqrt{6 } \gamma_I^{3/2} |W_0|^2 \tau_I^{5/4}}{2 \mathfrak{a}_I \tau_b^{21/4}}  \nonumber\\
&\quad +\frac{22 \sqrt{6\gamma_I} |W_0|^2 \mathfrak{a}_I \tau_I^{7/4}}{ \tau_b^{15/4}} - \frac{36 \sqrt{6\gamma_I} |W_0|^2 \tau_I^{3/4}}{ \tau_b^{15/4}} -\frac{18 \sqrt{6 \gamma_I} |W_0|^2 \mathfrak{a}_I \tau_I^{7/4}}{ \tau_b^{15/4}}.
\end{align}
The dominating contributions are
\begin{equation}
V_{m n p} P_{mI} P_{nb} P_{pb} \approx V_{I b b} P_{II} P_{bb} P_{bb} + V_{I I I} P_{II} P_{Ib} P_{Ib} = \frac{4 \sqrt{6\gamma_I} |W_0|^2 \mathfrak{a}_I \tau_I^{7/4}}{ \tau_b^{15/4}}.
\end{equation}
Inserting this into \eqref{coupling_term_potential_bbI_2_moduli}, we arrive at
\begin{equation}
\mathcal{L}^{(\phi_I \rightarrow \phi_b \phi_b)}_\text{int,pot} = - \frac{ \sqrt{3\gamma_I} |W_0|^2 \mathfrak{a}_I \tau_I^{7/4}}{ \tau_b^{15/4}} \delta \phi_I \delta \phi_b \delta \phi_b.
\end{equation}

We can also conclude that
\begin{equation}
\frac{\mathcal{L}^{(\phi_I \rightarrow \phi_b \phi_b)}_\text{int,kin}}{\mathcal{L}^{(\phi_I \rightarrow \phi_b \phi_b)}_\text{int,pot}} \approx \frac{\mathfrak{a}_I \tau_I}{2} \gg 1,
\end{equation}
which confirms our estimate~\eqref{amplitude_ratio_kinetic_potential_decay_into_volume}.

\subsection{Decay into volume axion}

\subsubsection*{Diagonalization of fields}

For the decay into the volume axion, we proceed analogously as for the decay into the volume modulus. The second derivative matrix w.r.t. the axions at leading order is given by
\begin{equation}
\langle V^{(c)}_{i j} \rangle \equiv \left\langle \frac{\partial^2 V_\text{LVS}}{\partial c_i \partial c_j} \right\rangle = \begin{pmatrix}
0 & 0 \\
0 & \frac{3 \gamma_I |W_0|^2 \mathfrak{a}_I^2 \tau_I^{3/2} }{\tau_b^{9/2}}
\end{pmatrix},
\end{equation}
where we have again used the relations \eqref{replacement_relations} after applying the second derivatives. The transformation to canonical fields is given by
\begin{equation}
\begin{pmatrix}
\delta c_b \\
\delta c_I
\end{pmatrix}
=
\begin{pmatrix}
\vec{w}_b \vphantom{\begin{pmatrix}
\delta c_b \\
\delta c_I
\end{pmatrix}}
\end{pmatrix} \frac{\delta a_b}{\sqrt{2}}
+
\begin{pmatrix}
\vec{w}_I \vphantom{\begin{pmatrix}
\delta c_b \\
\delta c_I
\end{pmatrix}}
\end{pmatrix} \frac{\delta a_I}{\sqrt{2}} 
\end{equation}
or $\delta c_i = Q_{ij} \delta a_j / \sqrt{2}$ where $Q$ is the matrix that contains the vectors $\vec{w}_j$ as columns. They are the eigenvectors of the matrix $(M^2_{(c)})_{ij} \equiv \langle (K^{-1})_{ik} V^{(c)}_{kj} \rangle/2$ whose eigenvalues are the axion masses. The eigenvectors fulfill the normalization condition
\begin{equation}
\label{eigenvector_normalisation_axion_2_moduli}
\vec{w}^\mathsf{T} _i \cdot \langle K \rangle \cdot \vec{w}_j \equiv Q_{k i} \langle K_{k l} \rangle Q_{l j} = \delta_{ij}.
\end{equation}
The $M^2_{(c)}$ matrix at leading order is given by
\begin{equation}
(M^2_{(c)})_{ij} \approx \begin{pmatrix}
0 & \frac{6 \gamma_I |W_0|^2 \mathfrak{a}_I^2 \tau_I^{5/2}}{\tau_b^{7/2}} \\
0 & \frac{4 |W_0|^2 \mathfrak{a}_I^2 \tau_I^2}{\tau_b^3}
\end{pmatrix}.
\end{equation}
The corresponding eigenvalues and eigenvectors are
\begin{align}
m_{c_b}^2 &= 0, && \vec{w}_1 =
\begin{pmatrix}
1 \\
0
\end{pmatrix}, \\
m_{c_I}^2 &\approx \frac{4 |W_0|^2 \mathfrak{a}_I^2 \tau_I^2}{\tau_b^3}, && \vec{w}_2 =
\begin{pmatrix}
\frac{3 \gamma_I \sqrt{\tau_I}}{2 \sqrt{\tau_b}} \\
1
\end{pmatrix}. \label{inflaton_axion_mass}
\end{align}
After rescaling to fulfill the normalization condition \eqref{eigenvector_normalisation_axion_2_moduli}, the normalized eigenvectors read
\begin{align}
\vec{w}_b &\equiv \frac{\vec{w}_1}{\sqrt{\vec{w}^\mathsf{T}_1 \cdot \langle K \rangle \cdot \vec{w}_1}} \approx \frac{2 \tau_b}{\sqrt{3}} \vec{w}_1 = \begin{pmatrix}
\frac{2 \tau_b}{\sqrt{3}} \\
0
\end{pmatrix},  \\
\vec{w}_I &\equiv \frac{\vec{w}_2}{\sqrt{\vec{w}^\mathsf{T}_2 \cdot \langle K \rangle \cdot \vec{w}_2}} \approx  \frac{2 \sqrt{2} \tau_b^{3/4} \tau_I^{1/4}}{\sqrt{3\gamma_I}} \vec{w}_2 = \begin{pmatrix}
\sqrt{6 \gamma_I} \tau_I^{3/4} \tau_b^{1/4} \\
\frac{ 2 \sqrt{2} \tau_I^{1/4} \tau_b^{3/4}}{\sqrt{3 \gamma_I}}
\end{pmatrix} .
\end{align}

\subsubsection*{Coupling terms}

The kinetic and potential trilinear coupling terms are respectively given by
\begin{align}
\mathcal{L}_{\text{int,kin},(c)} &= \langle \partial_{\tau_m} K_{np} \rangle \delta \tau_m \partial_\mu \delta c_n \partial^\mu \delta c_p \\
&= \frac{1}{2^{3/2}} K_{mnp} P_{mi} Q_{nj} Q_{pk} \delta \phi_i \partial_\mu \delta a_j \partial^\mu \delta a_k, \label{coupling_kinetic_general_axion_2_moduli} \\
\mathcal{L}_{\text{int,pot},(c)} &= -\frac{1}{2} \left\langle \frac{\partial^3 V_\text{LVS}}{\partial \tau_m \partial c_n \partial c_p} \right\rangle \delta \tau_m \delta c_n \delta c_p \\
&= -\frac{1}{2^{5/2}} \left\langle \frac{\partial^3 V_\text{LVS}}{\partial \tau_m \partial c_n \partial c_p} \right\rangle P_{mi} Q_{nj} Q_{pk} \delta \phi_i \delta a_j \delta a_k \,.\label{coupling_potential_general_axion_2_moduli}
\end{align}

Let us first argue that the potential coupling to the volume axion vanishes: Since $V_\text{LVS}$ does not depend on $c_b$ but only on $c_I$, the indices $n$ and $p$ in \eqref{coupling_potential_general_axion_2_moduli} must both take on the value ``$I$". However, the component $Q_{Ib}$ vanishes, so that there is no potential coupling $\sim \delta \phi_I \delta a_b \delta a_b$.

Focusing now on the kinetic coupling, we have
\begin{equation}
\mathcal{L}^{(\phi_I \rightarrow a_b a_b)}_{\text{int,kin},(c)} = \frac{1}{2^{5/2}} K_{mnp} P_{mI} Q_{nb} Q_{pb} m_{\tau_I}^2 \delta \phi_I \delta a_b \delta a_b,
\end{equation}
where we have again eliminated the partial derivatives via \eqref{eliminate_partial_derivatives}. Since $Q_{Ib} = 0$, the indices $n$ and $p$ must take on the value $b$, so that we have
\begin{align}
\mathcal{L}^{(\phi_I \rightarrow a_b a_b)}_{\text{int,kin},(c)} &= \frac{1}{2^{5/2}} K_{mbb} P_{mI} Q_{bb} Q_{bb} m_{\tau_I}^2 \delta \phi_I \delta a_b \delta a_b \\
 &= \frac{1}{2^{5/2}} (K_{bbb} P_{bI} + K_{Ibb} P_{II}) Q_{bb} Q_{bb} m_{\tau_I}^2 \delta \phi_I \delta a_b \delta a_b \\
&= \frac{1}{2^{5/2}} \left[ \left(-\frac{3}{2 \tau_b^3} \right) \sqrt{6 \gamma_I} \tau_I^{3/4} \tau_b^{1/4} + \frac{45 \gamma_I \sqrt{\tau_I}}{16 \tau_b^{7/2}} \frac{2 \sqrt{2} \tau_b^{3/4} \tau_I^{1/4}}{\sqrt{3 \gamma_I}} \right] \left(\frac{2 \tau_b}{\sqrt{3}} \right)^2 m_{\tau_I}^2 \delta \phi_I \delta a_b \delta a_b \\
&= \frac{ \sqrt{3\gamma_I} |W_0|^2 \mathfrak{a}_I^2 \tau_I^{11/4}}{2 \tau_b^{15/4}}  \delta \phi_I \delta a_b \delta a_b.
\end{align}
Comparing this to~\eqref{coupling_kinetic_bbI_final_2_moduli}, we see that the kinetic decay rate of the inflaton into the volume axion is indeed equal to the one into the volume modulus.

\subsection{Decay of inflaton axion}

Finally, we consider the kinetic decay $a_I \rightarrow \phi_b a_b$, which stems also from the term $\mathcal{L}_{\text{int,kin},(c)}$ as given in \eqref{coupling_kinetic_general_axion_2_moduli},
however, this time we have $i = b$ whereas $j$ and $k$ can take on either the value ``$b$" or ``$I$" with $j \neq k$. Since there are exactly two possibilities for that, we will w.l.o.g. set $j = I$ and $k = b$ and assign a factor 2
\begin{equation}
\mathcal{L}^{(a_I \rightarrow \phi_b a_b)}_{\text{int,kin},(c)} = \frac{1}{\sqrt{2}} K_{mnp} P_{mb} Q_{nI} Q_{pb} \delta \phi_b \partial_\mu \delta a_I \partial^\mu \delta a_b.
\end{equation}
Eliminating the derivatives, this becomes
\begin{equation}
\mathcal{L}^{(a_I \rightarrow \phi_b a_b)}_{\text{int,kin},(c)} = -\frac{1}{2\sqrt{2}} K_{mnp} P_{mb} Q_{nI} Q_{pb} m_{c_I}^2 \delta \phi_b \delta a_I \delta a_b.
\end{equation}
Since $Q_{Ib} = 0$, the index $p$ is forced to take on the value ``$b$". We obtain
\begin{equation}
\mathcal{L}^{(a_I \rightarrow \phi_b a_b)}_{\text{int,kin},(c)} = -\frac{1}{2\sqrt{2}} K_{mnb} P_{mb} Q_{nI} Q_{bb} m_{c_I}^2 \delta \phi_b \delta a_I \delta a_b.
\end{equation}
The contraction reads
\begin{align}
K_{mnb} P_{mb} Q_{nI} &= K_{bbb} P_{bb} Q_{bI} + K_{bIb} P_{bb} Q_{II} + K_{Ibb} P_{Ib} Q_{bI} + K_{IIb} P_{Ib} Q_{II} \\
&=  -\frac{3 \sqrt{2 \gamma_I} \tau_I^{3/4}}{ \tau_b^{7/4}}  + \frac{15 \sqrt{2\gamma_I} \tau_I^{3/4}}{4 \tau_b^{7/4}} + \frac{135 \sqrt{2} \gamma_I^{3/2}  \tau_I^{5/4}}{16 \mathfrak{a}_I \tau_b^{13/4}}  - \frac{9 \sqrt{2\gamma_I}}{8 \mathfrak{a}_I \tau_I^{1/4} \tau_b^{7/4}} \\
&\approx K_{bbb} P_{bb} Q_{bI} + K_{bIb} P_{bb} Q_{II} \\
&= \frac{3 \sqrt{2\gamma_I} \tau_I^{3/4}}{4 \tau_b^{7/4}}.
\end{align}
Inserting this and \eqref{inflaton_axion_mass} into $\mathcal{L}^{(a_I \rightarrow \phi_b a_b)}_{\text{int,kin},(c)}$, we obtain
\begin{equation}
\mathcal{L}^{(a_I \rightarrow \phi_b a_b)}_{\text{int,kin},(c)} = - \frac{ \sqrt{3\gamma_I} |W_0|^2 \mathfrak{a}_I^2 \tau_I^{11/4}}{\tau_b^{15/4}}  \delta \phi_b \delta a_I \delta a_b.
\end{equation}

\section{Dynamics of the three-moduli system $\tau_b$, $\tau_I$ and $\tau_L$}\label{appendix_field_dynamics}

In this section we proceed in analogy to Appendix~\ref{sec:2_moduli_system}, however, we do not only consider the inflaton and volume moduli but instead the effective three-moduli system including also the loop-stabilized cycle as well as all corresponding axionic superpartners. This is again a simplified system, which does not take into account the additional small cycles $\tau_{s,i}$ which must be present to ensure the stability of the volume during inflation. However, as we will argue below, we do not expect that the inclusion of said small cycles, which play a role very similar to the inflaton $\tau_I$ except that they are not initially excited, would change our findings of this section significantly. As before, we follow the methodology of~\cite{Cicoli:2010ha} in that we first expand $V$ and $K$ about the LVS vacuum in the $\delta \tau_i$ and $\delta c_i$ and then diagonalize and canonically normalize the fields so that we can read off the coupling strengths. We set $M_P=1$ throughout this appendix.

\subsection{Basic definitions}

The volume and the K\"ahler potential have the form
\begin{equation}
\mathcal{V} = \tau_b^{3/2} - \gamma_I \tau_I^{3/2} - \gamma_L \tau_L^{3/2}, \quad K = - 2 \ln \left( \mathcal V + \frac{\xi}{2} \right) - \ln (S + \bar{S}) + K_\text{cs},
\end{equation}
where $S$ is the axio-dilaton, $K_\text{cs}$ depends on the complex-structure moduli and we have absorbed a factor $g_s^{-3/2}$ into $\xi$. The resulting K\"ahler metric and its inverse at leading order are given by
\begin{equation}
K_{ij} = \frac{\partial^2 K}{\partial T_i \partial \bar{T}_j} \approx \begin{pmatrix}
 \frac{3 }{4 \tau _b^2} & -\frac{9 \gamma _I
   \sqrt{\tau _I} }{8 \tau _b^{5/2}} & -\frac{9 \gamma _L \sqrt{\tau
   _L}}{8 \tau _b^{5/2}} \\
 -\frac{9 \gamma _I \sqrt{\tau _I} }{8 \tau _b^{5/2}} & \frac{3 \gamma
   _I }{8  \sqrt{ \tau _I} \tau _b^{3/2}} & \frac{9 \gamma _I
   \gamma _L \sqrt{\tau _I \tau _L}}{8 \tau _b^3} \\
 -\frac{9 \gamma _L \sqrt{\tau _L} }{8 \tau _b^{5/2}} & \frac{9 \gamma
   _I \gamma _L \sqrt{\tau _I \tau _L}}{8 \tau _b^3} & \frac{3 \gamma _L }{8 \sqrt{\tau _L} \tau _b^{3/2}}
\end{pmatrix}, \quad
(K^{-1})^{i j} \approx
\begin{pmatrix}
\frac{4 \tau_b^2}{3} & 4 \tau_b \tau_I & 4 \tau_b \tau_L \\
 4 \tau_b \tau_I & \frac{8 \sqrt{ \tau_I} \tau _b^{3/2}}{3 \gamma_I} & 4 \tau_I \tau_L \\
 4 \tau_b \tau_L & 4 \tau_I \tau_L & \frac{8 \sqrt{ \tau_L} \tau _b^{3/2}}{3 \gamma_L}
\end{pmatrix}, \label{33m}
\end{equation}
where $T_i = \tau_i + {\rm i} c_i$ with $i \in \{b,I,L\}$.
Since the model is constructed so that $\tau_L$ is not stabilized by non-perturbative effects, the superpotential is again only corrected by D3-brane instantons on $\tau_I$ and hence looks the same as in~\eqref{superpotential_2_moduli}
\begin{equation}
W = W_0 + A_I \text{e}^{- \mathfrak{a}_I T_I},
\end{equation}
with $\mathfrak{a}_I = 2 \pi$.

The total scalar potential consist of the usual LVS contribution as given in~\eqref{LVS_potential_full}, which is generated through non-perturbative effects on the superpotential and $\alpha'$ corrections to the K\"ahler potential, and a contribution induced by loop effects given in~\eqref{lco},
\begin{equation}
V = V_\text{LVS} (\mathcal{V}, \tau_I, c_I) + V_\text{loop} (\mathcal{V},\tau_L).
\end{equation}
The individual contributions read \cite{Balasubramanian:2005zx,Cicoli:2012sz}
\begin{align}
V_\text{LVS} &= \mathcal{V}^{-2} \left[ \frac{8 \tau_b^{3/2} \sqrt{\tau_I}}{3 \gamma_I} \mathfrak{a}_I^2 |A_I|^2 \text{e}^{-2\mathfrak{a}_I \tau_I} +4 \mathfrak{a}_I \tau_I \text{e}^{-\mathfrak{a}_I \tau_I} |A_I W_0| \cos\left( \mathfrak{a}_I c_I \right) \right] + \frac{3 |W_0|^2 \xi}{4 \mathcal{V}^3}, \\
V_\text{loop} &= \left( \frac{\mu_1}{\sqrt{\tau_L}} - \frac{\mu_2}{\sqrt{\tau_L} - \mu_3} \right) \frac{|W_0|^2}{\mathcal{V}^3}.
\end{align}
The relevant, perturbed lagrangian up to cubic order looks the same as in the two-moduli case but with the indices $i,j,k \in \{ b, I, L \}$,
\begin{align}
\mathcal L &= \langle K_{ij} \rangle \partial_\mu \delta \tau_i \partial^\mu \delta \tau_j + \langle \partial_{\tau_i} K_{jk} \rangle \delta \tau_i \partial_\mu \delta \tau_j \partial^\mu \delta \tau_k + \langle K_{ij} \rangle \partial_\mu \delta c_i \partial^\mu \delta c_j + \langle \partial_{\tau_i} K_{jk} \rangle \delta \tau_i \partial_\mu \delta c_j \partial^\mu \delta c_k   \nonumber \\
&\quad - \langle V \rangle - \frac{1}{2} \left\langle \frac{\partial^2 V}{\partial \tau_i \partial \tau_j} \right\rangle \delta \tau_i \delta \tau_j - \frac{1}{6} \left\langle \frac{\partial^3 V}{\partial \tau_i \partial \tau_j \partial \tau_k} \right\rangle \delta \tau_i \delta \tau_j \delta \tau_k  - \frac{1}{2} \left\langle \frac{\partial^2 V}{\partial c_i \partial c_j} \right\rangle \delta c_i \delta c_j \nonumber \\
&\quad - \frac{1}{2} \left\langle \frac{\partial^3 V}{\partial \tau_i \partial c_j \partial c_k} \right\rangle \delta \tau_i \delta c_j \delta c_k. \label{perturbed_lagrangian}
\end{align}

\subsection{Decay into moduli fields}

\subsubsection*{Diagonalization of fields}

Following~\cite{Cicoli:2010ha}, we again proceed by first transforming the $\delta \tau_i$ into canonical fields. The second derivative matrix w.r.t. the moduli $V_{ij} \equiv \partial^2 V / (\partial \tau_i \partial \tau_j)$ at leading order is given by
\begin{equation}
\left\langle V_{i j} \right\rangle \approx \begin{pmatrix}
\frac{9 (11 |W_0|^2 (\mu_1 \tilde{\mu}+\mu_2 \sqrt{\tau_L}) + 3 |W_0|^2 \gamma_I \tau_I^{3/2} \sqrt{\tau_L} \tilde{\mu}}{4 \sqrt{\tau_L} \tilde{\mu} \tau_b^{13/2}} &  -\frac{9 |W_0|^2 \gamma_I \mathfrak{a}_I \tau_I^{3/2}}{2 \tau_b^{11/2}} & \frac{9 |W_0|^2 (\mu_1 \tilde{\mu}^2 - \mu_2 \tau_L )}{4 \tilde{\mu}^2 \tau_L^{3/2} \tau_b^{11/2}} \\
\sim & \frac{3 |W_0|^2 \gamma_I \mathfrak{a}_I^2 \tau_I^{3/2}}{ \tau_b^{9/2}} & \frac{9 \gamma_I \sqrt{\tau_I} (|W_0|^2  (\mu_2 \tau_L - \mu_1 \tilde{\mu}^2) - 3 |W_0|^2 \gamma_L \tau_L^2  \tilde{\mu}^2 )}{4 \tilde{\mu}^2 \tau_L^{3/2} \tau_b^{6}} \\
\sim & \sim & \frac{|W_0|^2 ( 3 \mu_1 \tilde{\mu}^3 - \mu_2 (\mu_3 - 3 \sqrt{\tau_L}) \tau_L)}{4 \tilde{\mu}^3 \tau_L^{5/2} \tau_b^{9/2}}
\end{pmatrix}, \label{second_derivative_matrix}
\end{equation}
where $\tilde{\mu} \equiv \mu_3 - \sqrt{\tau_L}$.
Here we have again used the relations~\eqref{replacement_relations} after applying the second derivatives.

The transformation to the canonical fields reads
\begin{equation}
\begin{pmatrix}
\delta \tau_b \\
\delta \tau_I \\
\delta \tau_L
\end{pmatrix}
=
\begin{pmatrix}
\vec{v}_b \vphantom{\begin{pmatrix}
\delta \tau_b \\
\delta \tau_I \\
\delta \tau_L
\end{pmatrix}}
\end{pmatrix} \frac{\delta \phi_b}{\sqrt{2}}
+
\begin{pmatrix}
\vec{v}_I \vphantom{\begin{pmatrix}
\delta \tau_b \\
\delta \tau_I \\
\delta \tau_L
\end{pmatrix}}
\end{pmatrix} \frac{\delta \phi_I}{\sqrt{2}}
+
\begin{pmatrix}
\vec{v}_L \vphantom{\begin{pmatrix}
\delta \tau_b \\
\delta \tau_I \\
\delta \tau_L
\end{pmatrix}}
\end{pmatrix} \frac{\delta \phi_L}{\sqrt{2}} \label{transformation_to_canonical}
\end{equation}
or $\delta \tau_i = P_{i j} \delta \phi_j / \sqrt{2}$ where $P$ is the matrix that contains the vectors $\vec{v}_j$ as columns. These vectors are the eigenvectors of the matrix $(M^2)_{ij} \equiv \left\langle \left(K^{-1}\right)_{i k} V_{k j} \right\rangle / 2$, whose eigenvalues $m_i^2$ are the masses of the canonical fields $\delta \phi_i$, and they fulfill the normalization condition
\begin{equation}
\label{eigenvector_normalisation}
\vec{v}^\mathsf{T} _i \cdot \langle K \rangle \cdot \vec{v}_j \equiv P_{k i} \langle K_{k l} \rangle P_{l j} = \delta_{ij}.
\end{equation}

Next we have to calculate the eigenvectors $\vec{v}_j$. The $M^2$ matrix at leading order in the small parameter $\epsilon \equiv 1/\sqrt{\tau_b}$ is given by,
\begin{equation}
\left( M^2 \right)_{ij} \approx \begin{pmatrix}
m_{11} \epsilon^9 & m_{12} \epsilon^7 & m_{13} \epsilon^7 \\
m_{21} \epsilon^8 & m_{22} \epsilon^6 & m_{23} \epsilon^9 \\
m_{31} \epsilon^8 & m_{32} \epsilon^9 & m_{33} \epsilon^6
\end{pmatrix},
\end{equation}
where the $m_{ij}$ are expressions which do not depend on $\tau_b$ and which are given by
\begin{align}
m_{11} &= \frac{3 \left[ - 6 |W_0|^2 \tilde{\mu}^2 \sqrt{\tau_L} \gamma_I \mathfrak{a}_I \tau_I^{5/2}  + |W_0|^2 \left( 14 \tilde{\mu}^2 \mu_1 + 11 \tilde{\mu} \mu_2 \sqrt{\tau_L} - 3 \mu_2 \tau_L \right) \right]}{2 \tilde{\mu}^2 \sqrt{\tau_L}}, \label{expression_m11}\\
m_{12} &= 6 |W_0|^2 \gamma_I \mathfrak{a}_I^2 \tau_I^{5/2}, \\
m_{13} &= \frac{|W_0|^2 \left( 6 \tilde{\mu}^3 \mu_1 - 3 \tilde{\mu} \mu_2 \tau_L + \mu_2 \tau_L \left( -\mu_3 + 3 \sqrt{\tau_L} \right) \right)}{2 \tilde{\mu}^3 \tau_L^{3/2}}, \\
m_{21} &= - 6 |W_0|^2 \mathfrak{a}_I \tau_I^2, \\
m_{22} &= 4 |W_0|^2 \mathfrak{a}_I^2 \tau_I^2, \\
m_{23} &= - \frac{\tau_I \left[ 18 |W_0|^2 \gamma_L \tilde{\mu}^3 \tau_L^2 + |W_0|^2 \left( - 12 \tilde{\mu}^3 \mu_1 + \mu_2 \tau_L (\mu_3 - 3 \sqrt{\tau_L}) + 3 \tilde{\mu} (2 \tilde{\mu}^2 \mu_1 + \mu_2 \tau_L)  \right) \right]}{2 \tilde{\mu}^3 \tau_L^{3/2}}, \\
m_{31} &= \frac{3 |W_0|^2 \left( \tilde{\mu}^2 \mu_1 - \mu_2 \tau_L \right)}{\gamma_L \tilde{\mu}^2 \tau_L}, \\
m_{32} &= \frac{3 \gamma_I \left[ 2 |W_0|^2 \gamma_L   \tau_L^2 \tilde{\mu}^2 \mathfrak{a}_I^2  \tau_I^{5/2}  + |W_0|^2 \sqrt{\tau_I} \left( \mu_2 \tau_L - \mu_1 \tilde{\mu}^2 \right) \right]}{\gamma_L \tilde{\mu}^2 \tau_L}, \\
m_{33} &= \frac{|W_0|^2 \left( 3 \tilde{\mu}^3 \mu_1 + \mu_2 \tau_L \left( - \mu_3 + 3 \sqrt{\tau_L} \right) \right)}{3 \gamma_L \tilde{\mu}^3 \tau_L^2}. \label{expression_m33}
\end{align}
The eigenvalues and eigenvectors of $M^2$ at leading order in $\epsilon$ are given by
\begin{align}
m_{\tau_b}^2 &= \frac{- m_{13} m_{22} m_{31} - m_{12} m_{21} m_{33} + m_{11} m_{22} m_{33}}{m_{22} m_{33}} \epsilon^9, &&\vec{v}_1 =
\begin{pmatrix}
- \frac{m_{33}}{m_{31}} \epsilon^{-2} \\
\frac{m_{21} m_{33}}{m_{22} m_{31}} \\
1
\end{pmatrix}, \label{volume_modulus_mass}\\
m_{\tau_I}^2 &= m_{22} \epsilon^6, &&\vec{v}_2 =
\begin{pmatrix}
\frac{m_{12} (m_{22} - m_{33})}{m_{12} m_{31} + m_{22} m_{32}} \epsilon^{-2} \\
\frac{m_{22} (m_{22} - m_{33})}{m_{12} m_{31} + m_{22} m_{32}} \epsilon^{-3} \\
1
\end{pmatrix}, \label{inflaton_mass} \\
m_{\tau_L}^2 &= m_{33} \epsilon^6, &&\vec{v}_3 =
\begin{pmatrix}
\frac{m_{13}}{m_{33}} \epsilon \\
\frac{m_{13} m_{21} + m_{23} m_{33}}{m_{33} (-m_{22} + m_{33})} \epsilon^3 \\
1 
\end{pmatrix}. \label{loop_modulus_mass}
\end{align}
Note that $m_{\tau_I}^2 / m_{\tau_L}^2 = m_{22}/m_{33} \sim \mathfrak{a}_I^2 \tau_I^2 \tau_L^2 \gg 1$.
To fulfill the normalization conditions \eqref{eigenvector_normalisation}, we rescale the above eigenvectors
\begin{align}
\vec{v}_b &\equiv \frac{\vec{v}_1}{\sqrt{\vec{v}^\mathsf{T}_1 \cdot \langle K \rangle \cdot \vec{v}_1}} \approx - \frac{2 m_{31}}{\sqrt{3} m_{33}} \vec{v}_1 = \begin{pmatrix}
\frac{2 \tau_b}{\sqrt{3}} \\
-\frac{2 m_{21}}{\sqrt{3} m_{22}} \\
- \frac{2 m_{31}}{\sqrt{3} m_{33}}
\end{pmatrix},  \\
\vec{v}_I &\equiv \frac{\vec{v}_2}{\sqrt{\vec{v}^\mathsf{T}_2 \cdot \langle K \rangle \cdot \vec{v}_2}} \approx  \frac{4 (m_{12} m_{31} + m_{22} m_{32}) \tau_I^{1/4}}{\sqrt{6 \gamma_I} m_{22} (m_{22} - m_{33}) \tau_b^{3/4}} \vec{v}_2 = \begin{pmatrix}
\frac{4 m_{12} \tau_b^{1/4} \tau_I^{1/4}}{\sqrt{6 \gamma_I} m_{22}} \\
\frac{4 \tau_b^{3/4} \tau_I^{1/4}}{\sqrt{6 \gamma_I}} \\
\frac{4 (m_{12} m_{31} + m_{22} m_{32}) \tau_I^{1/4}}{\sqrt{6 \gamma_I} m_{22} (m_{22} - m_{33} ) \tau_b^{3/4}}
\end{pmatrix}, \label{transformation_vector_v_I}  \\
\vec{v}_L &\equiv \frac{\vec{v}_3}{\sqrt{\vec{v}^\mathsf{T}_3 \cdot \langle K \rangle \cdot \vec{v}_3}} \approx  \frac{4 \tau_b^{3/4} \tau_L^{1/4}}{\sqrt{6 \gamma_L}} \vec{v}_2 = \begin{pmatrix}
\frac{4 m_{13} \tau_b^{1/4} \tau_I^{1/4}}{\sqrt{6 \gamma_L} m_{33}} \\
\frac{4 (m_{13} m_{21} + m_{23} m_{33}) \tau_L^{1/4}}{\sqrt{6 \gamma_L} m_{33} (m_{33} - m_{22} ) \tau_b^{3/4}} \\
\frac{4 \tau_b^{3/4} \tau_L^{1/4}}{\sqrt{6 \gamma_L}}
\end{pmatrix} ,
\end{align}
where we used $m_{22}>m_{33}$ to specify some signs. Note that the first two components of $\vec{v}_b$ and $\vec{v}_I$ correspond exactly to the respective eigenvectors~\eqref{eigenvector_v_b_2_moduli_system} and~\eqref{eigenvector_v_I_2_moduli_system} of the 2-moduli system.

\subsubsection*{Coupling terms}

The kinetic and potential trilinear coupling terms are respectively given by
\begin{align}
\mathcal{L}_\text{int,kin} &= K_{mnp} \delta \tau_m (\partial_\mu \delta \tau_n) (\partial^\mu \delta \tau_p), \label{kinetic_coupling_general}\\
\mathcal{L}_\text{int,pot} &= -\frac{1}{6} V_{mnp} \delta \tau_m \delta \tau_n \delta \tau_p\,. \label{potential_coupling_general}
\end{align}
The third order derivatives $ K_{ijk} \equiv \langle \partial_{\tau_i} K_{j k} \rangle$ and $ V_{ijk} \equiv \langle \partial_{\tau_i} \partial_{\tau_j} \partial_{\tau_k} V \rangle$ at leading order read
\begin{align}
K_{bbb} &= -\frac{3}{2 \tau_b^3} ,\quad
K_{bbI} = \frac{45 \gamma_I \sqrt{\tau_I}}{16 \tau_b^{7/2}} , \quad
K_{bbL} = \frac{45 \gamma_L \sqrt{\tau_L}}{16 \tau_b^{7/2}} , \quad
K_{bII} = - \frac{9 \gamma_I}{16 \sqrt{\tau_I} \tau_b^{5/2}} , \nonumber \\
K_{bIL} &= - \frac{27 \gamma_I \gamma_L \sqrt{\tau_I \tau_L}}{8 \tau_b^4} , \quad
K_{bLL} = - \frac{9 \gamma_L}{16 \sqrt{\tau_L} \tau_b^{5/2}} , \quad
K_{III} = - \frac{3 \gamma_I}{16 \tau_I^{3/2} \tau_b^{3/2}}, \\
K_{IIL} &= \frac{9 \gamma_I \gamma_L \sqrt{\tau_L}}{16 \sqrt{\tau_I} \tau_b^3} ,\quad 
K_{ILL} = \frac{9 \gamma_I \gamma_L \sqrt{\tau_I}}{16 \sqrt{\tau_L} \tau_b^3} ,\quad
K_{LLL} = - \frac{3 \gamma_L}{16 \tau_L^{3/2} \tau_b^{3/2}} , \nonumber
\end{align}

\begin{align}
V_{bbb} &=  -\frac{9 \left[ 143 |W_0|^2 \left( \mu_1 \tilde{\mu} + \mu_2 \sqrt{\tau_L} \right) + 72 |W_0|^2 \gamma_I \tilde{\mu} \tau_I^{3/2} \sqrt{ \tau_L}  \right]}{8 \tilde{\mu} \sqrt{\tau_L}  \tau_b^{15/2}} , \nonumber\\
V_{bbI} &= \frac{99 \gamma_I |W_0|^2 \mathfrak{a}_I \tau_I^{3/2}}{4 \tau_b^{13/2}} , \nonumber\\
V_{bbL} &= -\frac{99 |W_0|^2 \left( \mu_1 \tilde{\mu}^2 - \mu_2 \tau_L \right)}{8 \tilde{\mu}^2 \tau_L^{3/2} \tau_b^{13/2}} , \nonumber\\
V_{bII} &= - \frac{27 \gamma_I |W_0|^2 \mathfrak{a}_I \sqrt{\tau_I}}{2 \tau_b^{11/2}} , \nonumber\\
V_{bIL} &= - \frac{27 \gamma_I \left[ 2 |W_0|^2 \gamma_L \tilde{\mu}^2 \mathfrak{a}_I \tau_I^{3/2} \tau_L^2  + 2 |W_0|^2 \sqrt{\tau_I} \left( \mu_2 \tau_L - \mu_1 \tilde{\mu}^2 \right) \right]}{4 \tilde{\mu}^2 \tau_L^{3/2} \tau_b^7} , \\
V_{bLL} &= -\frac{9 |W_0|^2 \left[ 3 \mu_1 \tilde{\mu}^3 + \mu_2  (- \mu_3 + 3 \sqrt{\tau_L}) \tau_L \right]}{8 \tilde{\mu}^3 \tau_L^{5/2} \tau_b^{11/2}} , \nonumber\\
V_{III} &= -\frac{9 \gamma_I |W_0|^2 \mathfrak{a}_I^3 \tau_I^{3/2}}{\tau_b^{9/2}} , \nonumber\\
V_{IIL} &= \frac{9 \gamma_I \left[  8 |W_0|^2 \gamma_L \tilde{\mu}^2 \mathfrak{a}_I^2 \tau_I^2 \tau_L^2 - |W_0|^2 (\mu_1 \tilde{\mu}^2 - \mu_2 \tau_L) \right]}{8 \tilde{\mu}^2 \sqrt{\tau_I} \tau_L^{3/2} \tau_b^{6}} , \nonumber\\
V_{ILL} &= \frac{9 \gamma_I \left[ -3 |W_0|^2 \gamma_L \tilde{\mu}^3 \sqrt{\tau_I} \tau_L^2  + |W_0|^2 \left( 3 \mu_1 \sqrt{\tau_I} \tilde{\mu}^3 + \mu_2 \sqrt{\tau_I} \tau_L (- \mu_3 + 3 \sqrt{\tau_L}) \right) \right]}{8 \tilde{\mu}^3 \tau_L^{5/2} \tau_b^6} ,\nonumber \\
V_{LLL} &= - \frac{3 |W_0|^2 \left[ 5 \mu_1 \tilde{\mu}^4 - \mu_2 \tau_L \left( \mu_3^2 - 4 \mu_3 \sqrt{\tau_L} + 5 \tau_L \right) \right]}{8 \tilde{\mu}^4 \tau_L^{7/2} \tau_b^{9/2}} . \nonumber
\end{align}
Here we have again used the relations \eqref{replacement_relations}, however, this time only after forming the third derivatives.

To obtain the kinetic couplings, we can insert the canonical fields $\delta \tau_i = P_{ij} \delta \phi_j / \sqrt{2}$ into \eqref{kinetic_coupling_general},
\begin{equation}
\mathcal{L}_\text{int,kin} = \frac{1}{2^{3/2}} K_{m n p} P_{mi} P_{nj} P_{pk} \delta \phi_i (\partial_\mu \delta \phi_j) (\partial^\mu \delta \phi_k).
\end{equation}
Eliminating the derivatives via the relation~\eqref{eliminate_partial_derivatives}, we obtain
\begin{equation}
\mathcal{L}_\text{int,kin} = \frac{1}{2^{5/2}} K_{m n p} P_{mi} P_{nj} P_{pk} \left( m_i^2 - m_j^2 - m_k^2 \right) \delta \phi_i \delta \phi_j \delta \phi_k. \label{coupling_term}
\end{equation}
For the potential couplings, after inserting the canonical fields into \eqref{potential_coupling_general}, we have
\begin{equation}
\mathcal{L}_\text{int,pot} = -\frac{1}{12 \sqrt{2}} V_{m n p} P_{mi} P_{nj} P_{pk} \delta \phi_i \delta \phi_j \delta \phi_k. \label{coupling_term_potential}
\end{equation}

We can now calculate the individual coupling terms:
\begin{itemize}
\item Decay $\delta \phi_I \rightarrow \delta \phi_b \delta \phi_b$: \\
Relevant are those terms in $\mathcal{L}_\text{int,kin}$ for which one of the three indices $i$, $j$, $k$ is an ``$I$" while the other two are a ``$b$". As in the previous section, from \eqref{coupling_term}, we see that all factors in $\mathcal{L}_\text{int,kin}$ are invariant under permutation of these indices except for the factor $(m_i^2 - m_j^2 - m_k^2)$. Again, $m_{\tau_I}^2 \gg m_{\tau_b}^2$, and hence this factor is dominated by $m_{\tau_I}^2$. Therefore, it only changes by a minus sign under permutation of $i$, $j$ and $k$, depending on which of the three indices takes on the value ``$I$". Summing up the three terms, two of which have a minus sign, we obtain,
\begin{equation}
\mathcal{L}^{(\phi_I \rightarrow \phi_b \phi_b)}_\text{int,kin} = -\frac{1}{2^{5/2}} K_{m n p} P_{mI} P_{nb} P_{pb} m_{\tau_I}^2 \delta \phi_I \delta \phi_b \delta \phi_b. \label{coupling_term_kinetic_bbI}
\end{equation}
The contraction is given by
\begin{align}
K_{m n p} P_{mI} P_{nb} P_{pb} &\approx K_{bbb} P_{bI} P_{bb} P_{bb} + P_{Ibb} P_{II} P_{bb} P_{bb} \\
&\approx \frac{ \sqrt{6\gamma_I} \tau_I^{3/4} }{2 \tau_b^{3/4}} \,.
\end{align}
Inserting this and \eqref{inflaton_mass} into \eqref{coupling_term_kinetic_bbI}, we obtain
\begin{equation}
\mathcal{L}^{(\phi_I \rightarrow \phi_b \phi_b)}_\text{int,kin} \approx -\frac{\sqrt{3 \gamma_I} |W_0|^2 \mathfrak{a}_I^2 \tau_I^{11/4}}{2 \tau_b^{15/4}} \delta \phi_I \delta \phi_b \delta \phi_b.
\end{equation}

For the potential coupling, only those terms from \eqref{coupling_term_potential} contribute where one of the indices $i$, $j$, $k$ takes on the value $I$ while the other two take on the value $b$. Following the same argument as before there are in total three such terms that are all equal and can be accounted for by a factor of 3
\begin{equation}
\mathcal{L}^{(\phi_I \rightarrow \phi_b \phi_b)}_\text{int,pot} = -\frac{1}{4 \sqrt{2}} V_{m n p} P_{mI} P_{nb} P_{pb} \delta \phi_I \delta \phi_b \delta \phi_b. \label{coupling_term_potential_bbI}
\end{equation}
The contraction reads
\begin{align}
V_{mnp} P_{mI} P_{nb} P_{pb} &\approx V_{Ibb} P_{II} P_{bb} P_{bb} + V_{III} P_{II} P_{Ib} P_{Ib} \\
&= \frac{4 \sqrt{6 \gamma_I} |W_0|^2 \mathfrak{a}_I \tau_I^{7/4}}{\tau_b^{15/4}}\,.
\end{align}
Inserting this into \eqref{coupling_term_potential_bbI}, we arrive at
\begin{equation}
\mathcal{L}^{(\phi_I \rightarrow \phi_b \phi_b)}_\text{int,pot} = - \frac{ \sqrt{3\gamma_I} |W_0|^2 \mathfrak{a}_I \tau_I^{7/4}}{\tau_b^{15/4}} \delta \phi_I \delta \phi_b \delta \phi_b\,.
\end{equation}
From this we conclude
\begin{equation}
\frac{\mathcal{L}^{(\phi_I \rightarrow \phi_b \phi_b)}_\text{int,kin}}{\mathcal{L}^{(\phi_I \rightarrow \phi_b \phi_b)}_\text{int,pot}} \approx \frac{\mathfrak{a}_I \tau_I}{2} \gg 1.
\end{equation}
\item Decay $\delta \phi_I \rightarrow \delta \phi_L \delta \phi_L$: \\
Analogously to the decay $\delta \phi_I \rightarrow \delta \phi_b \delta \phi_b$, now those terms from $\mathcal{L}_\text{int,kin}$ contribute for which one of the three indices $i$, $j$, $k$ is an ``$I$" while the other two are an ``$L$". With $m_{\tau_I}^2 \gg m_{\tau_L}^2$, the factor $(m_i^2 - m_j^2 - m_k^2)$ is again dominated by $m_{\tau_I}^2$. Thus we arrive at
\begin{equation}
\mathcal{L}^{(\phi_I \rightarrow \phi_L \phi_L)}_\text{int,kin} = -\frac{1}{2^{5/2}} K_{m n p} P_{mI} P_{nL} P_{pL} m_{\tau_I}^2 \delta \phi_I \delta \phi_L \delta \phi_L. \label{coupling_term_kinetic_LLI}
\end{equation}
The contraction is given by
\begin{align}
K_{m n p} P_{mI} P_{nL} P_{pL} &\approx K_{bLL} P_{bI} P_{LL} P_{LL} + K_{ILL} P_{II} P_{LL} P_{LL} + K_{LLL} P_{LI} P_{LL} P_{LL} \\
&\approx -\frac{ \sqrt{6\gamma_I} \tau_I^{3/4} }{ \tau_b^{3/4}},
\end{align}
where we used that $\mathfrak{a}_I \tau_I, \tau_L \gg 1$ and assumed that there is no finetuning of the parameter $\tilde{\mu} = \mu_3 - \sqrt{\tau_L}$. Inserting this and \eqref{inflaton_mass} into \eqref{coupling_term_kinetic_LLI}, we obtain
\begin{equation}
\mathcal{L}^{(\phi_I \rightarrow \phi_L \phi_L)}_\text{int,kin} \approx \frac{\sqrt{3 \gamma_I} |W_0|^2 \mathfrak{a}_I^2 \tau_I^{11/4}}{\tau_b^{15/4}} \delta \phi_I \delta \phi_L \delta \phi_L.
\end{equation}

For the potential coupling, we obtain analogously to \eqref{coupling_term_potential_bbI},
\begin{equation}
\mathcal{L}^{(\phi_I \rightarrow \phi_L \phi_L)}_\text{int,pot} = -\frac{1}{4 \sqrt{2}} V_{m n p} P_{mI} P_{nb} P_{pb} \delta \phi_I \delta \phi_L \delta \phi_L. \label{coupling_term_potential_LLI}
\end{equation}
The contraction reads
\begin{align}
V_{mnp} P_{mI} P_{nL} P_{pL} &\approx  \frac{2 \sqrt{6 \gamma_I} \tau_I^{3/4} \left[ - 3 |W_0|^2 \gamma_L \tilde{\mu}^4 \tau_L^2 + |W_0|^2 \left(  -4 \mu_1 \tilde{\mu}^4 + \mu_2 ( \mu_3^2 - 4 \mu_3 \sqrt{\tau_L} + 4 \tau_L) \tau_L \right)  \right]}{\gamma_L \tilde{\mu}^4 \tau_L^2 \tau_b^{15/4}}\,.
\end{align}
The potential coupling is then given by
\begin{align}
\mathcal{L}^{(\phi_I \rightarrow \phi_L \phi_L)}_\text{int,pot} = &-\frac{ \sqrt{3 \gamma_I} \tau_I^{3/4} \left[ - 3 |W_0|^2 \gamma_L \tilde{\mu}^4 \tau_L^2 + |W_0|^2 \left(  -4 \mu_1 \tilde{\mu}^4 + \mu_2 ( \mu_3^2 - 4 \mu_3 \sqrt{\tau_L} + 4 \tau_L) \tau_L \right)  \right]}{2 \gamma_L \tilde{\mu}^4 \tau_L^2 \tau_b^{15/4}} \nonumber\\ 
&\times \delta \phi_I \delta \phi_L \delta \phi_L.
\end{align}
Note that the term $\sim |W_0|^2$, which stems from $V_\text{LVS}$, is larger than the term $\sim |W_0|^2$, which stems from $V_\text{loop}$, by a factor $\sim \tau_L^2$. This confirms the correctness of our estimation~\eqref{inflaton_to_loop_decay_scaling_LVS_vs_loop_potential}.
Again, the kinetic decay dominates the potential one
\begin{equation}
\frac{\mathcal{L}^{(\phi_I \rightarrow \phi_L \phi_L)}_\text{int,kin}}{\mathcal{L}^{(\phi_I \rightarrow \phi_L \phi_L)}_\text{int,pot}} \sim \mathfrak{a}_I^2 \tau_I^2 \gg 1.
\end{equation}
Furthermore, we see that the potential couplings into the volume modulus and loop modulus differ by a factor
\begin{equation}
\frac{\mathcal{L}^{(\phi_I \rightarrow \phi_b \phi_b)}_\text{int,pot}}{\mathcal{L}^{(\phi_I \rightarrow \phi_L \phi_L)}_\text{int,pot}} \sim \mathfrak{a}_I \tau_I \gg 1.
\end{equation}

\item Decay $\delta \phi_I \rightarrow \delta \phi_b \delta \phi_L$:

For this decay, the relevant terms are those with the indices $i=I$, $j=b$ and $k=L$ as well as all permutations thereof. In total, there are $3! = 6$ permutations, which have all the same absolute value but with four of them coming with a minus sign compared to the other two. Thus, w.l.o.g. we fix $i=I$, $j=b$ and $k=L$ and assign a factor $2-4=-2$,
\begin{equation}
\mathcal{L}^{(\phi_I \rightarrow \phi_b \phi_L)}_\text{int,kin} = -\frac{1}{2^{3/2}} K_{m n p} P_{mI} P_{nb} P_{pL} m_{\tau_I}^2 \delta \phi_I \delta \phi_b \delta \phi_L. \label{coupling_term_kinetic_bLI}
\end{equation}
The contraction scales as
\begin{equation}
K_{m n p} P_{mI} P_{nb} P_{pL} \sim \tau_b^{-3/2},
\end{equation}
so that the total coupling term scales like
\begin{equation}
\mathcal{L}^{(\phi_I \rightarrow \phi_b \phi_L)}_\text{int,kin} \sim \tau_b^{-9/2} \delta \phi_I \delta \phi_b \delta \phi_L.
\end{equation}
This, the kinetic decay $\delta \phi_I \rightarrow \delta \phi_b \delta \phi_L$ is suppressed compared to the kinetic decays $\delta \phi_I \rightarrow \delta \phi_b \delta \phi_b$ and $\delta \phi_I \rightarrow \delta \phi_L \delta \phi_L$.

For the potential coupling, all 6 permutations of $i=I$, $j=b$ and $k=L$ are the same so that we obtain:
\begin{equation}
\mathcal{L}^{(\phi_I \rightarrow \phi_b \phi_L)}_\text{int,pot} = -\frac{1}{2 \sqrt{2}} V_{m n p} P_{mI} P_{nb} P_{pL} \delta \phi_I \delta \phi_b \delta \phi_L. \label{coupling_term_potential_bLI}
\end{equation}
Calculating the contractions, it turns out that we have
\begin{equation}
\mathcal{L}^{(\phi_I \rightarrow \phi_b \phi_L)}_\text{int,pot} \sim \tau_b^{-9/2} \delta \phi_I \delta \phi_b \delta \phi_L,
\end{equation}
which is also suppressed compared to the potential decays $\delta \phi_I \rightarrow \delta \phi_b \delta \phi_b$ and $\delta \phi_I \rightarrow \delta \phi_L \delta \phi_L$.
\end{itemize}

Note that the inclusion of the other small cycles $\tau_{s,i}$ does not alter the results for the couplings to moduli fields because it only changes the expression for $\xi$ in \eqref{replacement_relations}, which then becomes a sum over all small cycles including $\tau_I$. However, $\xi$ appears only in the components $V_{bb}$, $(M^2)_{11}$ and $V_{bbb}$ at leading order. Even though this induces a slight shift of the volume modulus mass \eqref{volume_modulus_mass}, none of these three components enters the trilinear coupling terms and hence they remain unaltered.

\subsection{Decay into axion fields}

\subsubsection*{Diagonalization of fields}

For the decay into the volume axion, we proceed analogously as for the decay into volume modulus. The second derivative matrix w.r.t. the axions at leading order is given by
\begin{equation}
\langle V^{(c)}_{i j} \rangle \equiv \left\langle \frac{\partial^2 V}{\partial c_i \partial c_j} \right\rangle = \begin{pmatrix}
0 & 0 & 0 \\
0 & \frac{3 \gamma_I |W_0|^2 \mathfrak{a}_I^2 \tau_I^{3/2} }{\tau_b^{9/2}} & 0 \\
0 & 0 & 0
\end{pmatrix},
\end{equation}
where we have, again, used the relations \eqref{replacement_relations} after applying the second derivatives. The transformation to canonical fields is given by
\begin{equation}
\begin{pmatrix}
\delta c_b \\
\delta c_I \\
\delta c_L
\end{pmatrix}
=
\begin{pmatrix}
\vec{w}_b \vphantom{\begin{pmatrix}
\delta c_b \\
\delta c_I \\
\delta c_L
\end{pmatrix}}
\end{pmatrix} \frac{\delta a_b}{\sqrt{2}}
+
\begin{pmatrix}
\vec{w}_I \vphantom{\begin{pmatrix}
\delta c_b \\
\delta c_I \\
\delta c_L
\end{pmatrix}}
\end{pmatrix} \frac{\delta a_I}{\sqrt{2}}
+
\begin{pmatrix}
\vec{w}_L \vphantom{\begin{pmatrix}
\delta c_b \\
\delta c_I \\
\delta c_L
\end{pmatrix}}
\end{pmatrix} \frac{\delta a_L}{\sqrt{2}} 
\end{equation}
or $\delta c_i = Q_{ij} \delta a_j / \sqrt{2}$ where $Q$ is the matrix that contains the vectors $\vec{w}_j$ as columns. They are the eigenvectors of the matrix $(M^2_{(c)})_{ij} \equiv \langle (K^{-1})_{ik} V^{(c)}_{kj} \rangle/2$ whose eigenvalues are the axion masses. The eigenvectors fulfill the normalization condition
\begin{equation}
\label{eigenvector_normalisation_axion}
\vec{w}^\mathsf{T} _i \cdot \langle K \rangle \cdot \vec{w}_j \equiv Q_{k i} \langle K_{k l} \rangle Q_{l j} = \delta_{ij}.
\end{equation}
The $M^2_{(c)}$ matrix at leading order is given by
\begin{equation}
(M^2_{(c)})_{ij} \approx \begin{pmatrix}
0 & \frac{6 \gamma_I |W_0|^2 \mathfrak{a}_I^2 \tau_I^{5/2}}{\tau_b^{7/2}} & 0 \\
0 & \frac{4 |W_0|^2 \mathfrak{a}_I^2 \tau_I^2}{\tau_b^3} & 0 \\
0 & \frac{6 \gamma_I |W_0|^2 \mathfrak{a}_I^2 \tau_I^{5/2} \tau_L}{\tau_b^{9/2}} & 0
\end{pmatrix}\,.
\end{equation}
The corresponding eigenvalues and eigenvectors are
\begin{align}
m_{c_b}^2 &= 0, && \vec{w}_1 =
\begin{pmatrix}
1 \\
0 \\
0
\end{pmatrix}, \\
m_{c_I}^2 &= \frac{4 |W_0|^2 \mathfrak{a}_I^2 \tau_I^2}{\tau_b^3}, && \vec{w}_2 =
\begin{pmatrix}
\tau_b/\tau_L \\
\frac{2 \tau_b^{3/2}}{3 \gamma_I \sqrt{\tau_I} \tau_L} \\
1
\end{pmatrix}, \\
m_{c_L}^2 &= 0, && \vec{w}_3 =
\begin{pmatrix}
0 \\
0 \\
1
\end{pmatrix}.
\end{align}
After rescaling to fulfill the normalization condition \eqref{eigenvector_normalisation_axion}, the normalized eigenvectors read
\begin{align}
\vec{w}_b &\equiv \frac{\vec{w}_1}{\sqrt{\vec{w}^\mathsf{T}_1 \cdot \langle K \rangle \cdot \vec{w}_1}} \approx \frac{2 \tau_b}{\sqrt{3}} \vec{w}_1 = \begin{pmatrix}
\frac{2 \tau_b}{\sqrt{3}} \\
0 \\
0
\end{pmatrix},  \\
\vec{w}_I &\equiv \frac{\vec{w}_2}{\sqrt{\vec{w}^\mathsf{T}_2 \cdot \langle K \rangle \cdot \vec{w}_2}} \approx  \frac{\sqrt{6 \gamma_I} \tau_I^{3/4} \tau_L}{\tau_b^{3/4}} \vec{w}_2 = \begin{pmatrix}
\sqrt{6 \gamma_I} \tau_I^{3/4} \tau_b^{1/4} \\
\frac{ 2 \sqrt{2} \tau_I^{1/4} \tau_b^{3/4}}{\sqrt{3 \gamma_I}} \\
\frac{\sqrt{6 \gamma_I} \tau_I^{3/4} \tau_L}{\tau_b^{3/4}}
\end{pmatrix},  \\
\vec{w}_L &\equiv \frac{\vec{w}_3}{\sqrt{\vec{w}^\mathsf{T}_3 \cdot \langle K \rangle \cdot \vec{w}_3}} \approx  \frac{2 \sqrt{2} \tau_L^{1/4} \tau_b^{3/4}}{\sqrt{3\gamma_L}} \vec{w}_3 = \begin{pmatrix}
0 \\
0 \\
\frac{2 \sqrt{2} \tau_L^{1/4} \tau_b^{3/4}}{\sqrt{3\gamma_L}}
\end{pmatrix} .
\end{align}

\subsubsection*{Coupling terms}

The kinetic and potential trilinear coupling terms are respectively given by
\begin{align}
\mathcal{L}_{\text{int,kin},(c)} &= \langle \partial_{\tau_m} K_{np} \rangle \delta \tau_m \partial_\mu \delta c_n \partial^\mu \delta c_p \nonumber \\
&= \frac{1}{2^{3/2}} K_{mnp} P_{mi} Q_{nj} Q_{pk} \delta \phi_i \partial_\mu \delta a_j \partial^\mu \delta a_k, \label{coupling_kinetic_general_axion} \\
\mathcal{L}_{\text{int,pot},(c)} &= -\frac{1}{2} \left\langle \frac{\partial^3 V}{\partial \tau_m \partial c_n \partial c_p} \right\rangle \delta \tau_m \delta c_n \delta c_p \nonumber \\
&= -\frac{1}{2^{5/2}} \left\langle \frac{\partial^3 V}{\partial \tau_m \partial c_n \partial c_p} \right\rangle P_{mi} Q_{nj} Q_{pk} \delta \phi_i \delta a_j \delta a_k \,. \label{coupling_potential_general_axion}
\end{align}
Let us first argue that the potential couplings to the volume and loop axions vanish: Since $V$ does not depend on $c_b$ or $c_L$ but only on $c_I$, the indices $n$ and $p$ in \eqref{coupling_potential_general_axion} must both take on the value ``$I$". However, the components $Q_{Ib}$ and $Q_{IL}$ vanish, so that there are no potential couplings $\sim \delta \phi_I \delta a_b \delta a_b$ or $\sim \delta \phi_I \delta a_L \delta a_L$.

The individual coupling terms are then calculated as:
\begin{itemize}
\item Decay $\delta \phi_I \rightarrow \delta a_b \delta a_b$: \\
Eliminating the derivatives by using \eqref{eliminate_partial_derivatives}, the kinetic coupling term becomes
\begin{equation}
\mathcal{L}^{(\phi_I \rightarrow a_b a_b)}_{\text{int,kin},(c)} = \frac{1}{2^{5/2}} K_{mnp} P_{mI} Q_{nb} Q_{pb} m_{\tau_I}^2 \delta \phi_I \delta a_b \delta a_b.
\end{equation}
Since $Q_{Ib} = Q_{Lb} = 0$, the indices $n$ and $p$ must take on the value $b$, so that we have
\begin{align}
\mathcal{L}^{(\phi_I \rightarrow a_b a_b)}_{\text{int,kin},(c)} &= \frac{1}{2^{5/2}} K_{mbb} P_{mI} Q_{bb} Q_{bb} m_{\tau_I}^2 \delta \phi_I \delta a_b \delta a_b \\
 &= \frac{1}{2^{5/2}} (K_{bbb} P_{bI} + K_{Ibb} P_{II} + K_{Lbb} P_{LI}) Q_{bb} Q_{bb} m_{\tau_I}^2 \delta \phi_I \delta a_b \delta a_b \\
&\approx \frac{ \sqrt{3\gamma_I} |W_0|^2 \mathfrak{a}_I^2 \tau_I^{11/4}}{2 \tau_b^{15/4}}  \delta \phi_I \delta a_b \delta a_b.
\end{align}

\item Decay $\delta \phi_I \rightarrow \delta a_L \delta a_L$: \\
Analogously to before we have
\begin{equation}
\mathcal{L}^{(\phi_I \rightarrow a_L a_L)}_{\text{int,kin},(c)} = \frac{1}{2^{5/2}} K_{mnp} P_{mI} Q_{nL} Q_{pL} m_{\tau_I}^2 \delta \phi_I \delta a_L \delta a_L.
\end{equation}
Since $Q_{bL} = Q_{IL} = 0$, the indices $n$ and $p$ must take on the value $L$, so that we have
\begin{align}
\mathcal{L}^{(\phi_I \rightarrow a_L a_L)}_{\text{int,kin},(c)} &= \frac{1}{2^{5/2}} K_{mLL} P_{mI} Q_{LL} Q_{LL} m_{\tau_I}^2 \delta \phi_I \delta a_L \delta a_L \\
 &= \frac{1}{2^{5/2}} (K_{bLL} P_{bI} + K_{ILL} P_{II} + K_{LLL} P_{LI}) Q_{LL} Q_{LL} m_{\tau_I}^2 \delta \phi_I \delta a_L \delta a_L \\
&= -\frac{ \sqrt{3\gamma_I} |W_0|^2 \mathfrak{a}_I^2 \tau_I^{11/4}}{\tau_b^{15/4}}  \delta \phi_I \delta a_L \delta a_L.
\end{align}

\item Decay $\delta \phi_I \rightarrow \delta a_b \delta a_L$: \\
For this decay, we have
\begin{equation}
\mathcal{L}^{(\phi_I \rightarrow a_b a_L)}_{\text{int,kin},(c)} = \frac{1}{2^{3/2}} K_{mnp} P_{mI} Q_{nb} Q_{pL} m_{\tau_I}^2 \delta \phi_I \delta a_b \delta a_L,
\end{equation}
where we have also assigned a factor 2 because their are two possibilities how $\delta \phi_i \delta a_j \delta a_k$ can contribute to this decay.
Again, since $Q_{Ib} = Q_{Lb} = Q_{bL} = Q_{IL} = 0$, the indices are forced to take on the values $n=b$ and $p=L$ so that we have
\begin{align}
\mathcal{L}^{(\phi_I \rightarrow a_b a_L)}_{\text{int,kin},(c)} &= \frac{1}{2^{3/2}} K_{mbL} P_{mI} Q_{bb} Q_{LL} m_{\tau_I}^2 \delta \phi_I \delta a_b \delta a_L \\
 &= \frac{1}{2^{3/2}} (K_{bbL} P_{bI} + K_{IbL} P_{II} + K_{LbL} P_{LI}) Q_{bb} Q_{LL} m_{\tau_I}^2 \delta \phi_I \delta a_b \delta a_L \\
&\approx 0.
\end{align}
Note that this zero only holds at leading order under the approximation that $\mathfrak{a}_I \tau_I, \tau_L \gg 1$ and that there is no fine-tuning of the parameter $\tilde{\mu} = \mu_3 - \sqrt{\tau_L}$. At the next-to-leading order, we would get a contribution that scales as,
\begin{equation}
\mathcal{L}^{(\phi_I \rightarrow a_b a_L)}_{\text{int,kin},(c)} \sim \tau_b^{-9/2} \delta \phi_I \delta a_b \delta a_L,
\end{equation}
which is suppressed compared to $\delta \phi_I \rightarrow \delta a_b \delta a_b$ and $\delta \phi_I \rightarrow \delta a_L \delta a_L$.
\end{itemize}

\subsection{Decays of the inflaton axion}

The trilinear couplings of the inflaton axion always involve exactly one other axion and one modulus field. The relevant coupling terms are given in~\eqref{coupling_kinetic_general_axion} and~\eqref{coupling_potential_general_axion}. In analogy to the argument above, the potential coupling terms~\eqref{coupling_potential_general_axion} vanish because the indices $n$ and $p$ must both take on the value ``$I$'' while on of the indices $j$ and $k$ must either take on the value ``$b$'' or ``$L$''. This gives rise to either a factor ``$Q_{Ib}$'' or ``$Q_{IL}$'', both of which are zero.

From the kinetic coupling terms of the inflaton axion are induced from~\eqref{coupling_kinetic_general_axion}. There are always two possibilities how $\delta \phi_i \partial_\mu \delta a_j \partial^\mu \delta a_k$ can contribute to a decay of $a_I$ corresponding to $j=I$ or $k=I$. Eliminating the derivatives using~\eqref{eliminate_partial_derivatives}, the individual coupling terms are given as follows:
\begin{itemize}
    \item Decay $\delta a_I \rightarrow \delta \phi_b \delta a_b$: \\
    Here we have
    \begin{equation}
\mathcal{L}^{(a_I \rightarrow \phi_b a_b)}_{\text{int,kin},(c)} = -\frac{1}{2^{3/2}} K_{mnp} P_{mb} Q_{nI} Q_{pb} m_{c_I}^2 \delta \phi_b \delta a_I \delta a_b.
\end{equation}
Since $Q_{Lb} = Q_{Ib} = 0$, the index $p$ is forced to take on the value ``$b$'' so that we obtain
    \begin{align}
        \mathcal{L}^{(a_I \rightarrow \phi_b a_b)}_{\text{int,kin},(c)} &= -\frac{1}{2^{3/2}} K_{mnb} P_{mb} Q_{nI} Q_{bb} m_{c_I}^2 \delta \phi_b \delta a_I \delta a_b \\
        &\approx -\frac{1}{2^{3/2}} \left(K_{bbb} P_{bb} Q_{bI} + K_{bIb} P_{bb} Q_{II} \right) Q_{bb} m_{c_I}^2 \delta \phi_b \delta a_I \delta a_b \\
        &\approx -\frac{\sqrt{3\gamma_I} |W_0|^2 \mathfrak{a}_I^2 \tau_I^{11/4}}{\tau_b^{15/4}} \delta \phi_b \delta a_I \delta a_b \,.
    \end{align}
    \item Decay $\delta a_I \rightarrow \delta \phi_b \delta a_L$: \\
    This decay is given by
    \begin{equation}
\mathcal{L}^{(a_I \rightarrow \phi_b a_L)}_{\text{int,kin},(c)} = -\frac{1}{2^{3/2}} K_{mnp} P_{mb} Q_{nI} Q_{pL} m_{c_I}^2 \delta \phi_b \delta a_I \delta a_L.
    \end{equation}
    Since $Q_{IL} = Q_{bL} = 0$, the index $p$ is forced to take on the value ``$L$'' so that we have
    \begin{align}
        \mathcal{L}^{(a_I \rightarrow \phi_b a_L)}_{\text{int,kin},(c)} &= -\frac{1}{2^{3/2}} K_{mnp} P_{mb} Q_{nI} Q_{LL} m_{c_I}^2 \delta \phi_b \delta a_I \delta a_L \\
        &\sim \tau_b^{-9/2} \delta \phi_b \delta a_I \delta a_L\,.
    \end{align}
    \item Decay $\delta a_I \rightarrow \delta \phi_L \delta a_b$: \\
    The coupling terms read
    \begin{equation}
\mathcal{L}^{(a_I \rightarrow \phi_L a_b)}_{\text{int,kin},(c)} = -\frac{1}{2^{3/2}} K_{mnp} P_{mL} Q_{nI} Q_{pb} m_{c_I}^2 \delta \phi_L \delta a_I \delta a_b.
    \end{equation}
    Here the index $p$ is again forced to take on the value ``$b$'' and we obtain
    \begin{align}
        \mathcal{L}^{(a_I \rightarrow \phi_L a_b)}_{\text{int,kin},(c)} &= -\frac{1}{2^{3/2}} K_{mnp} P_{mL} Q_{nI} Q_{pb} m_{c_I}^2 \delta \phi_L \delta a_I \delta a_b \\
        &\sim \tau_b^{-9/2} \delta \phi_L \delta a_I \delta a_b \,.
    \end{align}
    \item Decay $\delta a_I \rightarrow \delta \phi_L \delta a_L$: \\
    For this decay we have
    \begin{equation}
\mathcal{L}^{(a_I \rightarrow \phi_L a_L)}_{\text{int,kin},(c)} = -\frac{1}{2^{3/2}} K_{mnp} P_{mL} Q_{nI} Q_{pL} m_{c_I}^2 \delta \phi_L \delta a_I \delta a_L.
    \end{equation}
    The index $p$ must take on the value ``$L$'' and the coupling terms are given by
    \begin{align}
        \mathcal{L}^{(a_I \rightarrow \phi_L a_L)}_{\text{int,kin},(c)} &= -\frac{1}{2^{3/2}} K_{mnp} P_{mL} Q_{nI} Q_{LL} m_{c_I}^2 \delta \phi_L \delta a_I \delta a_L \\
        &\approx - \frac{1}{2^{3/2}} \left( K_{LbL} P_{LL} Q_{bI} + K_{LIL} P_{LL} Q_{II} + K_{LLL} P_{LL} Q_{LI} \right) Q_{LL} m_{c_I}^2 \delta \phi_L \delta a_I \delta a_L \\
        &\approx \frac{2 \sqrt{3 \gamma_I} |W_0|^2 \mathfrak{a}_I^2 \tau_I^{11/4}}{\tau_b^{15/4}} \delta \phi_L \delta a_I \delta a_L \,.
    \end{align}
\end{itemize}

\subsection{Decay rates}

To obtain the corresponding decay rates, we use the standard formula
\begin{equation}
    \Gamma = \frac{1}{S} \int \frac{|\mathcal{M}|^2}{2E} d\text{LIPS},
\end{equation}
where $S$ is the symmetry factor, $E$ is the energy of the decaying particle, $|\mathcal{M}|^2$ is the matrix element squared and $d\text{LIPS}$ is an element of Lorentz invariant phase space. The decays we consider can be grouped into two categories, either with two identical decay products or with two different ones. The corresponding interaction terms are schematically of the form
\begin{equation}
    \mathcal{L}_\text{A} \supset g_\text{A} \varphi_\text{A} \psi_\text{A}^2, \quad \mathcal{L}_\text{B} \supset g_\text{B} \varphi_\text{B} \psi_\text{B} \chi_\text{B},
\end{equation}
where we assume that the decaying particle $\varphi$ is much heavier than the decay products $\psi$ and $\chi$, i.e. $m_{\varphi_\text{A}} \gg 2 m_{\psi_\text{A}}$ and $m_{\varphi_\text{B}} \gg m_{\psi_\text{B}} + m_{\chi_\text{B}}$. A crucial difference between the two categories lies in their respective symmetry factors and matrix elements. For category A, we have $S=2$ and $|\mathcal{M}|^2 = 4 g_\text{A}^2$ whereas for category B, we have $S=1$ and $|\mathcal{M}|^2 = g_\text{B}^2$. This results in the following decay rates for the two categories,
\begin{equation}
    \Gamma_{\varphi_\text{A} \rightarrow \psi_\text{A} \psi_\text{A}} = \frac{g_\text{A}^2}{8 \pi m_{\varphi_\text{A}}}, \quad \Gamma_{\varphi_\text{B} \rightarrow \psi_\text{B} \chi_\text{B}} = \frac{g_\text{B}^2}{16 \pi m_{\varphi_\text{B}}}.
\end{equation}
By reading off the respective couplings $g$ from the trilinear coupling terms above, we can easily obtain the corresponding decay rates.

The relevant decays of the inflaton fall into category A. For the kinetic decay into the volume modulus we have
\begin{equation}
|\mathcal M_1|^2 = \frac{3 \gamma_I |W_0|^4 \mathfrak{a}_I^4 \tau_I^{11/2}}{\tau_b^{15/2}}, \quad m_{\tau_I}^2 = \frac{4 |W_0|^2 \mathfrak{a}_I^2 \tau_I^2}{\tau_b^3},
\end{equation}
and thus obtain
\begin{equation}
\Gamma_1 \equiv \Gamma^\text{kin}_{\phi_I \rightarrow \phi_b \phi_b} \approx \frac{3 \gamma_I |W_0|^3 \mathfrak{a}_I^3 \tau_I^{9/2}}{64 \pi \mathcal V^4}.
\end{equation}
Analogously, for the potential decay into the loop modulus, the matrix element squared is given by
\begin{equation}
|\mathcal{M}_2|^2 = \left( -\frac{ \sqrt{3 \gamma_I} \tau_I^{3/4} \left[ - 3 |W_0|^2 \gamma_L \tilde{\mu}^4 \tau_L^2 + |W_0|^2 \left(  -4 \mu_1 \tilde{\mu}^4 + \mu_2 ( \mu_3^2 - 4 \mu_3 \sqrt{\tau_L} + 4 \tau_L) \tau_L \right)  \right]}{\gamma_L \tilde{\mu}^4 \tau_L^2 \tau_b^{15/4}} \right)^2
\end{equation}
and the decay rate by
\begin{equation}
\Gamma_2 \equiv \Gamma^\text{pot}_{\phi_I \rightarrow \phi_L \phi_L} = \frac{ 3 \gamma_I \sqrt{\tau_I} \left[ - 3 |W_0|^2 \gamma_L \tilde{\mu}^4 \tau_L^2 + |W_0|^2 \left(  -4 \mu_1 \tilde{\mu}^4 + \mu_2 ( \mu_3^2 - 4 \mu_3 \sqrt{\tau_L} + 4 \tau_L) \tau_L \right)  \right]^2}{64 \pi \gamma_L^2 |W_0| \tilde{\mu}^8  \mathfrak{a}_I \tau_L^4 \mathcal{V}^4}.
\end{equation}
Note that $\Gamma_1 /\Gamma_2 \sim \mathfrak{a}_I^4 \tau_I^4 \sim (\ln \mathcal{V})^4 \gg 1$.
Comparing the coupling functions, all other decay rates of inflaton decays can be related to $\Gamma_1$ and $\Gamma_2$ as given in Table~\ref{tab:decay_rates}.

Likewise, the decay rates of the inflaton axion fall into category $B$ and can also be related to $\Gamma_1$ as given in Table~\ref{tab:decay_rates}.

\section{Parameter dependence of cosmological results}
\label{app:parameters}

In this appendix we collect the dependence of the results of Section~\ref{sec:cosmofinal} on those parameters of the model which are were set to unity in Section~\ref{sec:cosmofinal} to simplify the discussion.

Defining the parameter
\begin{equation}
    \alpha=(2 \pi)^{3/2} \frac{\gamma_{I} |W_0|^3 \mathfrak{a}_{I}^{3}\tau^{9/2}_{I}}{(\ln[\mathcal{V}/W_0])^{9/2}},
\end{equation}
the total decay rate of the inflaton~\eqref{eq:totdecay} becomes
\begin{equation}
\label{eq:alpha}
    \Gamma_{\phi_I}^{\rm tot}= \alpha \frac{159}{64 \sqrt{2} \pi^{5/2}}{\mathcal{V}}^{-4}\left(\ln[{\mathcal{V}}/W_{0}]  \right)^{9/2} M_{P}, 
\end{equation}
where we have again chosen $N_g = 12$. Moreover, to take into account the model-dependent $\mathcal{O}(1)$-factor in~\eqref{axion_decay_constant}, we introduce $\sigma$, which is defined by the relation
\begin{equation}
    f_{a}\equiv \sigma\frac{M_{P}}{\sqrt{2}\pi \tau_L^{1/4}\sqrt{\mathcal{V}}}, \label{eq_fa_explicit}
\end{equation}
and according to~\eqref{fasl} is given by
\begin{equation}
    \sigma=\sqrt{\frac{3\gamma_{v}}{8}}.
\end{equation}
As in Section~\ref{sec:cosmo}, we also use the expression for the inflation scale
\begin{equation}
H_{I}=\kappa\frac{M_{P}}{{\mathcal{V}}^{3/2}}, \label{eq_HI_explicit}
\end{equation}
with $\kappa^2 \equiv \beta |W_0|^2$.

In Section~\ref{sec:cosmofinal} we have set the above parameters to $\alpha = \sigma = \kappa = 1$. To get an impression of the dependence on these parameters, in the following we calculate the power-like dependence of the various phenomenological bounds on $\alpha$, $\sigma$, $\kappa$. We neglect all logarithmic effects. To achieve this, we first disregard the logarithmic dependence in~\eqref{eq:alpha}, finding that the reheating temperature scales as
\begin{equation}
    T_r\sim\left(\frac{g_*\pi^2}{90}\right)^{-1/4}\sqrt{\Gamma_{\phi_I}^{\rm tot} M_P}\sim \sqrt{\alpha} / \mathcal{V}^{2}.
\end{equation}
Together with~\eqref{eq_fa_explicit} and~\eqref{eq_HI_explicit}, we can use this to equip the bounds~\eqref{eq_constraints_final_high_V} -- \eqref{eq_constraints_final_high_theta} and~\eqref{eq_constraints_final_low_V} -- \eqref{eq_constraints_final_low_theta} with an approximate scaling in the parameters $\alpha, \sigma$ and $\kappa$.

For the high-reheating-temperature case with $T_r \gg 1 \, \text{GeV}$ we then have the more general expressions
\begin{alignat}{2}
    (\sigma^{-10/31}\kappa^{24/31})\,1\times 10^7 \;\;&\lesssim {\mathcal{V}}&&\lesssim \;\;2\times 10^{10}\, (\alpha^{1/4})\\
    (\sigma^{36/31}\kappa^{-12/31})\,9\times 10^{13}\,{\rm GeV} \;\;& \gtrsim f_{a} && \gtrsim\;\; 2\times 10^{12}\,{\rm GeV}\, (\alpha^{-1/8}\sigma)\\
   (\sigma^{-36/31}\kappa^{12/31})\, 6\times 10^{-8}\,{\rm eV}\;\;&\lesssim m_{a}&&\lesssim\;\;  3\times 10^{-6}\,{\rm eV}\,(\alpha^{1/8}\sigma^{-1})\\
  (\alpha^{1/2}\sigma^{20/31}\kappa^{-48/31})\,  2\times 10^6\,{\rm GeV} \;\;&\gtrsim T_r &&\gtrsim\;\;  1\,{\rm GeV}\\
      (\sigma^{15/31} \kappa^{-5/31})\,7\times 10^{7}\,{\rm GeV} \;\;&\gtrsim H_I &&\gtrsim\;\;  1 \times 10^3\,{\rm GeV}\, (\alpha^{-3/8} \kappa), \\
    (\sigma^{-21/31}\kappa^{7/31})\, 0.1 \;\;&\lesssim \theta&&\lesssim\;\; 0.5 \, (\alpha^{7/96}\sigma^{-7/12}).
\end{alignat}
As before the left-hand side is the constraint from isocurvature and the right-hand side the boundary of the regime where the axion starts oscillating in a radiation dominated era.
The volume modulus mass at the upper end of the volume range is
\begin{equation}
    m_{b}\sim 660 \,{\rm GeV}\,(\alpha^{-3/8}). 
\end{equation}

\bigskip

For lower reheating temperatures $T_R \ll 300 \, \text{MeV}$ we find
\begin{alignat}{2}
    (\alpha^{1/4})\, 5\times 10^{10} \;\;&\lesssim {\mathcal{V}}&&\lesssim\;\; 8\times 10^{10}\,(\alpha^{4/19}\sigma^{6/19})\\
    (\alpha^{-1/8}\sigma)\,1.4\times 10^{12} \,{\rm GeV}\;\;& \gtrsim f_{a} && \gtrsim\;\; 1.0 \times 10^{12}\,{\rm GeV}\,(\alpha^{-2/19}\sigma^{16/19}) \\
    (\alpha^{1/8}\sigma^{-1})\,4\times 10^{-6}\,{\rm eV}\;\;&\lesssim m_{a}&&\lesssim\;\; 6 \times 10^{-6}\,{\rm eV}\,(\alpha^{2/19}\sigma^{-16/19})\\
    300\,{\rm MeV} \;\;&\gtrsim T_r &&\gtrsim\;\; 150 \,{\rm MeV}\,(\alpha^{3/38}\sigma^{-12/19}) \\
    (\alpha^{-3/8} \kappa) \, 250\,{\rm GeV} \;\;&\gtrsim H_I &&\gtrsim\;\; 100\,{\rm GeV} \, (\alpha^{-6/19} \sigma^{-9/19} \kappa), \\
    (\alpha^{3/32} \sigma^{-3/4})\, 1 \;\;&\lesssim \theta&&\lesssim\;\;  3 .
\end{alignat}
Now the left-hand side ensures that we are in the regime where oscillations start during the phase where the equation of state is matter-like. The right-hand side arises from achieving the full DM density without tuning the initial value $\theta_i$.
At the largest volume the mass of the volume modulus is given by
\begin{equation}
    m_{b}\sim 110 \,{\rm GeV} (\alpha^{-6/19}\sigma^{-9/19}).
\end{equation}

\begin{appendix}

\end{appendix}

\renewcommand{\em}{}
\bibliographystyle{utphys}
\addcontentsline{toc}{section}{References}
\bibliography{refs}

\end{document}